\documentclass[%
 reprint,
superscriptaddress,
nofootinbib,
 amsmath,amssymb,
 aps,
]{revtex4-1}

\usepackage{graphicx}
\usepackage{dcolumn}
\usepackage{bm}
\usepackage{lipsum}
\usepackage{amsmath,amssymb}
\usepackage{caption}
\usepackage{subcaption}

\usepackage[utf8]{inputenc}
% ...
\usepackage[T1]{fontenc}
% ...

\usepackage{verbatim}
\usepackage[left=2cm, right=2cm, top=2cm]{geometry}

\usepackage{epigraph}
\usepackage{siunitx}
\usepackage[english]{babel}
\usepackage{listings}
\usepackage{graphicx}
\usepackage[utf8]{inputenc}
\usepackage{listings}
\usepackage{matlab-prettifier}
\usepackage{xcolor}
\usepackage{amsfonts}
\usepackage{amsmath}
\usepackage{amssymb}
\usepackage{amsthm}
\usepackage{array}
\newcolumntype{P}[1]{>{\centering\arraybackslash}p{#1}}
\newcolumntype{M}[1]{>{\centering\arraybackslash}m{#1}}
\usepackage{caption}
\usepackage{subcaption}
\usepackage{epigraph}
\usepackage{csquotes}
\usepackage{braket}
\usepackage{xcolor}
\usepackage[colorlinks = true,
            linkcolor = blue,
            urlcolor  = blue,
            citecolor = blue,
            anchorcolor = blue]{hyperref}
\usepackage{comment}
\usepackage{hyperref}
\usepackage{color}

\newcommand{\mt}[1]{\textrm{\tiny #1}}
\newcommand{\tr}{\tilde{r}}
\newcommand{\tit}{\tilde{t}}

\newcommand{\Eps}{\dot{\mathcal{E}}}
\newcommand{\ob}[1]{\textcolor{blue}{\bf [[[Ollie: #1]]]}}

\newcommand{\js}[1]{\textcolor{red}{\bf [[[Joan: #1]]]}}

\newcommand{\BHPT}{\href{http://bhptoolkit.org}{BHPT}}

\begin{document}

\preprint{APS/123-QED}

\title{Constraining the spin parameter of near-extremal black holes using LISA}

\author{Ollie Burke}
\email[]{ollie.burke@aei.mpg.de}
\author{Jonathan R. Gair}

\affiliation{%
Max Planck Institute for Gravitational Physics (Albert Einstein Institute), Am M\"{u}hlenberg 1, Potsdam-Golm 14476, Germany}
\affiliation{%
  School of Mathematics, University of Edinburgh, James Clerk Maxwell Building, Peter Guthrie Tait Road, Edinburgh EH9 3FD, UK}
  \author{Joan Sim\'on}
 \affiliation{%
  School of Mathematics, University of Edinburgh, James Clerk Maxwell Building, Peter Guthrie Tait Road, Edinburgh EH9 3FD, UK}   
  \author{Matthew C. Edwards}
 \affiliation{%
  School of Mathematics, University of Edinburgh, James Clerk Maxwell Building, Peter Guthrie Tait Road, Edinburgh EH9 3FD, UK}   
\affiliation{%
  Department of Statistics,
  University of Auckland,
  38 Princes Street, Auckland 1010, New Zealand}

\begin{abstract}
 
We describe a model that generates first order adiabatic EMRI waveforms for quasi-circular equatorial inspirals of compact objects into rapidly rotating (near-extremal) black holes. Using our model, we show that LISA could measure the spin parameter of near-extremal black holes (for $a \gtrsim 0.9999$) with extraordinary precision,  $\sim$ 3-4 orders of magnitude better than for moderate spins, $a \sim 0.9$. Such spin measurements would be one of the tightest measurements of an astrophysical parameter within a gravitational wave context. Our results are primarily based off a Fisher matrix analysis, but are verified using both frequentist and Bayesian techniques. We present analytical arguments that explain these high spin precision measurements. The high precision arises from the spin dependence of the radial inspiral evolution, which is dominated by geodesic properties of the secondary orbit, rather than radiation reaction. High precision measurements are only possible if we observe the exponential damping of the signal that is characteristic of the near-horizon regime of near-extremal inspirals. Our results demonstrate that, if such black holes exist, LISA would be able to successfully identify rapidly rotating black holes up to $a = 1-10^{-9}$ , far past the Thorne limit of $a = 0.998$.  
\end{abstract}

\pacs{Valid PACS appear here}
\maketitle

\section{Introduction}

Extreme mass ratio inspirals (EMRIs) are one of the most exciting possible sources of gravitational radiation for the space-based detector LISA \cite{danzmann1996lisa}, but also one of the most challenging to model and extract from the data stream. An EMRI involves the slow inspiral of a stellar-origin compact object (CO) of mass $\mu \sim 10M_{\odot}$ into a massive black hole in the centre of a galaxy. For a central black hole with mass $M \sim 10^{(5-7)}M_{\odot}$, EMRIs emit gravitational waves (GWs) in the mHz frequency band and so are prime sources for the LISA detector. EMRIs begin when, as a result of scattering processes in the stellar cluster surrounding the massive black hole, the CO becomes gravitationally bound to the primary. The subsequent inspiral of the CO towards the horizon of the primary is driven by radiation reaction through the emission of gravitational waves. EMRI waveforms are very complicated and EMRIs can be present in the LISA frequency band for several years prior to plunge, so modelling the full observable signal is a complex task \cite{amaro2015research}. EMRI orbits are expected to be both eccentric and inclined even up to the last few cycles before plunging into the primary black hole. For these reasons, EMRIs pose a challenging problem for both waveform modellers~\cite{barack2004lisa,chua2017augmented,amaro2015research} and data analysts.

This same complexity also makes EMRIs one of the richest sources of gravitational waves. Typically an EMRI will be observable for $1/(\text{mass ratio}) \sim 10^{5-7}$ cycles before plunge and the emitted gravitational waves thus provide a very precise map of the spacetime geometry of the primary hole \cite{hughes2006sort,gair2013testing,glampedakis2006mapping,barack2007using}. Through accurate detection and parameter inference, one can conduct tests of general relativity to very high precision \cite{yunes2012gravitational,gair2013testing}. 

It is well known that the information about the source is carried through the time evolution of the phase in a gravitational wave \cite{PhysRevLett.70.2984, 2000PhRvD..62l4021F}. The slow evolution of EMRIs means that a large number of cycles can be observed during the inspiral, which will provide constraints on the parameters of the source with remarkable precision \cite{amaro2007intermediate,amaro2012low}. Previous work has indicated that LISA will be able to place constraints on the dimensionless Kerr spin parameter, $a$, of the primary black hole in an EMRI, at the level of 1 part in $10^{4}$ for moderately spinning, $a\sim 0.9$, primaries~\cite{gair2008black,chua2017augmented,barack2004lisa,babak2007kludge}. In this paper, we explore how well LISA will be able to measure the spin parameter for very rapidly rotating black holes, i.e., systems in which the spin parameter is close to the maximum value allowed by general relativity.

Super massive black holes with a large spin parameter are abundant throughout our universe. Observations indicate that massive BHs reside in the centres of most galaxies, where these black holes are known to accrete  matter and hence are predicted to have very high spins \cite{reynolds2013spin,risaliti2013rapidly,2000PhRvD..62l4021F,trakhtenbrot2014most,mcclintock2006spin,gou2011extreme}. The dimensionless Kerr spin parameter of a Kerr black hole, cannot exceed $1$, since the resulting spacetime contains a naked singularity no longer encased within a well defined horizon.  Thorne~\cite{1974ApJ...191..507T} showed that a moderately spinning black hole cannot be spun up by thin-disc accretion above a spin of $a\approx 0.998$. However, in principle primordial black holes could be formed with spins exceeding that value~\cite{arbey2020evolution}. ``Near-extremal'' black holes with spins close to the limit of $a=1$ have interesting properties and we focus our attention on these here.


This past decade, researchers \cite{2018CQGra..35j4002C,2016CQGra..33o5002G,2015PhRvD..92f4029G,2018arXiv180403704C,van2015near,Porfyriadis:2014fja,Hadar:2014dpa,Hadar:2015xpa,van2015near,Hadar:2016vmk,compere2018_NHEK} have explored the rich properties of near-extremal EMRIs. The gravitational radiation emitted from these systems is unique, and would prove a smoking gun for the existence of these near-extremal systems (see \cite{2016CQGra..33o5002G}). In this paper, we show qualitatively that the inspiraling dynamics of the compact object into an near-extremal massive black hole is very different from that into a moderately spinning black hole, and these differences are reflected in the emitted gravitational waves. As such, in order to detect and correctly perform parameter estimation on these near-extremal sources, it is essential to update our family of waveform models to include them. We will argue throughout this work that, if observed, near-extreme black holes offer \emph{significantly} greater precision measurements on the Kerr spin parameter than moderately spinning systems. In particular, LISA will have the capability to successfully conclude whether the central object in an EMRI system is truly a near-extremal black hole. Thus, if near-extremal black holes exist, LISA observations of EMRIs may be one of the best ways to find them.

In this paper we will consider only EMRIs on circular and equatorial orbits around near-extremal primary black holes. This choice is made primarily for computational convenience, but there are also astrophysical scenarios that produce such systems. As discussed in~\cite{Levin:2006uc}, compact objects can form within accretion disks around massive black holes. When these objects fall into the central black hole, the resultant EMRI will be circular and equatorial. Super-Eddington accretion can provide a means to spin up a black hole past the Thorne limit \cite{skadowski2011spinning}, and so it is not unreasonable to expect that this EMRI formation channel would be more important for near-extremal systems. The standard EMRI formation channel, involving capture of a compact object via scattering interactions, tends to form EMRIs with moderate initial eccentricities. However, this eccentricity decreases during the inspiral due to the emission of gravitational radiation~\cite{peters1963gravitational}. This decrease in eccentricity continues until the orbit reaches a critical radius at which is starts to increase again~\cite{glampedakis2002zoom,kennefick1998stability}. The critical radius moves closer to the last stable orbit as the spin parameter increases and for near-extremal systems is located within the regime where transition from inspiral to plunge occurs~\cite{buonanno2000transition,ori2000transition}. Additionally, the increase in eccentricity is a subdominant effect throughout the transition regime~\cite{burke2019transition}. As the spin increases, we therefore expect that for an object captured at a fixed radius, the amount of eccentricity dissipated before the critical radius increases, and the eccentricity gained after the critical radius decreases. Therefore, even in the standard capture picture it is reasonable to assume the eccentricity is small at the end of the inspiral. We will show in this paper that very precise measurements of spin for near-extremal systems are possible, but this precision comes from observation of features~\cite{2016CQGra..33o5002G} in the final phase of the inspiral, which is where the near-circular assumption is most likely to be valid.

The main results of the paper are given in figures \ref{fig:Fisher_Matrix_Results} and \ref{fig:Fisher_Matrix_Whole_Param_Plots} in section \ref{sec:Fish_Matrix_Numerics}. Readers who wish to understand why near-extremal systems offer greater  precision spin measurements than moderate spin systems should direct their attention to section \ref{subsec:FisherMatrixParadigm}.

This paper is organised as follows. In section \ref{sec:KerrGeodesicsEquatorialPlane}, we set notation  and discuss the trajectory of a compact object on a circular and equatorial orbit around a near-extremal Kerr BH. In section \ref{subsec:FisherMatrixParadigm}, we show that the spin dependence of kinematical quantities appearing in the radial evolution rather than radiation-reactive effects dominate the spin precision measurements for near-extremal EMRI systems. Our Teukolsky based waveform generation schemes are outlined in section \ref{sec:Waveform_Gen}. We discuss prospects for detection in section \ref{sec:Detectability}, arguing that LISA is more sensitive to heavier mass systems $M\sim 10^{7}M_{\odot}$ than lighter systems $M\sim10^{6}M_{\odot}$. Our Fisher Matrix results are presented in section \ref{sec:Fish_Matrix_Numerics}. Here we show that we can constrain the spin parameter $\Delta a \sim 10^{-10}$, even when correlations amongst other parameters are taken into account. Finally, in section~\ref{MCMC}, we perform a Bayesian analysis to verify our Fisher matrix results, before finishing with conclusions and outlooks in section~\ref{conclusion}. 

\section{Background}\label{sec:KerrGeodesicsEquatorialPlane}

We consider the inspiral of a secondary test particle of mass $\mu$ on a circular, equatorial orbit around a primary super massive Kerr black hole with mass $M$ and Kerr spin parameter $a\lesssim 1$ where the mass ratio is assumed small $\eta = \mu/M \ll 1$. The secondary is on a prograde orbit aligned with the rotation of the primary black hole with $a > 0$ and dimensionful angular momentum $L > 0$. Unless stated otherwise, throughout this paper any quantity with an over-tilde is \emph{dimensionless}, e.g., $\tilde{r} = r/M$ and $\tilde{t} = t/M$ etc. The one exception is the dimensionless spin parameter, which we denote by $a$ \emph{without a tilde}. Quantities with an over-dot will denote coordinate time derivatives, e.g., $\dot{r}={\rm d}r/{\rm d}t$. We use geometrised units such that $G = c = 1$. 

In Boyer-Lindquist \cite{boyer1967maximal} coordinates, the metric of a Kerr black hole for $\theta = \pi/2$ is given by
\begin{multline}\label{Metric:KerrMetric}
g = -\left(1-\frac{2}{\tilde{r}}\right)d\tilde{t}^{2} + \frac{\tilde{r}^2}{\tilde{\Delta}}d\tilde{r}^{2} + \\  \left(\tilde{r}^{2} + a^{2} + \frac{2a^{2}}{\tilde{r}}\right) d\phi^{2}  - \frac{4a}{\tilde{r}} d\tilde{t}d\phi,
\end{multline}
where $\tilde{\Delta} = \tilde{r}^2 - 2\tilde{r} + a^2$ and $a$ is the dimensionless spin parameter introduced earlier. This is related to the mass, $M$, and angular momentum, $J$, of the Kerr black hole via $a=J/M$ and lies in the range $a \in [0,1]$. The event horizon is located on the surface defined by $\tilde{\Delta} = 0$, when
\begin{equation}
\tilde{r}_{+} = 1 + \sqrt{1-a ^{2}}.
\end{equation}
Introducing an extremality parameter $\epsilon \ll 1$
\begin{equation}\label{extremalityparameter}
\epsilon = \sqrt{1- a^{2}},
\end{equation}
the event horizon is at
\begin{equation}\label{eventhoriz}
\tilde{r}_{+} = 1 + \epsilon .
\end{equation}
The trajectory of the secondary confined to the equatorial plane of a central Kerr hole is governed by the Kerr geodesic equations \cite{carter1968global}
\begin{align}
\left(\tilde{r}^{2}\frac{d\tilde{r}}{d\tilde{\tau}}\right)^{2} &= [\tilde{E}(\tilde{r}^2 + a^2) - a\tilde{L}]^{2} - \Delta [(\tilde{L} - a\tilde{E})^2 -  \tilde{r}^2] \nonumber \\
\tilde{r}^{2}\frac{d\phi}{d\tilde{\tau}} &= -(a\tilde{E} - \tilde{L}) + \frac{a}{\tilde{r}}(\tilde{E}[\tilde{r}^2 + a^2] - a\tilde{L})\nonumber \\
\tilde{r}^{2}\frac{d\tilde{t}}{d\tilde{\tau}} & = -a(a\tilde{E}-\tilde{L}) + \frac{\tilde{r}^2 + a^2}{\Delta}(\tilde{E}[\tilde{r}^2 + a^2] - a\tilde{L})\nonumber ,
\end{align}
in which $\tilde{\tau}$ denotes the proper-time coordinate for the inspiraling object. The dimensionless conserved quantities $\tilde{E} = E/\mu$ and $\tilde{L} = L/(M\mu)$ are related to the energy, $E$, and angular momentum, $L$, measured at infinity. For the circular and equatorial orbits considered here, the energy $\tilde{E}$ and angular frequency $\tilde{\Omega}$ can be expressed analytically
\begin{align}
\tilde{E} &=  \frac{1 - 2/\tilde{r} + \tilde{a}/\tilde{r}^{3/2}}{\sqrt{1-3/\tilde{r} + 2a/\tilde{r}^{3/2}}} \label{Energy} \\
\tilde{\Omega}  &  = \frac{1}{\tilde{r}^{3/2} + a}, \label{AngularMomentum} 
\end{align}
in which the dimensionless angular frequency $\tilde{\Omega}$ is defined through $\Omega = \tilde{\Omega}/M = d\phi/d t$. 

Circular orbits exist only outside the innermost stable circular orbit (ISCO). For radii smaller than the ISCO, the secondary will start to plunge towards the horizon of the primary. The ISCO for equatorial orbits is at~\cite{bardeen1972rotating}
\begin{subequations}\label{AllEqns}
\begin{align}
\tilde{r}_{\text{isco}} &= 3 + Z_{2} - [(3-Z_{1})(3+Z_{1}+2Z_{2})]^{1/2} \label{iscolocationr1}\\
Z_{1} &= 1 + (1-a^2)^{1/3}[(1+ a)^{1/3} + (1-a)^{1/3}]\\
Z_{2} &= (3 a^{2} + Z_{1}^{2})^{1/2}. \label{iscolocationr3}
\end{align}
\end{subequations}
For near extremal orbits, using \eqref{extremalityparameter} and \eqref{AllEqns}, and expanding for $\epsilon \ll 1$, we obtain
\begin{equation}
\tilde{r}_{\text{isco}} = 1 + 2^{1/3}\epsilon^{2/3} + \mathcal{O}(\epsilon^{4/3}),\end{equation}
and deduce
\begin{equation}\label{Orbital Separation}
|\tilde{r}_{\text{isco}} - \tilde{r}_{+}| = \mathcal{O}(\epsilon^{2/3}), \ \text{for} \ \epsilon \ll 1.
\end{equation}
The radial coordinate separation between the ISCO and horizon is determined by the spin parameter. In the limit, $\epsilon \rightarrow 0$, then $\tilde{r}_{\text{isco}} \rightarrow \tilde{r}_{+} \rightarrow 1$ in Boyer-Lindquist coordinates.

\subsection{Radiation Reaction}
To compute circular and equatorial \emph{adiabatic} inspirals, a detailed knowledge of the radial self force is required (see, for example, \cite{barack2018self} for a detailed review). In this paper, we will work at leading order, including the radiative (dissipative) part of the radial self force at first order, but neglecting first order conservative effects and all second order in mass-ratio effects. The first order dissipative force can be computed by solving the Teukolsky equation~\cite{1973ApJ...185..635T}. The rate of emission of energy is given by
\begin{align}\label{FluxExpression}
\langle-\dot{\tilde{E}}\rangle = \langle\dot{\tilde{E}}_{GW}\rangle &= \langle\dot{\tilde{E}}^{\infty}\rangle + \langle\dot{\tilde{E}}^{H}\rangle \nonumber\\
& = 2\sum_{l = 2}^{\infty}\sum_{m = 1}^{l}(\langle\dot{\tilde{E}}^{\infty}_{ lm}\rangle + \langle\dot{\tilde{E}}^{H}_{ lm}\rangle) .
\end{align}
where $\langle\dot{\tilde{E}}_{GW}\rangle = \langle-(u^{t})^{-1}F_{t}\rangle$, $\langle \cdot \rangle$ denotes coordinate time averaging over several periods of the orbit. 
The quantity $u^{t}$ is the $t$ component of the four velocity and $F_{t}$ the $t$ component of the gravitational self force at first order in the mass ratio $\eta$. This expression is valid only if $\eta \ll 1$, that is, when orbits evolve \emph{adiabatically} such that the timescale on which the orbital parameters evolve is much longer than the orbital period.

The fluxes $\langle\dot{\tilde{E}}^{\infty}\rangle$ and $\langle\dot{\tilde{E}}^{H}\rangle$ denote the (dimensionless and orbit averaged) dissipative fluxes of gravitational radiation emitted towards infinity and towards the horizon respectively. From here on, we shall drop the angular brackets $\langle \dot{E} \rangle \rightarrow \dot{E}$, to avoid cumbersome notation. The quantities $|m| \leq l$ are angular multiples which appear in the decomposition of the emitted radiation into a sum of spheroidal harmonics. 
The components of the fluxes $\dot{\tilde{E}}$ are obtained by numerical solution of the Teukolsky equation sourced by a point particle (the secondary). There exists an open source code in the Black Hole Perturbation Toolkit (\BHPT) \cite{BHPToolkit} to do this for circular and equatorial orbits - specifically the \hyperlink{https://bhptoolkit.org/Teukolsky/}{Teukolsky} package.

The enhancement of symmetry in the near horizon geometry of extreme Kerr \cite{Bardeen:1999px} provides an additional tool to compute the fluxes $\dot{\tilde{E}}$ analytically from first principles. See \cite{2018CQGra..35j4002C,Porfyriadis:2014fja,Hadar:2014dpa,Hadar:2015xpa,2015PhRvD..92f4029G,2015PhRvD..92f4029G,van2015near,Hadar:2016vmk,2016CQGra..33o5002G,2018arXiv180403704C,compere2018_NHEK} for a description of this work.  For circular equatorial orbits near the horizon of a near-extremal black hole, there is a remarkably simple approximation for the total flux~\cite{2015PhRvD..92f4029G}, which takes the form 
\begin{equation}\label{NielsEnergy}
\dot{\tilde{E}}^{\text{NHEK}}_{\text{GW}}= \eta(\tilde{C}_{H} + \tilde{C}_{\infty})(\tilde{r}-\tilde{r}_{+})/\tilde{r}_{+}, \quad \frac{\tilde{r} - \tilde{r}_{+}}{\tilde{r}_{+}} \ll 1.
\end{equation}
The quantities $\tilde{C}_{H}$ and $\tilde{C}_{\infty}$ are constants representing the emission towards the horizon and infinity respectively. These constants are given analytically in equations (76) and (77) of \cite{2015PhRvD..92f4029G} and codes in the \BHPT \ can be used to evaluate them.  Numerically evaluating them and summing the contribution of the first $|m| \leq l = 30$ modes gives $\tilde{C}_{H} \approx 0.987$ and $\tilde{C}_{\infty} \approx -0.133$.  Eq.(\ref{NielsEnergy}) is useful when working within the near-horizon geometry of the rapidly rotating hole, but it breaks down far from the horizon and extra terms would be required to compute reliable fluxes.

All the numerical work presented in this paper, which is found in section~\ref{sec:Detectability} onwards, will use the exact fluxes obtained from \BHPT. However, to understand our numerical results, we develop a set of new analytic tools in sections~\ref{sec:linear_ODE_section} and ~\ref{subsec:analytic_fish}. These will partially make use of the leading contribution to \eqref{NielsEnergy}
\begin{equation}\label{eq:approx:NHEK}
    \dot{\tilde{E}}^{\text{NHEK}}_{GW} \approx \eta (\tilde{C}_{H} + \tilde{C}_{\infty})x \ , \quad x = \tilde{r} - 1 \ll 1.
\end{equation}
This differs from \eqref{NielsEnergy} by $\mathcal{O}(\epsilon)$ contributions since $\tr - 1$ measures the BL radial distance to the extremal horizon and \emph{not} the radial distance to the near-extremal horizon $\tr_+$. The approximation~\eqref{eq:approx:NHEK} can be derived from first principles by solving the Teukolsky equation in the NHEK region\footnote{This follows by measuring the radial distance to the extremal horizon by $\lambda$, defined through  $\tr = \frac{\tr_++\tr_-}{2} + \lambda \tr$, and then taking the decoupling limit $\lambda\to 0$.}. Our numerical analysis based on the \BHPT, suggests the spin dependence of certain observables, to be discussed in section \ref{subsec:analytical_traj}, is better captured by \eqref{eq:approx:NHEK}. Table \ref{CH:8 Table 2:ComparisonNHEKFormula} compares the flux at $\tilde{r}_{\text{isco}}$ computed using \BHPT \  to that obtained from the near-extremal approximations of Eq.(\ref{NielsEnergy}) and Eq.(\ref{eq:approx:NHEK}).
This table corroborates that \eqref{eq:approx:NHEK} is a good approximation to the total energy flux, particularly in the limit as $a \to 1$, where it outperforms the full expression, \eqref{NielsEnergy}.

\begin{table*}
\centering
\begin{tabular}{cccccc}
   &  &  \\ \hline
 $a$ & $\dot{\tilde{E}}_{\text{Exact}}/\eta$ &  $\dot{\tilde{E}}^{+}_{\text{NHEK}}/\eta$ &  $\dot{\tilde{E}}_{\text{NHEK}}/\eta$
&
 $|\dot{\tilde{E}}^{+}_{\text{NHEK}}-\dot{\tilde{E}}_{\text{Exact}}|/\eta$ & $|\dot{\tilde{E}}_{\text{NHEK}}-\dot{\tilde{E}}_{\text{Exact}}|/\eta$ \\ \hline
 $1-10^{-5}$ & 0.0264197 & 0.0261523 & 0.0300885 & 0.0002674 & 0.0036688 \\
 $1-10^{-6}$ & 0.0129344 & 0.0125200 & 0.0137455 & 0.0004143 & 0.0008111\\
 $1-10^{-7}$ & 0.0061516 & 0.0059484 &0.006333 & 0.0002031 & 0.0001814\\
 $1-10^{-8}$ & 0.0028875 & 0.0028082 & 0.0029294 & 0.0000793 &  0.0000419\\
 $1-10^{-9}$ & 0.0013472 & 0.0013193 & 0.0013575 & 0.0000280 & 0.0000103\\
 $1-10^{-10}$ &0.0006273 & 0.0006176  & 0.0006296 & 0.0000097 & 0.0000023\\
 $1-10^{-11}$ &0.0002915 & 0.0002883 & 0.0002922 & 0.0000031 & 0.0000007\\
 $1-10^{-12}$& 0.0001354 & 0.0001344 &  0.0001356 & 0.0000009 & 0.0000002 \\ \hline
\end{tabular}
\caption{NHEK fluxes at the ISCO computed using the approximations Eq.~\eqref{NielsEnergy} (denoted $\dot{\tilde{E}}^{+}_{\text{NHEK}}$) and Eq.~\eqref{eq:approx:NHEK} (denoted $\dot{\tilde{E}}_{\text{NHEK}}$), and computed exactly using \href{http://bhptoolkit.org}{BHPT} (denoted $\dot{\tilde{E}}_{\text{Exact}}$ and based on the first thirty $m$ and $l$ modes).}
\label{CH:8 Table 2:ComparisonNHEKFormula}
\end{table*}

\subsection{Inspiral and Waveform}
The radial evolution of the secondary can be found by taking a coordinate time derivative of the circular energy relation \eqref{Energy} 
\begin{equation}\label{eq1:orbital radius evolution}
\frac{d\tilde{r}}{d\tilde{t}} = - \frac{P_{\mt{GW}}}{\partial_{\tilde{r}}\tilde{E}}
\end{equation}
where we defined $P_{\mt{GW}} := \dot{\tilde{E}}_{\mt{GW}}$. As the ISCO is approached, the denominator  $\partial_{\tilde{r}}\tilde{E}$ tends to zero, marking a break down of the quasi-circular approximation. The ODE \eqref{eq1:orbital radius evolution} is easily numerically integrated given an expression for the flux $P_{\text{GW}}$.

The outgoing gravitational wave energy flux measured at infinity has a harmonic decomposition \cite{2000PhRvD..62l4021F}
\begin{equation}\label{Flux_Harmonic_m}
\dot{\tilde{E}}^{\infty}_{m} = \mathcal{A}_{m}\eta\tilde{\Omega}^{2 + 2m/3}\Eps^{\infty}_{m},
\end{equation}
where
\begin{equation}
\mathcal{A}_{m} = \frac{2(m+1)(m+2)(2m)!m^{2m-1}}{(m-1)[2^{m}m!(2m+1)!!]^{2}},
\end{equation}
and $(2m+1)!! = (2m+1)(2m-1)\ldots 3\cdot 1$. Here $\Eps^{\infty}_{m}$ is the relativistic correction to the Newtonian expression for the flux in harmonic $m$.

In this work, we shall consider two different waveform models. For the analytic discussion in section \ref{subsec:FisherMatrixParadigm}, we will use the waveform model in ~\cite{2000PhRvD..62l4021F}, whereas for the numerical analysis in later sections, we will use the full Teukolsky based waveform.

Let us first review the main features of the model discussed in ~\cite{2000PhRvD..62l4021F} for the waveform observed by the detector in the source frame. This model is written
\begin{equation}\label{WaveformAllHarmonics}
h(\tilde{t};\boldsymbol{\theta}) \approx \sum_{m = 2}^{\infty} h_{\text{o},m}\sin(2\pi \tilde{f}_{m} \tilde{t} + \phi_{0})\,.
\end{equation}
Some remarks are in order. First, we ignore the $m=1$ contribution since, as argued in ~\cite{2000PhRvD..62l4021F}, this is subleading to the $m\geq 2$ contributions. Second, the amplitude $h_{o,m} = \sqrt{\langle h_{+m}^2 + h_{\times m}^2\rangle}$ corresponds to the root mean square (RMS) amplitude of gravitational waves emitted towards infinity in harmonic $m$. These are averaged $\langle \cdot \rangle$ over the viewing angle\footnote{The (normalised) spheroidal harmonics $_{-2}S^{am\tilde{\Omega}}_{ml}$ are integrated out over the 2-sphere.} and over the period of the waves. Third, the oscillatory phase depends on the initial phase $\phi_{0}$ and the frequency $\tilde{f}_m$ of each waveform harmonic is given by
\begin{equation}\label{eq:frequency}
    \tilde{f}_{m} = \frac{m}{2\pi}\tilde{\Omega}\,.
\end{equation}
The relation between the RMS amplitude and the outgoing radiation flux in harmonic $m$ is
\begin{equation}\label{eq:rms_amplitude}
h_{o,m} = \frac{2\sqrt{\eta\dot{\tilde{E}}^{\infty}_{m}}}{m\tilde{\Omega}\tilde{D}}
\end{equation}
where $\tilde{D} = D/M$ is the distance to the source from earth. Using Eq.\eqref{Flux_Harmonic_m}, we can rewrite $h_{o,m}$ as
\begin{equation}\label{RMSStrain}
h_{o,m} = \sqrt{\frac{8(m+1)(m+2)(2m)!m^{2m-1}}{(m-1)[2^{m}m!(2m+1)!!]^{2}}\Eps^{\infty}_{m}} \frac{\sqrt{\eta}}{\tilde{D}}\tilde{\Omega}^{m/3}
\end{equation}
for $m \geq 2$. We note that the effect of the averaging is that this waveform model does not represent the waveform measured by any physical observer. However, it captures the main physical features of the waveform which encode information about the source parameters.

Given the nature of our orbits, our parameter space will only be six dimensional $\boldsymbol{\theta} = \{\tilde{r}_{0},a,\mu,M,\phi_{0},\tilde{D}\}$, where $\tilde{r}_0$ stands for the initial size of the circular orbit. We stress this waveform model does not include the LISA response functions, which affect the amplitude evolution of the signal and induce modulations, due to Doppler shifting, through the motion of the LISA spacecraft \cite{cutler1998angular,cornish2003lisa}. Since these response functions do not depend on the \emph{intrinsic} parameters of the system that we are most interested in, we omit these here and, consequently, they will also be omitted in our analytic discussion based on this waveform model.

Let us now review the full Teukolsky based waveform model that we will use in our numerical study. This is given by
\begin{equation}\label{eq:Teuk_Waveform_gen}
    h_{+} - ih_{\times} = \frac{\mu}{\tilde{D}}\sum_{ml}\frac{1}{m^{2}\tilde{\Omega}^{2}}\ G_{ml}\exp(-i[\phi_{0} + m\tilde{\Omega} \tilde{t}])
\end{equation}
where
\begin{equation}
    G_{ml} = _{-2}S^{am\tilde{\Omega}}_{ml}(\theta)\exp(i\phi)Z^{\infty}_{ml}(\tilde{r},a)
\end{equation}
depends on the radial Teukolsky amplitude at infinity, $Z^{\infty}_{ml}(\tilde{r},a)$, and the viewing angle $(\theta,\phi)$. The latter dependence is through the spin-weight minus 2 spheroidal harmonics $_{-2}S^{am\tilde{\Omega}}_{ml}(\theta,\phi) = _{-2}S^{am\tilde{\Omega}}_{ml}(\theta)\exp(i\phi)$. 
This work will consider two viewing angles: face on $(\theta,\phi) = (0,0)$ and edge on $(\theta,\phi) = (\pi/2,0)$. Using the identities
\begin{align}
    _{-2}S^{a(-m)\tilde{\Omega}}_{(-m)l}(\pi/2,0) &= (-1)^{l}_{-2}\bar{S}^{am\tilde{\Omega}}_{ml}(\pi/2,0) \\
    Z^{\infty}_{(-m)l} &= (-1)^{l}\bar{Z}^{\infty}_{ml}
\end{align}
where barred quantities are complex conjugates, we can write equation \eqref{eq:Teuk_Waveform_gen} as
\begin{equation}
   h_{+} = \frac{2\mu}{D}\left(\sum_{m = 1}^{\infty}\frac{1}{m^{2}\tilde{\Omega}^{2}}                     \exp(-i[\phi_{0} + m\tilde{\Omega} \tilde{t}])\sum_{l = m}^{\infty} G_{ml}\right),  
\end{equation}
for the edge-on case, and as
\begin{equation}
 h_{+} - ih_{\times} \approx \frac{\mu}{4\tilde{\Omega}^{2}D}  G_{22} \exp(-i[\phi_{0} + 2\tilde{\Omega} \tilde{t}]),
\end{equation}
for the face-on case. Note we have neglected higher order $l$ modes with $m = 2$ fixed in the last equation since the Teukolsky amplitudes $Z^{\infty}_{l2}$ for $l > 2$ are negligible in comparison to the dominant quadrupolar $l = m = 2$ mode. Figure 1. in~\cite{2015PhRvD..92f4029G} further justifies our claim that higher order $m$ modes when $l=2$ can be ignored for face-on sources. Furthermore, the only spheroidal harmonics that are non-vanishing at $\theta = 0$ are those with $m = -s$, or $m = 2$ ~\cite{hughes2000evolution,del20033}.


To perform our numerics, the spheroidal harmonics are calculated using the \hyperlink{https://bhptoolkit.org/SpinWeightedSpheroidalHarmonics/}{SpinWeightedSpheroidalHarmonics} mathematica package in the \BHPT, whereas the Teukolsky amplitudes $Z^{\infty}_{ml}$ are calculated using the \hyperlink{https://bhptoolkit.org/Teukolsky/}{Teukolsky} package from the same toolkit. For reasons discussed later, we generate both amplitudes and spheroidal harmonics for a fixed spin parameter $a = 1-10^{-9}$. For the remainder of this study, we will only consider the plus polarised signal $h(t;\boldsymbol{\theta})\equiv h_{+}(t;\boldsymbol{\theta})$ for the face-on and edge-on observations.

We finish this waveform discussion with a comment regarding the relation between the two models considered in this work. The (dimensionful) Teukolsky amplitudes are related to the energy flux for each $(l,m)$ mode by
\begin{equation}\label{eq:dotE_inf_Teuk_Amplitudes}
    \dot{E}^{\infty}_{lm} = \frac{|Z^{\infty}_{lm}|^{2}}{4\pi m^{2}\Omega^{2}}.
\end{equation}
Hence $|\tilde{Z}^{\infty}_{ml}| \sim M\tilde{\Omega}\sqrt{\eta\dot{\tilde{E}}_{lm}}.$ Averaging over the sky and ignoring the phase of the radial amplitude $Z^{\infty}_{ml}$, the Teukolsky waveform \eqref{eq:Teuk_Waveform_gen} reduces to \eqref{WaveformAllHarmonics}. Our numerical results indicate that the spin precision measurements are driven by the radial trajectory given by \eqref{eq1:orbital radius evolution}, which is common to both \eqref{WaveformAllHarmonics} and \eqref{eq:Teuk_Waveform_gen}, while not being largely influenced by the spin dependence on the waveform amplitude. Given this fact and since it is analytically much easier to analyse the waveform model \eqref{WaveformAllHarmonics}, this is the one being discussed in the analytics section \ref{subsec:FisherMatrixParadigm} to explain the increase in the spin precision measurement for near-extremal primaries.

\subsection{Gravitational Wave Data Analysis}

The data stream of a gravitational wave detector, $d(t) = h(t;\boldsymbol{\theta}) + n(t)$, is typically assumed to consist of probabilistic noise $n(t)$ and (one or more) deterministic signals, $h(t;\boldsymbol{\theta})$, with parameters $\boldsymbol{\theta}$. Assuming that the noise is a weakly stationary Gaussian random process with zero mean, the likelihood is \cite{whittle:1957}
\begin{equation}\label{log-likelihood-function}
    p(d|\boldsymbol{\theta}) \propto \exp\left[-\frac{1}{2}(d - h|d - h)\right]
\end{equation}
with inner product
\begin{equation}\label{eq:Inner_Prod}
(b|c) = 4\mathrm{Re}\int_{0}^{\infty}\frac{\hat{b}(f)\hat{c}^{*}(f)}{ S_{n}(f) }df. 
\end{equation}
Here $\hat{b}(f)$ is the continuous time fourier transform (CTFT) of the signal $b(t)$ and $S_{n}(f)$ the power spectral density (PSD) of the noise. Here we use the analytical PSD given by Eq.(1) in \cite{2018arXiv180301944C}. We do not include the galactic foreground noise in the PSD to ensure all noise realisations generated through $S_{n}(f)$ are stationary. This is not a serious restriction as for the sources we consider here, the majority of the GW emission is at higher frequencies where the galactic foreground lies below the level of instrumental noise in the detector.

The optimal signal to noise ratio (SNR) of a source is given by
\begin{equation}\label{continuous_SNR}
    \rho^{2} = (h|h).
\end{equation}
This is the SNR that would be realised in a matched filtering search and is a measure of the brightness, or ease of detectability, of a gravitational wave signal. Measures of the similarity of two template waveforms $h_{1}:=h(t;\boldsymbol{\theta}_{1})$ and $h_{2}:=h(t;\boldsymbol{\theta}_{2})$ are the \emph{overlap} $\mathcal{O}(h_{1},h_{2})\in[-1,1]$ and \emph{mismatch} $\mathcal{M}(h_{1},h_{2})$ functions
\begin{align}
    \mathcal{O}(h_{1},h_{2}) &= \frac{(h_{1}|h_{2})}{\sqrt{(h_{1}|h_{1})(h_{2}|h_{2})}} \label{eq:overlap}\\
    \mathcal{M}(h_{1},h_{2}) &= 1 - \mathcal{O}(h_{1},h_{2}) \label{eq:mismatch}.
\end{align}
If $\mathcal{O}(h_{1},h_{2}) =  1$ then the shape of the two waveforms matches perfectly. Waveforms with $\mathcal{O}(h_{1},h_{2}) = 0$ are orthogonal, being as much in phase as out of phase over the observation.

Consider $\boldsymbol{\theta} = \boldsymbol{\theta}_{0} + \Delta\boldsymbol{\theta}$ for $\Delta\boldsymbol{\theta}$ a small deviation around the true parameters $\boldsymbol{\theta}_{0}$. Assuming that the waveform $h(t;\boldsymbol{\theta})$ has a valid first order expansion\footnote{In the literature, this is called the \emph{linear signal approximation}. It is a good approximation for sufficiently small $\Delta \boldsymbol{\theta}$, such that $\Delta \boldsymbol{\theta}\,\partial_{\boldsymbol{\theta}}^{2} h   \ll \partial_{\boldsymbol{\theta}}\Delta  h$.} in $\Delta \boldsymbol{\theta}$, we substitute into \eqref{log-likelihood-function} and expand up to second order in $\Delta \boldsymbol{\theta}$
\begin{equation}\label{FisherProbDens}
p(d|\boldsymbol{\theta})  \propto \exp\left(-\frac{1}{2}\sum_{i,j}\Gamma_{ij}(\Delta\theta^{i}-\Delta\theta^{i}_{\rm bf})(\Delta\theta^{j}-\Delta\theta^{j}_{\rm bf})\right),
\end{equation}
where $\Delta\theta^{i}_{\rm bf}=(\Gamma^{-1})^{ij}(\partial_{j}h|n)$ and $\Gamma_{ij}$ is the \emph{Fisher Matrix} given by
\begin{equation}\label{Fisher_Matrix}
\Gamma_{ij} = \left(\frac{\partial h}{\partial \theta^{i}} \bigg\rvert \frac{\partial h}{\partial \theta^{j}}\right).
\end{equation}
The Fisher Matrix $\Gamma \sim \rho^{2}$ and therefore $\Delta \boldsymbol{\theta}$ scales like  $(\Gamma^{-1})^{ij}(\partial_{j}h|n) \sim \rho^{-1}$. The linear signal approximation is therefore valid for high SNR, $\rho \gg 1$.

The Fisher Matrix $\Gamma$, evaluated at the true parameters $\boldsymbol{\theta}_{0}$, provides an estimate of the width of the likelihood function \eqref{log-likelihood-function}. Hence, it can be used as a guide to how precisely you can measure parameters. The inverse of the Fisher matrix is an approximation to the variance-covariance matrix $\Sigma$ on parameter precisions $\Delta \theta^{i}$
\begin{equation}
    \text{Cov}\left(\Delta\theta^{i},\Delta \theta^{j}\right) \approx (\Gamma^{-1})^{ij}. 
\end{equation}
The square route of the diagonal elements of the inverse fisher matrix provide estimates on the precision of parameter measurements, accounting for correlations between the parameters.

\section{Analytic estimates of spin precision}
\label{subsec:FisherMatrixParadigm}
Before discussing numerical results on the measurement precisions for the parameters $\boldsymbol{\theta}$ of near-extremal EMRIs, we would like to develop some analytic tools that will allow us to understand the precisions we find numerically. In particular the fact that spin measurements for near-extremal primaries are noticeably \emph{tighter} than those obtained for more moderately rotating primaries. Throughout this section, we will use the waveform model \eqref{WaveformAllHarmonics} for analytical convenience. We will focus on the spin-spin component of the Fisher matrix
\begin{equation}\label{Fisher_Matrix_Formula_1D}
    \Gamma_{aa} = 4\int df \frac{|\partial \hat{h}(f,r(a),t;\boldsymbol{\theta})/\partial a|^{2}}{ S_{n}(f)} 
\end{equation}
in the following analytic discussion. Our numerical and statistical analysis will be more general and employ the Teukolsky based model \eqref{eq:Teuk_Waveform_gen}. In future work, we will extend this analytic considerations to multiple parameter study. 

If all other parameters were known perfectly, the estimated precision on the spin parameter would be
\begin{equation}\label{eq:precision}
    \Delta a \approx 1/\sqrt{\Gamma_{aa}}.
\end{equation}
Thus, to compare precisions between near-extremal (denoted \emph{ext}) and moderately rotating (denoted \emph{mod}) primaries one is led to study the ratio
\begin{equation}
    \frac{\Gamma_{aa}^{\text{ext}}}{\Gamma_{aa}^{\text{mod}}}\,.
\label{eq:ratio-fisher}
\end{equation}

Consider the (semi-analytic) gravitational wave amplitude \eqref{WaveformAllHarmonics}
\begin{equation}
  h(t) = \sum_m h_m(t) \approx \sum_m \frac{2\sqrt{\dot{E}^{\infty}_{ m}}}{m\tilde{\Omega}\tilde{D}}\,\sin (m\tilde{\Omega}\tit)\,,
\label{eq:simple-amplitude}
\end{equation}
where we have chosen the initial phase $\phi_{0}=0$ for simplicity. 
The Fisher Matrix depends on the PSD of the detector. In the numerical calculations presented later we will use the full frequency dependent PSD, but to derive our analytic results we will approximate $S_n(f) \approx S_n(f_\circ)$, a constant. The rationale for this is that EMRIs evolve quite slowly and so the total change in the PSD over the range of frequencies present in the signal is small. Between 1 mHz and 100 mHz, the (square root of the) LISA PSD changes by just one order of magnitude, which is much smaller than the three orders of magnitude improvement in spin measurement precision that we find numerically. Additionally, the difference in the ISCO frequencies across all combinations of mass and spin considered in our numerical analysis is less than a factor of $2.5$. PSD variations can not therefore explain the numerical results, and so we can ignore these in deriving the analytic results which do explain the numerics. Under this approximation
\begin{equation}
  \Gamma_{aa} \approx \frac{4}{S_n(f_\circ)} \int dt\,(\partial_a h(t))^2\,.
\label{eq:fish-approx}
\end{equation}
We additionally assume that the choice of $f_\circ$ does not depend on the spin, and therefore the ratio  \eqref{eq:ratio-fisher} is independent of $S_n(f_\circ)$. Again, this approximation could introduce at most an order of magnitude uncertainty, and most likely much less than that.
Once the Fisher matrix is written in the form \eqref{eq:fish-approx}, we can use the semi-analytic waveform model \eqref{eq:simple-amplitude} to evaluate it. In appendix \ref{app:fish-estimation}, we argue the dominant contribution can be approximated by
\begin{equation}
\begin{aligned}
  \Gamma_{aa} &\approx \frac{8M}{\tilde{D}^2\,S_n(f_\mt{o})} \sum_m \Gamma_{aa,m} \\
  \Gamma_{aa,m} &\approx  \int^{\tit_\mt{cut}}_{\tit_0} d\tit\,\dot{\tilde{E}}^{\infty}_{ m}\,(\tilde{\Omega}\tit)^2\,\left(1+\frac{3}{2}\sqrt{\tr}\,\partial_a\tr\right)^2\,.
\end{aligned}
\label{eq:fish-estimation}
\end{equation}
Here $\tilde{t}_{0}$ is the coordinate time at which the observation starts and $\tilde{t}_{\text{cut}}$ is the coordinate time at the end of the observation. For the results in this paper, we analyse $\sim 1$ year long signals and fix $\tilde{t}_{\text{cut}}$ independently of spin, such that all inspirals terminate before $\tilde{r}_{\text{isco}}$ is reached.

As seen in \eqref{eq:fish-estimation}, a proper understanding of the precision in the spin measurement requires quantifying the spin dependence of the inspiral trajectory of the secondary, i.e. $\partial_a \tr$. 

\subsection{Spin dependence on the radial evolution}
\label{sec:linear_ODE_section}

 Our primary goal here is to understand the spin dependence on the radial trajectory of the secondary ($\partial_{a}\tilde{r}$) for any spin parameter $a$ of the primary. 
 
 The trajectory of the secondary is the integral of the inspiral equation
\begin{equation}\label{ODE:radial_evolution}
    \partial_{\tilde{r}}\tilde{E}(\tilde{r},a)\frac{d\tilde{r}}{d\tilde{t}} = -P_{\mt{GW}}(\tr,a)\,.
\end{equation}
This follows from energy conservation, where $\tilde{E}(\tilde{r},a)$ is the energy of a circular orbit \eqref{Energy} and $P_{\mt{GW}}:=\dot{\tilde{E}}_{\mt{GW}}(\tr,a)$ is the energy rate carried away by gravitational waves \eqref{FluxExpression}. While $\tilde{E}(\tilde{r},a)$ is \emph{kinematic}, that is, derived through geodesic properties, $P_{\mt{GW}}$ is \emph{dynamic}, that is, it is a radiation reactive term determined by solving Teukolsky's equation for a point particle source. The former is under analytic control, whereas the latter typically requires numerical treatment.

The quantity $\partial_{a}\tilde{r}$ captures the change in the secondary's trajectory when the spin parameter $a$ of the primary varies, keeping the remaining primary and secondary parameters fixed, including $\tit$. More explicitly, the integral $\tr(\tr_0,a)$ of \eqref{ODE:radial_evolution} depends on the initial condition $\tilde{r}(\tilde{t}_0)=\tr_0$ and it depends on the spin parameter $a$ both through $(\partial_{\tilde{r}}\tilde{E})$ and $(P_{\mt{GW}})$ information, but \emph{not} through $\tilde{t}$, which is simply labelling the points in the trajectory. We will comment on the possible spin dependence on the initial condition $\tr_0$ below. 

One possibility to compute $\partial_a\tilde{r}$ is to integrate \eqref{ODE:radial_evolution} and to take the spin derivative explicitly afterwards. 
A second, equivalent, way is to observe $\tr$ is a monotonic function of $\tilde{t}$ at fixed spin and initial radius $\tr_0$. Hence, it can be used as the integration coordinate to study $\partial_a\tilde{r}(\tilde{r})$. To do this, notice that the total spin derivative of the kinematic and dynamic functions in \eqref{ODE:radial_evolution}, at fixed $\tr_0$ and $\tilde{t}$, is
\begin{equation}
\begin{aligned}
  \left.\frac{\partial}{\partial a} \partial_{\tr}\tilde{E}(\tr(a),a)\right|_{\tr_0,\tilde{t}} &= (\partial^2_{\tr}\tilde{E})\,\partial_a\tilde{r} + \partial^2_{a\tr}\tilde{E}\,, \\
  \left.\frac{\partial P_{\mt{GW}}}{\partial a}\right|_{\tr_0,\tilde{t}} &= \left(\partial_{\tr}P_{\mt{GW}}\right)\,\partial_a\tilde{r} + \partial_a P_\mt{GW}\,.
\end{aligned}
\label{eq:partiala}
\end{equation}
To ease our notation, all spin partial derivatives in the rhs, and in the forthcoming discussion, should be understood as computed at fixed $\tr_0$ and $\tilde{t}$. 
Defining $u = \partial_{a}\tilde{r}$ (to ease notation) and computing the total spin derivative of equation \eqref{ODE:radial_evolution}, we obtain
\begin{equation}
    \left[u\partial^{2}_{\tilde{r}}\tilde{E}  + \partial^{2}_{\tilde{ar}}\tilde{E} + \partial_{\tilde{r}}\tilde{E}\frac{du}{d\tilde{r}}\right]\frac{d\tilde{r}}{d\tilde{t}} = -\frac{dP_{\text{GW}}}{da}\,.
\end{equation}
Plugging in the radial velocity using \eqref{ODE:radial_evolution} one obtains
\begin{equation}\label{eq:Linear_ODE_general}
    \frac{du}{d\tilde{r}} + \left(\frac{\partial^{2}_{\tilde{r}}\tilde{E}}{\partial_{\tilde{r}}\tilde{E}} - \frac{\partial_{\tilde{r}}P_{\text{GW}}}{P_{\text{GW}}}\right)u = -\frac{\partial^{2}_{a\tilde{r}}\tilde{E}}{\partial_{\tilde{r}}\tilde{E}} + \frac{\partial_{a}P_{\text{GW}}}{P_{\text{GW}}}.
\end{equation}
This is a first order \emph{linear} ODE, valid for \emph{any} spin and for \emph{any} location of the secondary, whose solution describes the desired spin dependence in the radial trajectory $\partial_{a}\tilde{r}(\tr)$.

Its general solution is a sum of the homogeneous solution $u_{h}$ and a particular solution $u_{p}$. It will depend on an initial  condition $u(\tr_{0})$. The initial condition of the radial trajectory is
\begin{equation}\label{eq:r0BC}
    \tilde{r}(\tilde{r}_0, a, t=0) = \tilde{r}_0  \qquad \Rightarrow \quad \frac{\partial \tilde{r}}{\partial a}\bigg\rvert_{r_0, t=0} = 0,
\end{equation}
from which we deduce $u(t=0) = 0$\footnote{The initial condition $u(\tr_{0})$ can play an important role when gluing a numerical calculation for $\partial_a\tr$ with an analytic one in some specific piece of the trajectory where the information determining the solution to \eqref{eq:Linear_ODE_general} is under analytic control. We will be more explicit about this when we discuss $\partial_a\tr$ in the region close to ISCO.}.


The homogeneous version of equation \eqref{eq:Linear_ODE_general} is equivalent to
\begin{equation}
    \frac{du_{h}}{u_{h}} + d\log\left(\frac{\partial_{\tilde{r}}\tilde{E}}{P_{GW}}\right) = 0 \Rightarrow u_{h} = k_{0}\frac{P_{\text{GW}}}{\partial_{\tilde{r}}\tilde{E}}
\end{equation}
where $k_{0}$ is an arbitrary integration constant. 
We follow a standard approach and look for a particular solution of the form $u_{p} = k(\tilde{r},a)u_{h}$. Plugging this into \eqref{eq:Linear_ODE_general} gives
\begin{align}
    k(\tilde{r},a) &= \int \frac{\partial_{\tilde{r}}\tilde{E}}{P_{\mt{GW}}}\left(-\frac{\partial^{2}_{a\tilde{r}}\tilde{E}}{\partial_{\tilde{r}}\tilde{E}} + \frac{\partial_{a}P_{\mt{GW}}}{P_{\mt{GW}}}\right)d\tilde{r} \label{eq:spin_dependence_geodesic_flux_k} \\
    &= -\int \frac{\partial_{\tilde{r}}\tilde{E}}{P_{\mt{GW}}}\partial_{a}\log\left(\frac{\partial_{\tilde{r}}\tilde{E}}{P_{\mt{GW}}}\right)d\tilde{r} \label{eq:spin_dependence_k}.
\end{align}
Combining our results, we obtain
\begin{equation}\label{gen:Linear_ODE}
    \partial_{a}\tilde{r} = \frac{P_{\mt{GW}}}{\partial_{\tilde{r}}\tilde{E}}\left(k_0 - \int \frac{\partial_{\tilde{r}}\tilde{E}}{P_{\mt{GW}}}\partial_{a}\log\left(\frac{\partial_{\tilde{r}}\tilde{E}}{P_{\mt{GW}}}\right)d\tilde{r}\right).
\end{equation}
This is valid for \emph{any} spin, for \emph{any} location of the secondary and for \emph{any} flux $P_{\mt{GW}}$. This analytic result will allow us to determine what the dominant source of the spin dependence is in different regions of the trajectory. 

\begin{figure*}[t]
    \centering
    \includegraphics[width = \textwidth]{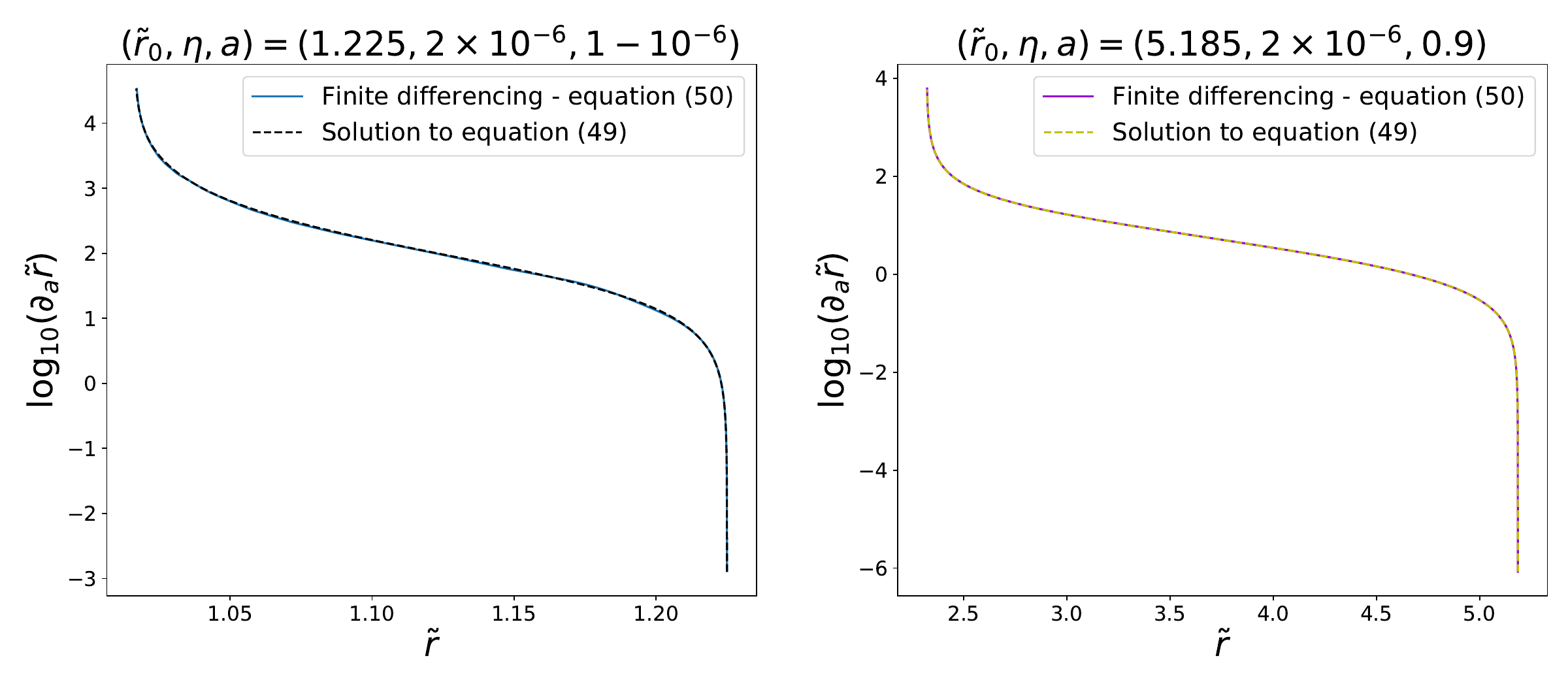}
    \caption{The dashed curves (black dashed and yellow dashed) on each figure is the solution to \eqref{gen:Linear_ODE} with $k_{0} = 0$ corresponding to $\partial_{a}\tilde{r}(\tilde{r_{0}}) = 0$. In both plots, the solid colours (blue and violet) are $\partial_{a}\tilde{r}$ calculated using a fifth order stencil method. In each plot, the intrinsic parameters given in the titles.}
    \label{fig:spin_derivative_general}
\end{figure*}

In figure \ref{fig:spin_derivative_general} we show the near perfect agreement between the solution to \eqref{gen:Linear_ODE} and our numerical calculation of $\partial_{a}\tilde{r}$ using finite difference method
\begin{equation}
   \partial_{a}\tilde{r} \approx \frac{\tilde{r}(a + \delta,\tilde{t},\dot{\mathcal{E}}(a + \delta)) - \tilde{r}(a - \delta,\tilde{t},\dot{\mathcal{E}}(a - \delta))}{2\delta}.
\label{eq:finite_difference_methods}
\end{equation}
the method used to calculate year-long trajectories used for our Fisher matrix results in later sections, for both moderately and rapidly rotating primaries.

Following~\cite{2000PhRvD..62l4021F}, we express the energy flux as a relativistic correction factor, $\Eps$, times the leading order Newtonian flux
\begin{equation}
    P_{\mt{GW}} = \frac{32}{5}\eta\,\tilde{\Omega}^{10/3}\Eps\,.
\label{eq:finn-thorne}
\end{equation}
Plugging this into Eq.~\eqref{gen:Linear_ODE} gives 
\begin{equation}
    \partial_{a}\tilde{r} = \frac{1}{\mathcal{Q}}\left(k_{0} - \int \mathcal{Q}\,\partial_{a}\log\mathcal{Q}\,d\tr\right),  \quad  \mathcal{Q} =   \frac{\partial_{\tilde{r}}\tilde{E}}{\tilde{\Omega}^{10/3}\Eps}.
\label{eq:ft-solution}
\end{equation}
Decomposing the source term 
\begin{equation}\label{eq:ODE_integrand_rewrite}
    \mathcal{Q}\,\partial_{a}\log\mathcal{Q} = \frac{\partial_{\tilde{r}}\tilde{E}}{\tilde{\Omega}^{10/3}\Eps}\left(\frac{\partial^{2}_{a\tilde{r}}\tilde{E}}{\partial_{\tilde{r}}\tilde{E}} - \frac{\partial_{a}\Eps}{\Eps} + \frac{10}{3}\tilde{\Omega}\right)\,,
\end{equation}
we see that the first and third terms are \emph{kinematic}, i.e., driven by geodesic physics, whereas the second is \emph{dynamical}, i.e., driven by the energy flux. Comparison between these terms at different stages of the inspiral, as a function of the spin, can help us to determine what the driving source of spin dependence is in each case. In the next subsection, we investigate the contribution of both the geodesic and radiation reactive terms to $\partial_{a}\tilde{r}$.

\subsection{Comparison of radial evolution for moderate and near-extremal black holes} \label{subsec:analytical_traj}
Despite the universality of \eqref{gen:Linear_ODE} or \eqref{eq:ft-solution}, the dependence on the energy flux makes it not feasible to analytically integrate $\partial_a\tr$ along the entire secondary trajectory. However, 
we can integrate \eqref{gen:Linear_ODE} in \emph{specific} regions of the secondary trajectory.

\begin{figure}
    \centering
    \includegraphics[height = 5cm,width = 8cm]{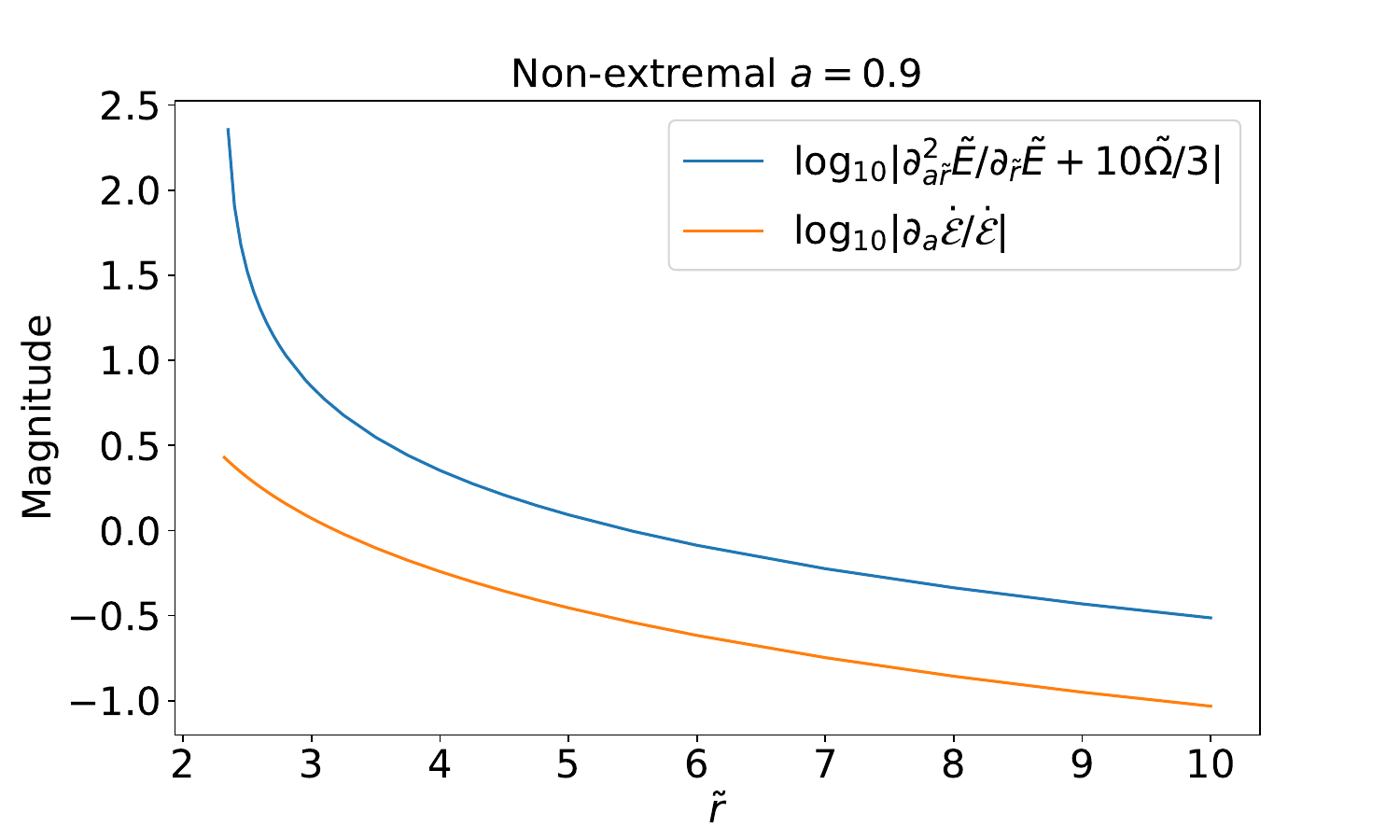}
    \includegraphics[height = 5cm,width = 8cm]{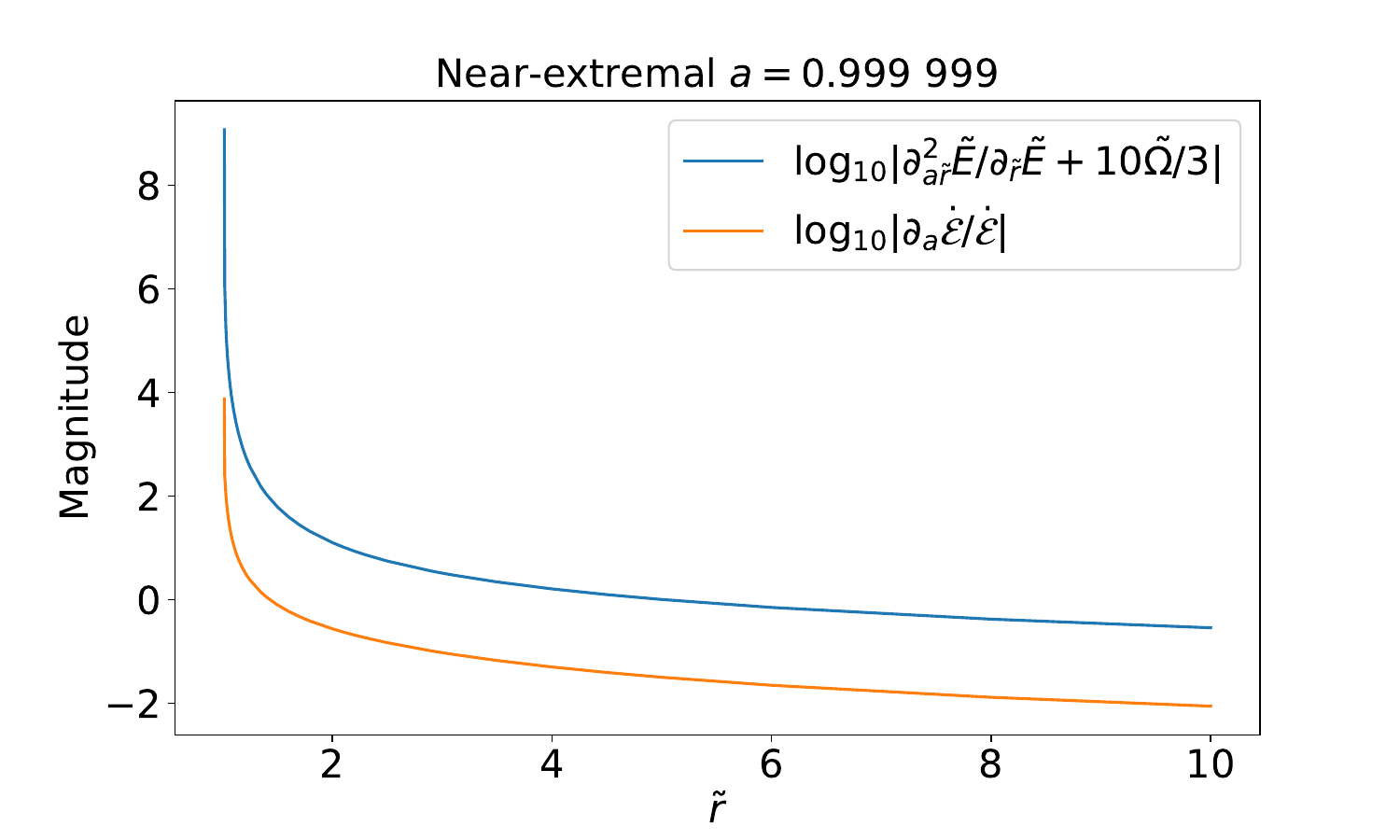}
    \caption{The top plot compares the kinematic and radiation reaction quantities given in \eqref{eq:ODE_integrand_rewrite} for a spin of $a = 0.999 999$. The bottom plot is the same but for a spin parameter of $a = 0.9$. Notice that in these two cases the kinematical quantities \emph{dominate} over the relativistic correction terms. }
    \label{fig:Kinematic_vs_spin}
\end{figure}

It is possible to prove that $d\partial_a\tr /d\tr < 0$ and hence that $\partial_a\tr$ grows monotonically over the inspiral. It is therefore natural to study the behaviour of $\partial_a\tr$ close to ISCO, where its contribution to the Fisher matrix \eqref{eq:fish-estimation} will be maximal. We first compare the kinematic and dynamical contributions to \eqref{eq:ODE_integrand_rewrite}. 
Using results from the \BHPT, we have numerically calculated the spin derivative of $\Eps$ for two primaries with spin parameters $a = 0.9$ and $a = 1-10^{-6}$. These are compared with the kinematic sources in \eqref{eq:ODE_integrand_rewrite} in figure \ref{fig:Kinematic_vs_spin}.
These figures show that
\begin{equation}\label{eq:kinematics_vs_GRCs_equation}
\bigg\rvert\frac{\partial^{2}_{a\tilde{r}}\tilde{E}}{\partial_{\tilde{r}}\tilde{E}} + \frac{10}{3}\tilde{\Omega}\bigg\rvert \gg \bigg\rvert\frac{\partial_{a}\Eps}{\Eps}\bigg\rvert, 
\end{equation}
for both spin parameters. This suggests it is the kinematic sources in \eqref{eq:ODE_integrand_rewrite} that drive the spin dependence of the secondary trajectory, particularly close to ISCO. 
Although we have only verified it for two choices of spin parameter, we will assume this approximation holds for \emph{any} spin parameter $a \geq 0.9$. 

We first consider moderately spinning black holes close to ISCO. Dropping the dynamical contribution to \eqref{eq:ODE_integrand_rewrite}, we can compare the two remaining terms. The angular velocity piece is bounded and order one, but $\partial_{\tilde{r}}\tilde{E}$ tends to zero at ISCO. This means that $\partial^{2}_{a\tilde{r}}\tilde{E}/\partial_{\tilde{r}}\tilde{E}$ dominates close to ISCO, allowing us to use the approximation
\begin{equation}
    \frac{\partial_{\tilde{r}}\tilde{E}}{\tilde{\Omega}^{10/3}\Eps}\left(\frac{\partial^{2}_{a\tilde{r}}\tilde{E}}{\partial_{\tilde{r}}\tilde{E}} - \frac{\partial_{a}\Eps}{\Eps} + \frac{10}{3}\tilde{\Omega}\right) \approx \frac{\partial^{2}_{a\tilde{r}}\tilde{E}}{\tilde{\Omega}^{10/3}\Eps}\,.
\end{equation}
Since, for moderate spins, the variation of $\tilde{\Omega}$ and $\dot{\mathcal{E}}$ with radius close to ISCO is negligible compared to the variation in $\partial^{2}_{a\tilde{r}}\tilde{E}$, we will approximate them by their values at $\tr_\mt{isco}$. This allows us to integrate \eqref{gen:Linear_ODE} to give the spin dependence of the radial trajectory for moderately spinning black holes
\begin{equation}\label{eq:soln_approximation_all_spins}
    \partial_{a}\tilde{r} \approx \frac{1}{\partial_{\tilde{r}}\tilde{E}}\left(k_{\mt{mod}}\tilde{\Omega}_{\mt{isco}}^{10/3}\Eps_{0}(a,\tilde{r}_{\text{isco}}) - \partial_{a}\tilde{E}\right)\,,
\end{equation}
where $k_\mt{mod}$ is an arbitrary constant. Since
\begin{equation}
     \partial_{\tr} \tilde{E} = \frac{\tr^2 -3a^2+8a\sqrt{\tr} -6\tr}{2\tr^{7/4}\left(\tr^{3/2}-3\sqrt{\tr} + 2a\right)^{3/2}}\,,
\label{eq:r-der-E}     
\end{equation}
it follows from eq. (A5) in \cite{burke2019transition} that
$\partial_{\tr}\tilde{E}(\tilde{r}_{\text{isco}})=0$. For moderately rotating primaries and near ISCO, we can expand $\tilde{E} \approx \tilde{E}(\tr_\mt{isco}) + \frac{1}{2}\,\partial^2_{\tr}\tilde{E}(\tr_\mt{isco})\,(\tr-\tr_\mt{isco})^2$ leading to
\begin{equation}
\begin{aligned}
  \partial_a \tilde{E} &\approx \partial_a \tilde{E}(\tr_\mt{isco}) + \partial^2_{\tr}\tilde{E}(\tr_\mt{isco})\,(\tr-\tr_\mt{isco})(-\partial_a\tr_\mt{isco}) \\
  \partial_{\tr} \tilde{E} &\approx \partial^2_{\tr}\tilde{E}(\tr_\mt{isco})\,(\tr-\tr_\mt{isco})
\end{aligned}
\end{equation}
Using these expansions in Eq.~\eqref{eq:soln_approximation_all_spins} we deduce $\partial_{a}\tilde{r} = \tilde{k}_{\mt{mod}}/(\tilde{r}-\tilde{r}_{\mt{isco}}) + \partial_a \tilde{r}_{\mt{isco}}$, where $\tilde{k}_{\mt{mod}}=k_{\mt{mod}}\tilde{\Omega}_{\mt{isco}}^{10/3}\Eps_{0}(a,\tilde{r}_{\text{isco}})/\partial^2_{\tr}\tilde{E}(\tr_\mt{isco})$. Assuming that $\tilde{r}_0$ is sufficiently close to $\tilde{r}_{\mt{isco}}$ that  this approximation holds throughout the range $[\tilde{r}_{\mt{isco}}, \tilde{r}_0]$, we can use the boundary condition~\eqref{eq:r0BC} to determine $\tilde{k}_{\mt{mod}}=\partial_a \tilde{r}_{\mt{isco}} (\tilde{r}_{\mt{isco}}-\tilde{r}_0)$ and hence 
\begin{equation}
    \partial_a \tr \approx (-\partial_a \tr_\mt{isco})\,\frac{\tr_0 - \tr}{\tr-\tr_\mt{isco}}\,.
\label{eq:near-isco-mod1}
\end{equation}

We now repeat this analysis for near-extremal primaries. Near ISCO, the energy flux can be approximated by the NHEK flux $(x\equiv \tr - 1 \ll 1)$
\begin{equation}\label{eq:NHEK_flux}
    P_{\mt{GW}} \approx \eta(\tilde{C}_{\infty} + \tilde{C}_{H}) x\,.
\end{equation}
Using this approximation, there is no explicit spin dependence and so the $\partial_a P_{\mt{GW}}$ term in Eq.~\eqref{eq:spin_dependence_geodesic_flux_k} vanishes. Expanding ~\eqref{eq:r-der-E} for $x=\tr -1\ll 1$ and $\epsilon \ll 1$, the denominator involves
\begin{equation*}
  \tr^{3/2}-3\sqrt{\tr} + 2a = \frac{3}{4}x^2 - \epsilon^2 - \frac{1}{4}x^3 + \frac{9}{64}x^4 + \mathcal{O}(x^5,x\epsilon^2,\epsilon^4),
\end{equation*}
while the numerator has the expansion
\begin{equation*}
  \tr^2 -3a^2+8a\sqrt{\tr} -6\tr = \frac{1}{2}x^3 - \epsilon^2 - \frac{5}{32}x^4 + \mathcal{O}(x^5,x\epsilon^2,\epsilon^4)\,.
\end{equation*}
We conclude
\begin{equation}
  \partial_{\tr} \tilde{E} \approx \frac{2}{3\sqrt{3}}\left(1-\frac{11}{8}x - \frac{x^3_\mt{isco}}{x^3}\right) + \mathcal{O}(x^2,\epsilon^2/x^2).
 \label{eq:damp-exp}
\end{equation}
Using this approximation in Eq.~\eqref{eq:spin_dependence_geodesic_flux_k} we find
\begin{align}
k(\tilde{r},a)&\approx -\int \frac{\partial^2_{a\tilde{r}} \tilde{E}}{P_{\mt{GW}}} \,{\rm d}\tilde{r}\\ &\approx \frac{2x_{\mt{isco}}^2}{\eta (\tilde{C}_{\infty} + \tilde{C}_{H}) \sqrt{3}} \frac{\partial x_{\mt{isco}}}{\partial a} \int x^{-4} {\rm d}x \\
&\approx\frac{8}{9\sqrt{3} \eta (\tilde{C}_{\infty} + \tilde{C}_{H})} \frac{1}{x^3}
\end{align}
where we have used $x_{\text{isco}} \approx 2^{1/3}\epsilon^{2/3}$ and $P_{\text{GW}}$ defined in \eqref{eq:NHEK_flux}. This is valid for $x\ll 1$ and includes the corrections due to $x\sim x_\mt{isco}$. Assuming $\tr_0$ is close to ISCO, so that the initial condition \eqref{eq:r0BC} holds, we conclude that the spin dependence in the near-ISCO region of a near-extremal black hole is
\begin{equation}
	\partial_a \tr \approx \frac{8}{9\sqrt{3}\,x^2\partial_{\tr} \tilde{E}} \left(1 - \frac{x^3}{x_0^3}\right)\,.
\label{eq:throat-integral}
\end{equation}

\begin{figure*}
    \centering
    \includegraphics[width = \textwidth]{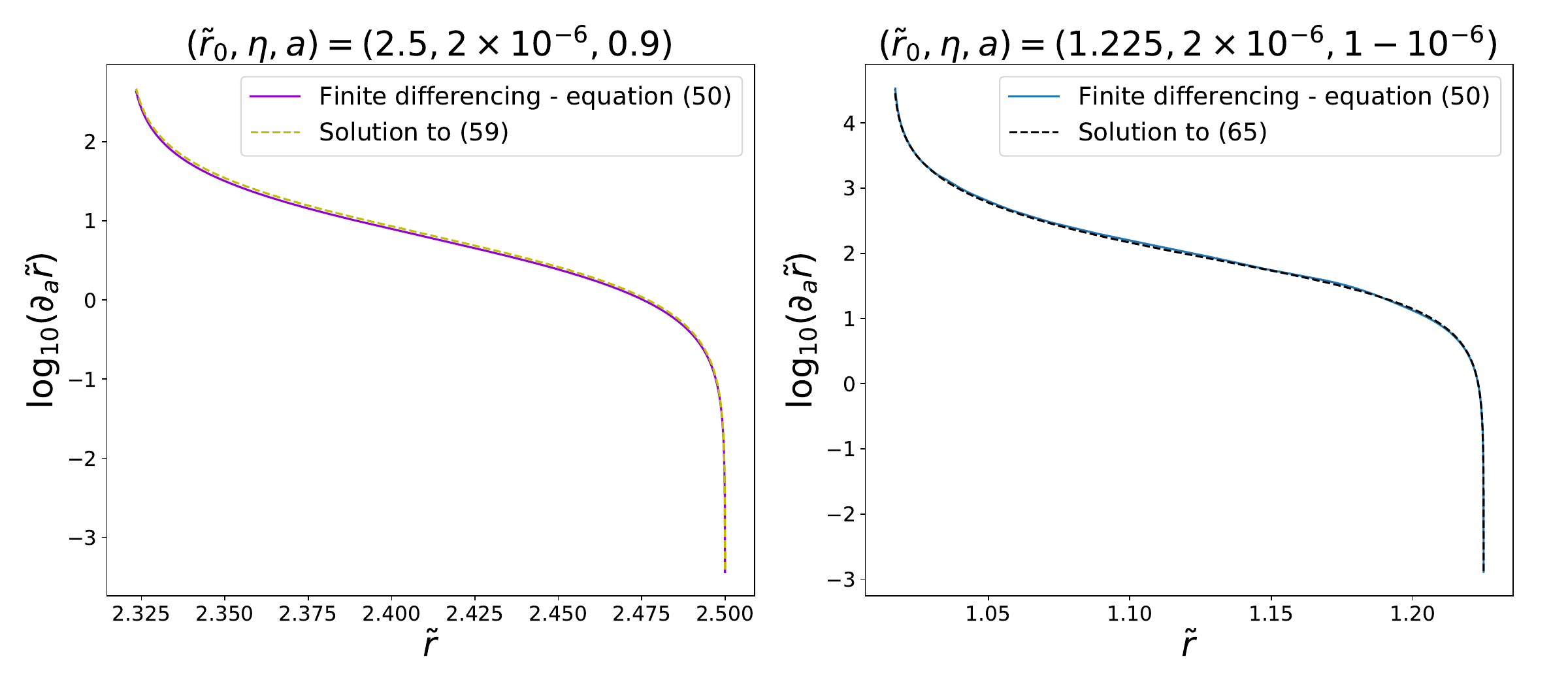}
    \caption{The yellow dashed and black dashed curves are solutions to \eqref{eq:near-isco-mod1} and \eqref{eq:throat-integral}. The purple and blue curves are the true solutions to $\partial_{a}\tilde{r}$ obtained numerically without near-ISCO simplifications. We see both approximations capture the leading order behaviour of the spin derivative of the radial trajectory very well.}
    \label{fig:Joans_approximation_derivs}
\end{figure*}

Figure \ref{fig:Joans_approximation_derivs} compares \eqref{eq:near-isco-mod1} and \eqref{eq:throat-integral} to the full $\partial_{a}\tilde{r}$ computed numerically without using the near-ISCO approximations. We see that the approximations are very accurate in the region close to the ISCO where they are valid.

Before continuing, we will comment further on the choice of flux \eqref{eq:approx:NHEK} instead of \eqref{NielsEnergy}. The latter has an \emph{explicit} dependence on $\tr_+=1+\epsilon$. Consequently, it carries an additional spin dependence. In particular, $\partial_a \tr_+ = -a/\epsilon$. Thus, for near-extremal primaries this spin dependence can induce extra diverging sources for $\partial_a\tr$ with a very specific sign. We can easily compute their effects by integrating the ODE with such an energy flux source. The result one finds does \emph{not} agree with the numerical evaluation of $\partial_a \tr$ generated from the \BHPT, which computes the exact flux\footnote{In particular, it is no longer the case that $\partial_a\tr$ is monotonically increasing all along the inspiral trajectory, whereas the \BHPT \ data is monotonically increasing, a feature our ODE with flux \eqref{eq:approx:NHEK} reproduces.}. We conclude that \eqref{eq:approx:NHEK} appears to capture the spin dependence of our observable (the amplitude of the gravitational wave) more accurately than ~\eqref{NielsEnergy}. This is in fact the reason we chose to work with \eqref{eq:approx:NHEK}.

Let us close this discussion with a brief comparison between the analytic results for moderate and near-extremal spins. We write $\tr - \tr_\mt{isco} \sim \delta > \eta^{2/5}$, the latter inequality ensuring that we avoid entering the transition region~\cite{burke2019transition,compre2019transition}. Expanding Eq.~\eqref{eq:damp-exp} we find for near-extremal black holes
\begin{equation*}
    \partial_{\tilde{r}} \tilde{E} \approx \frac{2}{3\sqrt{3}} \left( 3\frac{\delta}{x_{\mt{isco}}} -\frac{11}{8}\delta - \frac{11}{8} x_{\mt{isco}} + \cdots\right). \label{eq:NHEKdrE}
\end{equation*}
The first term is dominant unless |$\delta \lesssim x_{\mt{isco}}^2 \sim \epsilon^{4/3}$. The constraint $\delta > \eta^{2/5}$ ensures this is only violated if $\epsilon > \eta^{3/10}$. This will be satisfied for all the cases that we consider in this paper, but we emphasise this is not a physical constraint. When this constraint is violated, additional terms become important in the expansion which we have ignored, and these ensure that $\partial_{\tilde{r}} \tilde{E} \rightarrow 0$ at $\tilde{r}_{\mt{isco}}$. We conclude the scaling of $\partial_{\tilde{r}} \tilde{E}$ is $\delta/\epsilon^{2/3}$ for near-extremal black holes, compared to $\delta$ for moderate spins.

It follows using \eqref{eq:near-isco-mod1} and \eqref{eq:throat-integral} that
\begin{equation} \label{eq:spin_radial_evolv_two_cases}
    \partial_{a}\tilde{r} \propto 
    \begin{cases} 
     \frac{1}{\delta}, & \text{moderate spins}  \\
     \frac{\epsilon^{2/3}}{\delta (\delta+\epsilon^{2/3})^2}, & \text{near-extremal spins}
      \end{cases}.
\end{equation}
The spin dependence on the radial trajectory for near-extremal primaries is \emph{larger} than for moderately rotating ones. 



\subsection{Precision of spin measurement}\label{subsec:analytic_fish}
In the previous section we showed the effect that the spin parameter has on the radial trajectory. This was achieved by studying the \emph{general} linear ODE for $\partial_{a}\tilde{r}$, Eq.~\eqref{gen:Linear_ODE}. By arguing that the kinematic terms dominate the behaviour of $\partial_{a}\tilde{r}$, for \emph{both} the near-extreme and moderately spinning black holes, analytic solutions were found near the ISCO. We were able to conclude that $\partial_{a}\tilde{r}$ grows much more rapidly close to the ISCO for near-extreme black holes than for moderately spinning black holes. We also emphasise that Eq.~\eqref{eq:kinematics_vs_GRCs_equation} shows that corrections to $\partial_{a}\tilde{r}$ of the form $\partial_{a}\Eps$ are \emph{subdominant}. We now explore the consequences of these results for the precision of spin measurements, computed using the Fisher Matrix formalism.

Due to the large number of observable gravitational wave cycles that are generated while the secondary is within the strong field gravity region outside the primary Kerr black hole, extreme mass-ratio inspirals will provide measurements of the system parameters with unparalleled precision~\cite{danzmann1996lisa}. In particular, it has been shown that our ability to constrain the spin parameter $a$ is expected to be $\mathcal{O}(10^{-6})$ \cite{chua2017augmented,babak2007kludge,barack2004lisa,babak2017science}. It has also been shown that the measurements are more precise for prograde inspirals into more rapidly spinning black holes, when the secondary spends more orbits closer to the event horizon of the primary (see Fig.(11) in \cite{babak2017science}). In subsequent sections we show through numerical calculation that spin measurements are even more precise for EMRIs into near-extremal black holes. We now try to understand this result using Eq.~\eqref{eq:fish-estimation}.

Inspection of \eqref{eq:fish-estimation} suggests there are two main effects: the dependence on $\tit^2$ and the dependence on $(\partial_a\tr)^2$. First, 
the fact that $\tit \sim \mathcal{O}(\eta^{-1})$ follows from integrating \eqref{ODE:radial_evolution}, and therefore the contribution to the Fisher matrix due to $\tit^2$ is large and scales like $\eta^{-2}$. Second, $\partial_a\tr$ is monotonically increasing as the secondary spirals inwards. Thus, its maximal contribution comes from the region close to ISCO, which supports results in \cite{babak2017science}. Eq.~\eqref{eq:spin_radial_evolv_two_cases} shows this contribution is largest in the last stages \emph{before} entering into the transition regime. As changes in $\tilde{\Omega}$ close to ISCO are negligible, the factor $(\tilde{\Omega}\tit)^2$ is, approximately, the square of the number of cycles, a proxy widely used in the literature in discussions of the precision of measurements. Our estimate \eqref{eq:fish-estimation} confirms this intuition and shows the spin precision will be further increased by large values of the radial spin derivative, $\partial_a \tilde{r}$.

\subsection{Comparison of spin measurement precision for moderate and near-extremal black holes}

The Fisher matrix estimate \eqref{eq:fish-estimation} depends on the spin derivative of the radial evolution, on the duration of the inspiral and on the energy flux. Eq.~\eqref{eq:spin_radial_evolv_two_cases} shows that, at a fixed distance to the corresponding ISCO, $\partial_a \tr$ is larger for a near-extremal primary than for a moderately rotating primary. As a consequence of time dilation near the black hole horizon, $\dot{E}\to 0$ near the ISCO for near-extremal primaries, but remains finite for moderately rotating ones. This means that the energy flux for near-extremal inspirals is much smaller than that for moderate spins, but the duration of the inspiral is longer. 
However, we can write \eqref{eq:fish-estimation} as an integral over the BL radial coordinate $\tr$. In that case, the integrand is proportional to
\begin{equation*}
  \frac{\dot{\tilde{E}}^{\infty}_{m}}{\dot{\tilde{E}}_{\mt{GW}}}\,(\tit\,\tilde{\Omega})^2\,(\partial_{\tr} \tilde{E})\,(\partial_a\tr)^2
\end{equation*}
close to the relevant ISCO. While the energy fluxes are much smaller for near-extremal inspirals, the ratio of energy fluxes appearing above is an order one quantity for all spin parameters. The expression above is therefore a product of factors that have been argued to be either of comparable magnitude or much smaller for moderately rotating primaries. We therefore expect the precision of spin measurements to be much higher for near-extremal EMRIs.

A quantitative comparison between the near-extremal and the moderately spinning sources requires a precise calculation of the ratio \eqref{eq:ratio-fisher} computed along the entire respective trajectories. In general, this is a hard analytic task since both energy fluxes $\dot{\tilde{E}}_{\mt{GW}}$ and $\dot{\tilde{E}}^{\infty}_{m}$ must be handled through numerical means and long observations of inspirals (starting in the weak field) require calculations performed in the frequency domain where $S_{n}(f)$ shows non-trivial (non-constant) behaviour. This would be no more straightforward than direct numerical computation of the Fisher Matrix and so we do not pursue it here.

For any sources whose trajectory lies entirely lie in the near-ISCO region, these analytic approximations allow us to compute the ratio \eqref{eq:ratio-fisher} reliably. This can be exploited to obtain an analytic approximation to the Fisher Matrix for such sources and this calculation will be pursued elsewhere. Additionally, earlier arguments tell us that it is the near-ISCO regime that dominates the spin precision and so these expressions are sufficient to understand the increase in spin precision seen for near-extremal inspirals.

\begin{widetext}
\paragraph{Near-extremal source.} From \eqref{eq:throat-integral}, it follows 
\begin{equation}\label{eq:near_ext_trajectory_approx}
    \partial_a \tr \approx \frac{4}{3x_0^3}\,\frac{x}{x^3-x_\mt{isco}^3}\,(x_0^3-x^3)\,.
\end{equation}
Since $\sqrt{\tr}\partial_a \tr$ grows fast and the rate of change of $\tr$ and $\tilde{\Omega}$ is small, near-ISCO, we can approximate \eqref{eq:fish-estimation} by
\begin{equation}
\begin{aligned}
  \Gamma_{aa} &\approx 18\frac{\mu}{(\eta\tilde{D})^2\,S_n(f_\circ)}\,\tr_{\mt{ext}}\,\tilde{\Omega}_{\mt{ext}}^2 
  \sum_m \int^{\tit_{\mt{cut}}}_0 d(\eta\tit) \frac{d\tilde{E}^{\infty}_{ m}}{\eta d\tit}
  \,(\eta\tit)^2\,(\partial_a\tr)^2.
\end{aligned}
\end{equation}
Here, $\tit_{\mt{cut}}$ is the time at the end of the integration, where $x=x_{\mt{cut}}$. Our approximations break down when the transition regime breaks down, so we can assume $x_{\mt{cut}} \sim \eta^{2/5} + \epsilon^{2/3}$, which is a small quantity. Using $d\tilde{E}^{\infty}_{m}/d\tit = \eta\tilde{C}_{\infty m}\,x$  and assuming $x_0\geq x \gg x_\mt{isco}$, so that the trajectory can be approximated by $x(\tit) \approx x_0\,e^{-y}$ with $y=\alpha\,\eta \tit \equiv \frac{3\sqrt{3}}{2}\,(\tilde{C}_\mt{H} + \tilde{C}_\infty)\,\eta\tit$, the Fisher matrix reduces to 
\begin{equation}
\begin{aligned}
   \Gamma_{aa}^{\mt{ext}} \approx  \frac{64\mu}{(\eta\tilde{D})^2\,S_n(f_\circ)} \frac{\tr_{\mt{ext}}\,\tilde{\Omega}_{\mt{ext}}^2}{(3x_0\alpha)^3}
   \cdot G(y_{\mt{cut}})\,\left(\sum_m \tilde{C}_{\infty m}\right)\,,
\end{aligned}
\label{eq:fest-extremal}
\end{equation}
with $e^{y_\mt{cut}}=x_0/x_\mt{cut}$ and
\begin{equation}
   G(y_{\mt{cut}}) = -9y^3_\mt{cut} + (9y^2_\mt{cut}+2)\,\sinh 3y_\mt{cut} - 6y_\mt{cut}\,\cosh 3y_\mt{cut}\approx \frac{x_0^3}{2x^3_\mt{cut}}\left[(3\log \frac{x_0}{x_\mt{cut}} -1)^2+1\right]\,, 
\end{equation}
where in the last step we used $x_0 \gg x_\mt{cut}$.
\paragraph{Moderately spinning source.} Using the same kind of approximations as above, but taking into account the different energy flux and different trajectory
\begin{equation}
 (\tr - \tr_\mt{isco})^2 - (\tr_0 -\tr_\mt{isco})^2 \approx \frac{64}{5} \eta\,\tilde{\Omega}_\mt{isco}^{10/3}\,\frac{\dot{\mathcal{E}}_0(a)}{\partial^2_{\tr} \tilde{E}(\tr_\mt{isco})}\,\tit \,,
\end{equation}
one can approximate the Fisher matrix for moderate spins by 
\begin{equation}
\begin{aligned}
  \Gamma_{aa}^{\mt{mod}} &\approx 18 \left(\frac{5\partial^2_{\tr} \tilde{E}(\tr_\mt{isco})}{64\dot{\mathcal{E}}_0}\right)^3\,\frac{\mu}{(\eta\tilde{D})^2\,S_n(f_\circ)}\,\frac{\tr_\mt{isco}\,(\partial_a \tr_\mt{isco})^2}{\tilde{\Omega}_\mt{isco}^8}
  & (\tr_0 - \tr_\mt{isco})^6\,F(\delta) \left(\sum_m  \,\left.\frac{d\tilde{E}^{\infty}_{ m}}{\eta\,d\tit}\right|_{\mt{isco}}\,\right)\,,
\end{aligned}\label{eq:analytic_fish_non_ext_approx}
\end{equation}
where $\delta \equiv \frac{\tr_\mt{cut} - \tr_\mt{isco}}{\tr_0 - \tr_\mt{isco}} < 1$ and
\begin{equation}
\begin{aligned}
   F(\delta) &=-2\log \delta -4(1-\delta)- (1-\delta^2) + \frac{8}{3}(1-\delta^3) - \frac{1}{2}(1-\delta^4) - \frac{4}{5}(1-\delta^5) + \frac{1}{3}(1-\delta^6)\,. 
\end{aligned}\label{eq:non_ext_cutoff}
\end{equation}

\paragraph{Ratio of Fisher matrices.} Within these approximations, the ratio \eqref{eq:ratio-fisher} now reduces to 
\begin{equation}
\begin{aligned}
   \frac{\Gamma_{aa}^{\mt{ext}}}{\Gamma_{aa}^{\mt{mod}}} &\approx \frac{256}{9}\left(\frac{64}{45\sqrt{3}\,\partial^2_{\tr} \tilde{E}(\tr_\mt{isco})}\right)^3\,\frac{\tr_\mt{ext}\tilde{\Omega}^2_\mt{ext}}{\tr_\mt{isco}\tilde{\Omega}^2_\mt{isco}\,(\partial_a \tr_\mt{isco})^2}
   \frac{G(y_\mt{cut})}{x^3_0\,(\tr_0-\tr_\mt{isco})^6\,F(\delta)}\, {\cal T}\,, \\
   {\cal T} &= \frac{\sum_m \tilde{C}_{\infty m}}{(\tilde{C}_\mt{H} + \tilde{C}_\infty)^3} \frac{(\tilde{\Omega}^{10}_\mt{isco}\,\dot{\mathcal{E}}^3_0)}{\sum_m  \,\left.\frac{d\tilde{E}^{\infty}_{ m}}{\eta\,d\tit}\right|_{\mt{isco}}}
\end{aligned}
\label{eq:final-ratio}
\end{equation}
\end{widetext}
The most relevant feature for our current discussion is the quotient dependence 
\begin{equation}
   \frac{G(y_\mt{cut})}{x^3_0\,(\tr_0-\tr_\mt{isco})^6\,F(\delta)} \approx \frac{1}{x^3_\mt{cut}\,(\tr_0-\tr_\mt{isco})^6 F(\delta)} 
\end{equation}
The first two denominator factors increase the ratio, since $x_\mt{cut}\ll 1$ and $\tr_0 - \tr_\mt{isco} < 1$. The last could in principle be large, due to the logarithmic term. However $\delta$ and $x_{\mt{cut}}$ have similar scaling and therefore $x_{\mt{cut}} F(\delta) \ll 1$. We deduce that the spin component of the Fisher Matrix is \emph{much larger} for near-extremal inspirals than for moderate spins. This is confirmed by the numerical results that will be reported in subsequent sections.

We finish by noting that the Fisher matrices increase in magnitude as the trajectory is cut off closer to $\tilde{r}_{\text{isco}}$. In the case of moderate spin, we already noted the logarithmic dependence of $F(\delta)$ as $\delta \rightarrow 0$. This has previously been observed in the literature, see for example Fig.(11) in \cite{babak2017science}. For near-extremal EMRIs, if $x_{\text{cut}} \sim x_{\text{isco}} \sim \epsilon^{2/3}$, then for fixed $x_{0}$ and as $\epsilon \to 0$ the spin Fisher matrix scales as $\Gamma_{aa} \sim (\log(\epsilon)/\epsilon)^{2}$. We deduce that observing the latter stages of inspiral is important for precise parameter measurement, for any primary spin.

In summary, we have derived an analytic approximation, valid close to ISCO, for the spin component of the Fisher Matrix. This indicates that this component is much larger for near-extremal spins and therefore we expect much more precise measurements of the spin parameter in that case. The approximation depends sensitively on certain quantities, such as the cut-off radius, $x_{\mt{cut}}$, that are somewhat arbitrary. However, for any choice the near-extremal precision is a few orders of magnitude better. This provides support for the numerical results that we will obtain in Sec.(\ref{sec:Fish_Matrix_Numerics}), which show a similar trend.

\section{Waveform Generation}\label{sec:Waveform_Gen}
In this section we provide more details on how we construct the waveform model used to compute the Fisher Matrix in the next section. The waveform model was previously given in Eq.~\eqref{WaveformAllHarmonics} and Eq.~\eqref{eq:Teuk_Waveform_gen}. Here, we describe how the various terms entering these equations are evaluated.


\subsection{Energy Flux}
Both waveform models \eqref{WaveformAllHarmonics} and \eqref{eq:Teuk_Waveform_gen} depend on the radial trajectory $\tilde{r}(\tilde{t},a,\eta,\dot{\mathcal{E}})$. The amplitude evolution using the Teukolsky formalism depends on the spheroidal harmonics $_{-2}S^{am\tilde{\Omega}}_{ml}(\theta,\phi)$ and Teukolsky amplitudes at infinity $Z^{\infty}_{ml}(\tilde{r},a)$. The energy flux at infinity $\dot{\tilde{E}}^{\infty}_{ml}(\tilde{r},a)$ is related to the Teukolsky amplitudes $Z^{\infty}_{ml}$ through equation \eqref{eq:dotE_inf_Teuk_Amplitudes}. Thus, to accurately generate the waveforms \eqref{WaveformAllHarmonics} and \eqref{eq:Teuk_Waveform_gen} far from the horizon where near-extremal simplifications can not be made, the various radiation reactive terms $Z^{\infty}_{ml},\dot{\tilde{E}}(\dot{\mathcal{E}}),\dot{\tilde{E}}^{\infty}_{m}(\dot{\mathcal{E}}^{\infty}_{m})$ have to be handled numerically. This section outlines our numerical routines to do so.

We use the \BHPT \ to calculate the first order dissipative radial fluxes $\dot{\tilde{E}}_{\text{GW}}$ for $a = 1 - \{10^{-i}\}_{i=3}^{i = 9}$ from which $\Eps$ in Eq.~\eqref{eq:finn-thorne} can be computed. We used the \texttt{Teukolsky} mathematica script in the toolkit and tuned the numerical precision to $\sim 240$ decimal digits to avoid numerical instabilities when computing $\dot{\tilde{E}}_{\text{GW}}$ in the near-horizon regime for rapidly rotating holes. For moderately spinning holes $a \lesssim 0.999$, we used the tabulated data in Table II of \cite{2000PhRvD..62l4021F}.

\begin{figure}[!ht]
\begin{center}
\includegraphics[width=.49\textwidth]{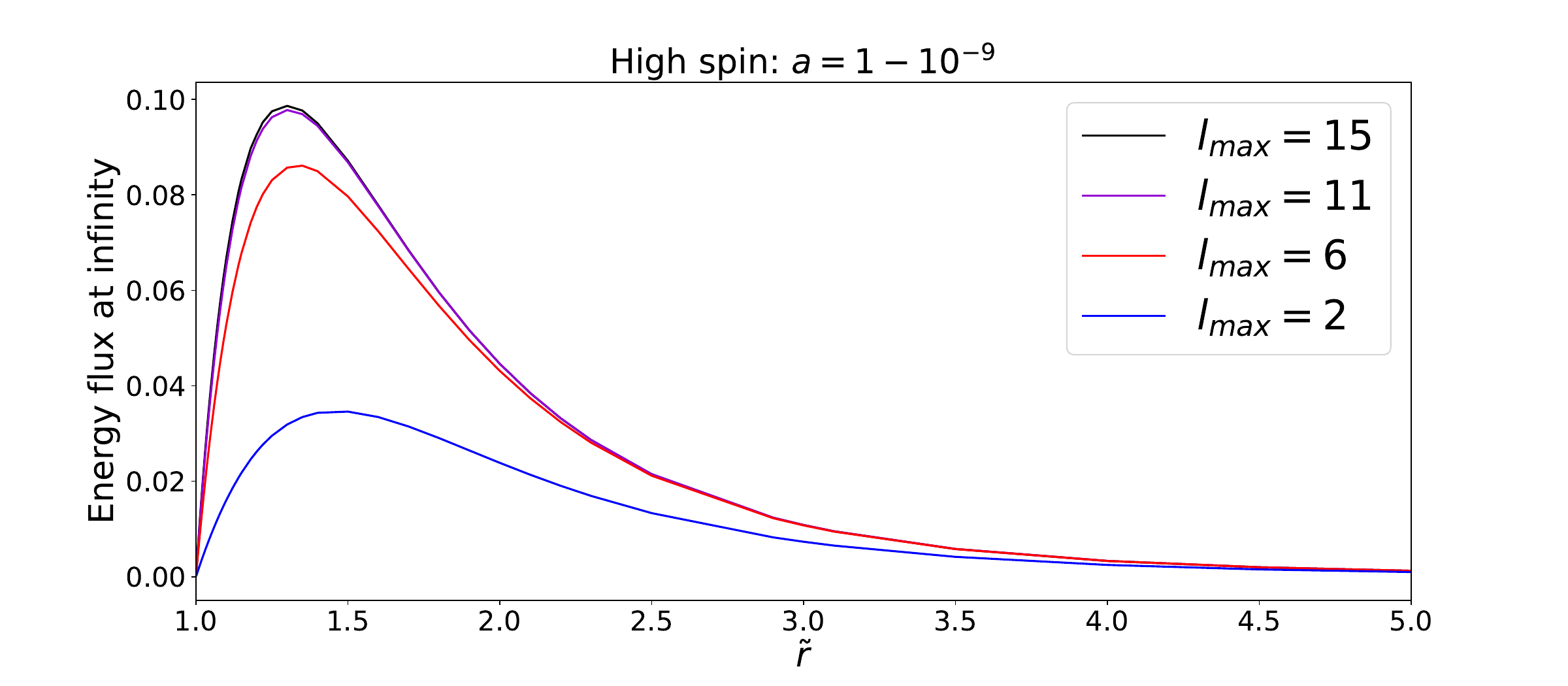}
\caption{Comparison of the total energy flux at infinity (black curve) including different harmonic $\dot{\tilde{E}}^{\infty}_{lm}$ contributions. Note that at $\tilde{r}\approx 1.3$, the $l=2$ harmonic energy flux  $\dot{\tilde{E}}^{\infty}_{2}$ contributes $\sim$ 32\% of the total energy flux, whereas including the first $l_{\text{max}} = 11$ harmonics (violet curve) contributes more than $\sim 98\%$.}
\label{fig:extra_harmonics}
\end{center}
\end{figure}

Each coefficient appearing in Eq.~\eqref{RMSStrain} is itself a sum over $l$ modes, $\dot{\tilde{E}}^{\infty}_{m} = \sum_{|l|=m}^{\infty}\dot{\tilde{E}}^{\infty}_{ml}$. Both the sum over $l$ and the sum over $m$ in Eq.~\eqref{RMSStrain} can be truncated without appreciable loss of accuracy.  As discussed in~\cite{compere2018_NHEK}, near-extremal EMRIs require a significant number of harmonics to be included to obtain an accurate representation of the gravitational wave signal. To illustrate for a high spin of $a = 1-10^{-9}$, we used the \BHPT \ to compute $\dot{\tilde{E}}^{\infty}_{ml}$ for harmonics $|m| \leq l\in\{2,\ldots,15\}$. Figure \ref{fig:extra_harmonics} illustrates the convergence as the number of harmonics is increased.

Based on these results, we go further by including harmonics with $l \leq l_{\mt{max}}=20$ to calculate
the total energy flux $\dot{\tilde{E}}_{\text{GW}}$ 
\begin{equation}
    \dot{\tilde{E}}_{\text{GW}} = \sum_{|l| = 2}^{l_{\mt{max}}}\sum_{|m| \leq l}(\dot{\tilde{E}}^{\infty}_{ml} + \dot{\tilde{E}}^{H}_{ml}),
\end{equation}
using the \hyperlink{https://bhptoolkit.org/Teukolsky/}{Teukolsky} package in the \BHPT. In the same numerical routine, we compute $\dot{\tilde{E}}^{\infty}_{m} = \sum_{|l|=m}^{l_{\mt{max}}}\dot{\tilde{E}}^{\infty}_{ml}$ using $l_{\mt{max}}=20$ for $m \leq 20$. These formulas are rearranged to obtain $\Eps$ and $\Eps^{\infty}_{m}$ using \eqref{eq:finn-thorne} and \eqref{Flux_Harmonic_m}.

Finally for our Teukolsky based waveforms used in numerics section \ref{sec:Fish_Matrix_Numerics}, we use the \BHPT \ to extract the Teukolsky amplitudes $Z^{\infty}_{ml}(\tilde{r},a)$ and build an interpolant over $r$ for each harmonic $m = \{1, \ldots , l_{\text{max}} = 20\}$
\begin{equation}\label{eq:Teuk_Interp}
    G_{m}(\tilde{r},a) = \sum_{l=m}^{\infty}   \ _{-2}S^{am\tilde{\Omega}}_{ml}(\theta)\exp(i\phi)Z^{\infty}_{ml}(\tilde{r},a)
\end{equation}
for each viewing angle $(\theta,\phi) = (\pi/2,0)$ and $(\theta,\phi) = (0,0).$
To summarise, we use \eqref{eq:Teuk_Interp} in \eqref{eq:Teuk_Waveform_gen} to compute Fisher matrices numerically in section \ref{sec:Fish_Matrix_Numerics}. To aid our analytic study, we use the computed $\dot{E}_{\infty,m}$ in the waveform model \eqref{WaveformAllHarmonics} when evaluating the ratio \eqref{eq:final-ratio}. 


\subsection{Radial trajectory \& Waveform}\label{subsec:Radial_Trajectory}

\begin{figure}[!ht]
\includegraphics[width = .49\textwidth]{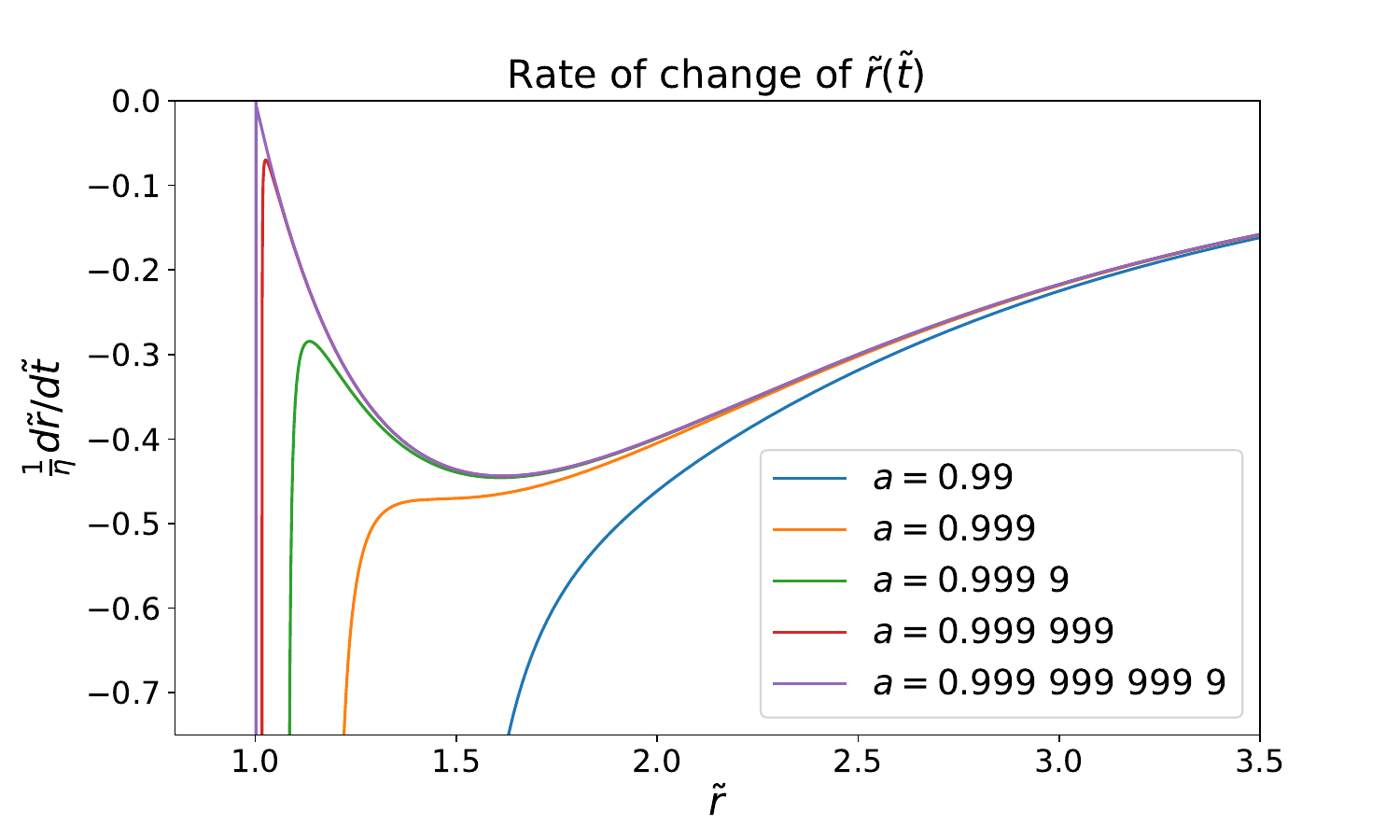}
\includegraphics[width = .49\textwidth]{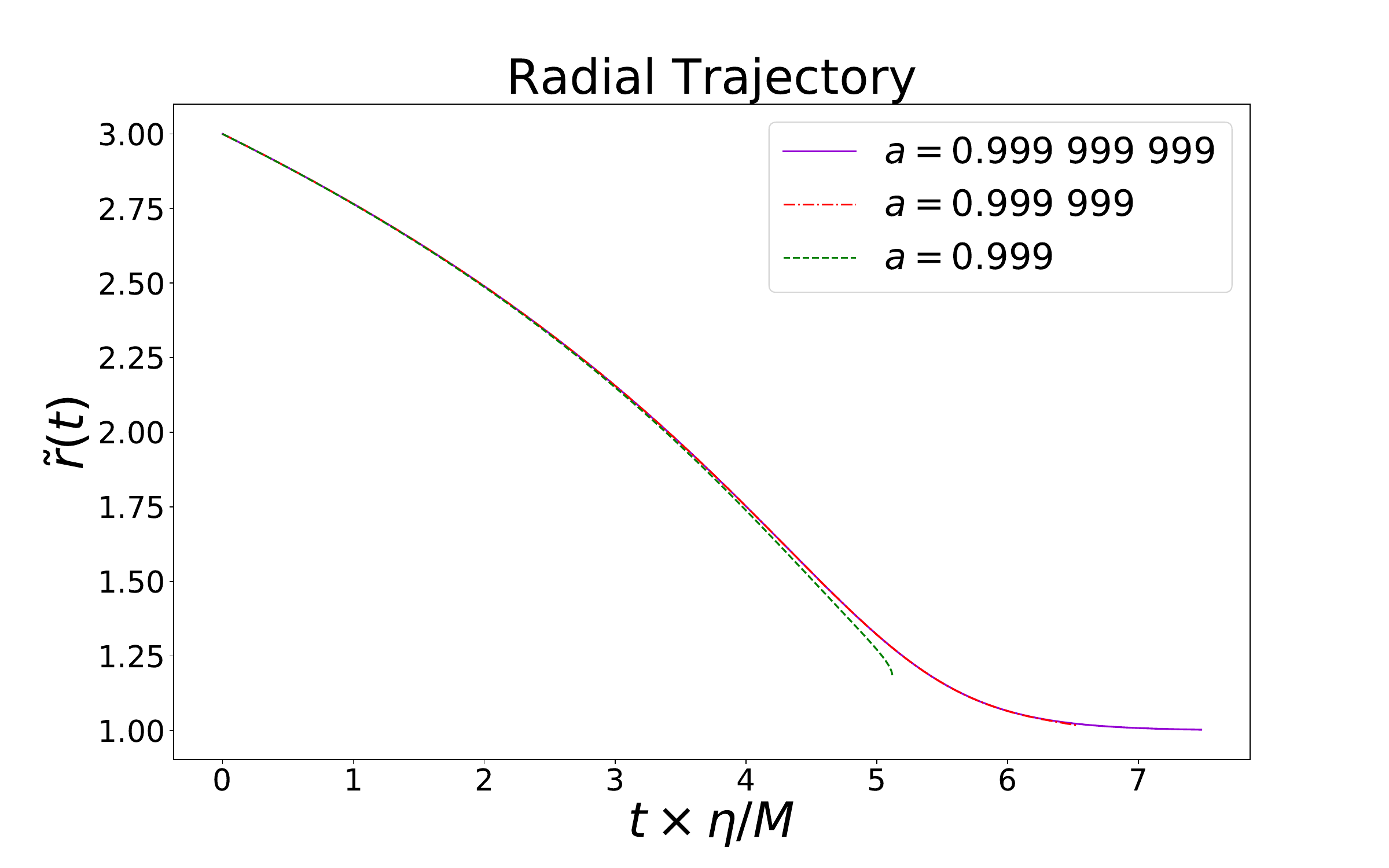}
\caption{The top panel shows how $d\tilde{r}/d\tilde{t}$ varies with $\tilde{r}$. The higher the spin parameter, the more time the secondary spends in the throat before plunge. The lower panel shows the corresponding inspiral trajectory. The dampening is clearly shown when the primary is near maximal spin, as seen in~\cite{2016CQGra..33o5002G}.}
\label{drdt_a_plot}
\end{figure}

The radial trajectory can be constructed by numerically integrating the ODE \eqref{eq1:orbital radius evolution} using an interpolant for $\Eps(\tilde{r})$ and suitable initial conditions. As before, we use the spin independent initial condition $\tilde{r}(\tilde{t}_{0} = 0) = \tilde{r}_{0}$. Fig.(\ref{drdt_a_plot}) shows some example radial trajectories for various spin parameters, computed using flux data from the \BHPT. In the high spin regime, the exponential decay of the radial coordinate is prominent as discussed in~\cite{2016CQGra..33o5002G,compere2019transition}. 
Throughout our simulations, the observation ends after a fixed amount of time, chosen such that this is before the transition to plunge for all parameter values used to compute the Fisher Matrix. This is important to avoid introducing artifacts from the termination of the waveform, given that the transition to plunge is not properly included in this waveform model. It is clear from Figure~(\ref{drdt_a_plot}) that larger the spin parameter, the longer the secondary spends in the dampening regime. See equation (22) of~\cite{2016CQGra..33o5002G} for further details.

\begin{figure}
    \centering
    \includegraphics[width = .49\textwidth]{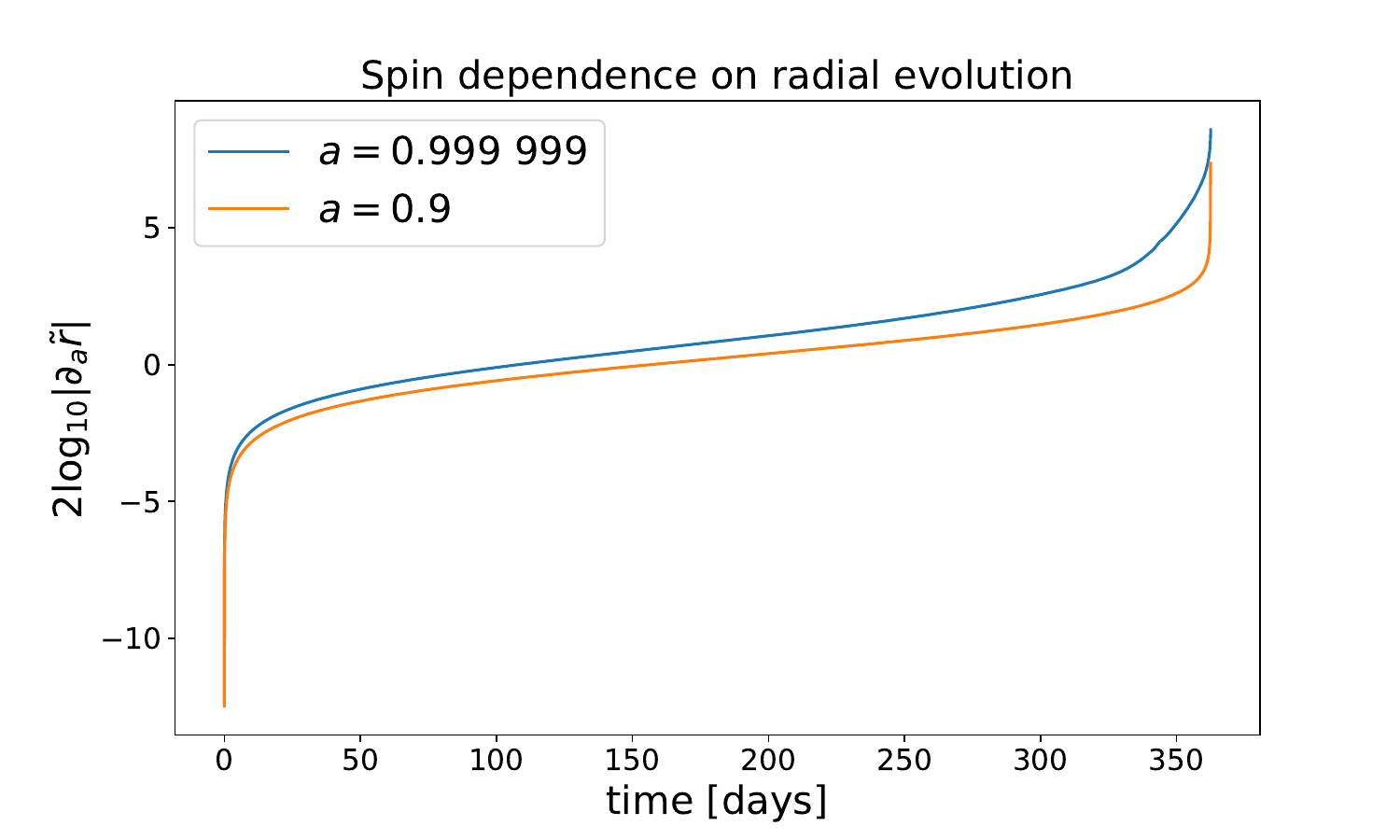}
    \caption{The blue curve is $\partial_{a}\tilde{r}$ for $a = 0.999999$. The orange curve is $\partial_{a}\tilde{r}$ for $a = 0.9$. Notice that the spin dependence on $r$ grows rapidly in the near-ISCO region of the rapidly rotating hole.}
    \label{fig:Spin_Dependence_Numerical}
\end{figure}

The spin dependence of the radial evolution can be calculated by integrating \eqref{eq1:orbital radius evolution} and then taking 
numerical derivatives. We consider two reference cases, both with component masses $\mu = 10M_{\odot}$ and $M = 2\times 10^{6}M_{\odot}$, but with different spin parameters $a = 0.9$ and $a = 1-10^{-6}$. We compute one year long trajectories, with $\tilde{r}(0) = 5.08$ in the first case and $\tilde{r}(0) = 4.315$ in the second. The spin derivative of the radial evolution can be calculated by perturbing the spin and using the symmetric difference formula for $\delta \ll 1$
\begin{equation}
    \frac{\partial \tilde{r}}{\partial a} \approx \frac{\tilde{r}(a + \delta,\tilde{t},\dot{\mathcal{E}}(a + \delta)) - \tilde{r}(a - \delta,\tilde{t},\dot{\mathcal{E}}(a - \delta))}{2\delta}.
\label{eq:def-spin-r}
\end{equation}
Figure \ref{fig:Spin_Dependence_Numerical} plots the quantity $|\partial_{a}\tilde{r}|^{2}$ appearing in the Fisher matrix estimation \eqref{eq:fish-estimation}. 
By inspection, it is clear that $|\partial_{a}\tilde{r}|^{2}$ is largest when the spin parameter is close to unity and when the radius is close to $\tilde{r}_{\text{isco}}$, matching our analytical conclusions using approximations \eqref{eq:throat-integral} and \eqref{eq:soln_approximation_all_spins}.

Using the semi-analytic model \eqref{WaveformAllHarmonics} we now evaluate the estimate \eqref{eq:final-ratio}, 
for the same two systems, but different $r_0$ to ensure that the assumptions made in deriving Eq.~\eqref{eq:final-ratio} still hold ($\tilde{r}_0=2.85$ for $a = 0.9$ and $\tilde{r}_0=1.2$ for $a=1-10^{-6}$). We choose termination points $\tilde{r}_{\text{cut}} = \tilde{r}_{\text{isco}} + \lambda$ with $\lambda \sim \{\lambda_{\text{ext}} = 10^{-4}, \lambda_{\text{mod}} = 10^{-2}\}$, just outside the transition region. 
Finally, the expression $\sum C_{\infty,m}$ was calculated using the \hyperlink{https://github.com/BlackHolePerturbationToolkit/MathematicaToolkitExamples}{high\_spin\_fluxes.nb} mathematica notebook in the  \BHPT, including harmonics up to $m = 10$.  We find the ratio to be
\begin{equation}\label{ratio_integrals}
    \frac{\Gamma^{\text{ext}}_{aa}}{\Gamma^{\text{mod}}_{aa}} \sim 500.
\end{equation}
giving a rough estimate that the spin precision increases by at least two orders of magnitude for these two sources. 

This verifies claims made in section \ref{subsec:analytic_fish}. When correlations with other parameters and the shape of the PSD are ignored, we predict a precision on the spin parameter roughly \emph{two} orders of magnitude higher than for moderately spinning black holes. 

\begin{figure*}
\includegraphics[width=\textwidth]{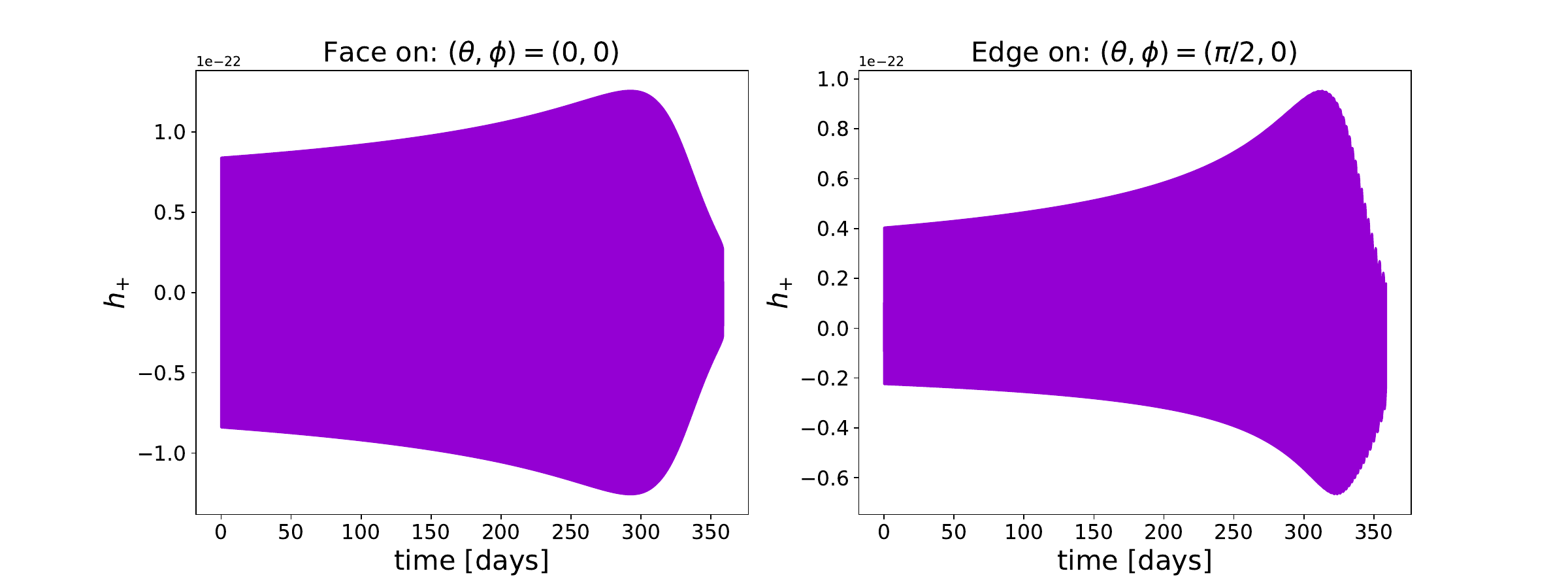}
\caption{A near-extremal waveform with parameters $\boldsymbol{\theta}_{\text{light}}$ viewed face-on (left) and edge-on (right). The dampening region lasts $\sim 55$ days. The edge-on case is asymmetric due to the large number of $l = 20$ modes and shows prominent relativistic beaming near the ISCO as observed in figure 3.b) of \cite{2016CQGra..33o5002G}.}
\label{TwoWaveforms}
\end{figure*}

To generate gravitational waveforms for the numerical study we use the Teukolsky waveform model ~\eqref{eq:Teuk_Waveform_gen}. The waveform depends on 
parameters $\boldsymbol{\theta}=\{a,\tilde{r}_0,\mu,M,\phi_{0},\tilde{D}\}$. 
We will consider two classes of near-extremal source, differentiated by the magnitude of their component masses and mass ratio. The first ``heavier" source has parameters
\begin{multline}\label{params:heavy}
    \boldsymbol{\theta}_{\text{heavy}} = \{\tilde{r}(t_{0} = 0) = 1.225, a = 1-10^{-6},\mu = 20M_{\odot}, \\ M = 10^{7}M_{\odot},\phi_{0} = \pi, \\
    D = \{D_{\text{edge}} = 1.8, D_{\text{face}} = 3\}\text{Gpc}\}
\end{multline}
and the second ``lighter'' source has
\begin{multline}\label{params:light}
    \boldsymbol{\theta}_{\text{light}} = \{\tilde{r}(t_{0} = 0) = 4.3, a = 1-10^{-6},\mu = 10M_{\odot}, \\ M = 2\times 10^{6}M_{\odot},\phi_{0} = \pi, \\
    D = \{D_{\text{edge}} = 1,D_{\text{face}} = 4\}\text{Gpc}\}.
\end{multline}
where $D_{\text{edge}}$ and $D_{\text{face}}$ refer to the distance if each source is viewed edge-on/face-on respectively. The distances are fine tuned\footnote{Strictly speaking, distance here is not a physical parameter since our waveform model does not include the LISA response to the strains $h_{+}$ and $h_{\times}$.} so that we achieve a signal to noise ratio of $\rho \sim 20.$ This is discussed later in section \ref{sec:Detectability}.
The lighter source is sampled with sampling interval $\Delta t_{s} \approx 4$ seconds and the heavier one with $\Delta t_{s} \approx 25$ seconds. We note here that $\Delta t_{s} = M\Delta\tilde{t}$ where $\Delta \tilde{t}$ is the dimensionless sampling interval used to integrate \eqref{eq1:orbital radius evolution}. The sampling interval is chosen from Shannon's sampling theorem such that $\Delta t_{s} < 1/(2f_{\text{max}})$, where
\begin{equation}
    f^{\text{edge}}_{\text{max}} = \frac{20}{2\pi} \frac{\tilde{\Omega}_{\text{isco}}}{M}, \quad 
    f^{\text{face}}_{\text{max}} =\frac{2}{2\pi} \frac{\tilde{\Omega}_{\text{isco}}}{M} 
\end{equation}
are the highest frequencies present in the waveform for the edge-on and face-on cases respectively. 
To illustrate, near-extremal waveforms with parameters $\boldsymbol{\theta}_{\text{light}}$ for both edge-on and face-on viewing angles are plotted in Fig.(\ref{TwoWaveforms}).

The lighter source is interesting because it exhibits both an ``inspiral" regime and a exponentially decaying regime that we will refer to as ``dampening". The heavier source is interesting because the dampening regime lasts more than one year and so the signal is in the dampening region for the entire duration of the observation. In the next section, we discuss detectability of these two types of sources by LISA.

\section{Detectability}\label{sec:Detectability}

The LISA PSD reaches a minimum around $3$mHz, and is fairly flat within the band from 1 to 100mHz. For an edge-on near-extremal inspiral with primary mass of $\sim 10^{7}M_{\odot}$, the dominant harmonic has a frequency of $\sim3.2$mHz at plunge, while the $m=20$ harmonic has frequency of $64$ mHz. Such heavy sources are thus ideal systems for observing the near-ISCO dynamics. For the lighter mass considered, $2 \times 10^6 M_{\odot}$, the near-ISCO dynamics are at frequencies a factor of $5$ higher, where the LISA PSD starts to rise. While the near-ISCO radiation will still be observable for these systems, its relative contribution to the signal will be relatively reduced. We therefore expect to obtain more precise spin measurements for the heavier of the two reference systems.

The discrete analogue of the optimal matched filtering SNR defined in Eq.~\eqref{continuous_SNR} 

\begin{equation}\label{eq:discrete_SNR}
\rho^{2} \approx \frac{4\Delta t_{s}}{N} \sum_{i=0}^{\left\lfloor N/2 + 1\right\rfloor}\frac{|\tilde{h}(f_{i})|^{2}}{ S_{n}(f_{i})}. 
\end{equation}
\newline

Here $N$ is the length of the time series, $\Delta t_{s}$ the sampling interval (in seconds) and $f_{i} = i/N\Delta t_{s}$ are the Fourier frequencies. 
In Eq.(\ref{eq:discrete_SNR}), the discrete time Fourier transform (DTFT) $\tilde{h}(f_{j})$ is related to the CTFT through $\hat{h}(f) = \Delta t_{s} \cdot \tilde{h}(f)$. 
To avoid problems with spectral leakage, prior to computing the Fourier transform, we smoothly taper the end points of our signals using the Tukey window
\begin{widetext}
\begin{equation}\label{Tukey_Window}
w[n] = \begin{cases} 
     \frac{1}{2}[1 + \cos(\pi(\frac{2n}{\alpha(N-1)} - 1))] & 0\leq n \leq \frac{\alpha(N-1)}{2} \\
      1 & \frac{\alpha(N-1)}{2}\leq n\leq (N-1)(1 - \alpha/2) \\
      \frac{1}{2}[1 + \cos(\pi(\frac{2n}{\alpha(N-1)} - \frac{2}{\alpha} + 1))] & (N-1)(1 - \alpha/2)\leq n \leq (N-1). 
      \end{cases}
      \end{equation}
   \end{widetext}
here $n$ is defined through $\tilde{t}_{n} = n\Delta \tilde{t}.$ The tunable parameter $\alpha$ defines the width of the cosine lobes on either side of the Tukey window. If $\alpha = 0$ then our window is a rectangular window offering excellent frequency resolution but is subject to high leakage (high resolution). If $\alpha = 1$ then this defines a Hann window, which has poor frequency resolution but has significantly reduced leakage (high dynamic range). For the heavier source, we use $\alpha = 0.25$ to reduce leakage effects significantly and frequency resolution is not a problem since the frequencies of the signal are contained within the LISA frequency band (for all harmonics). For the lighter source, we use $\alpha = 0.05$ to reduce edge effects while retaining the ability to \emph{resolve} the frequencies where the signal is dampened. We found that calculated SNRs and parameter measurement precisions are insensitive to the choice of $\alpha$ in the heavier system. The lighter system is more sensitive: for larger $\alpha$, more of the dampening regime is lost, with a corresponding impact on the measurement precisions. We believe that $\alpha = 0.05$ is large enough to reduce leakage but small enough to resolve as much of the dampening regime as possible.

After tapering, we zero pad our waveforms to an integer power of two in length, in order to facilitate rapid evaluation of the DTFT using the fast Fourier transform. 
Computing the SNR in this way gives $\rho \sim 20$ for the light and heavy sources respectively when viewed both edge-on and face-on under the configuration of parameters $\boldsymbol{\theta}_{\text{light}}$ and $\boldsymbol{\theta}_{\text{heavy}}$. 

In all cases we marginally exceed the threshold of $\rho \approx 20$ which is typically assumed to be required for EMRI detection in the literature~\cite{babak2017science,chua2017augmented}. 

As mentioned above, the lighter source exhibits two regimes of interest - the initial gradually chirping phase, where the waveform resembles those for moderately spinning primaries, and then the exponentially damped phase while the secondary is in the near-horizon regime. It is natural to ask what proportion of the SNR, and later what proportion of the spin measurement precision, is contributed by each regime. For both edge-on and face-on systems, we separate the two parts of the waveform using Tukey windows and compute the SNR contributed by each part to find 
\begin{align}\label{SNR:light_source}
    \rho_{\text{face-on}}^{2} &\sim \begin{cases} 83\% & \text{Outside Dampening region} \\
        17\% & \text{Dampening region}.
        \end{cases}  \\
    \rho_{\text{edge-on}}^{2} &\sim \begin{cases} 96\% & \text{Outside Dampening region} \\
        4\% & \text{Dampening region}.
        \end{cases}
\end{align}
For the face-on source, there is just a single dominant harmonic, and the frequency of this harmonic is such that it lies in the most sensitive part of the LISA frequency range. This helps to enhance the relative SNR contributed by the dampening region. The edge-on source, by contrast, has multiple contributing harmonics, which are spread over a range of frequencies, and the proportional contribution of the dampening region to the overall SNR is therefore diminished.

\begin{figure*}
    \centering
    \includegraphics[width = \textwidth]{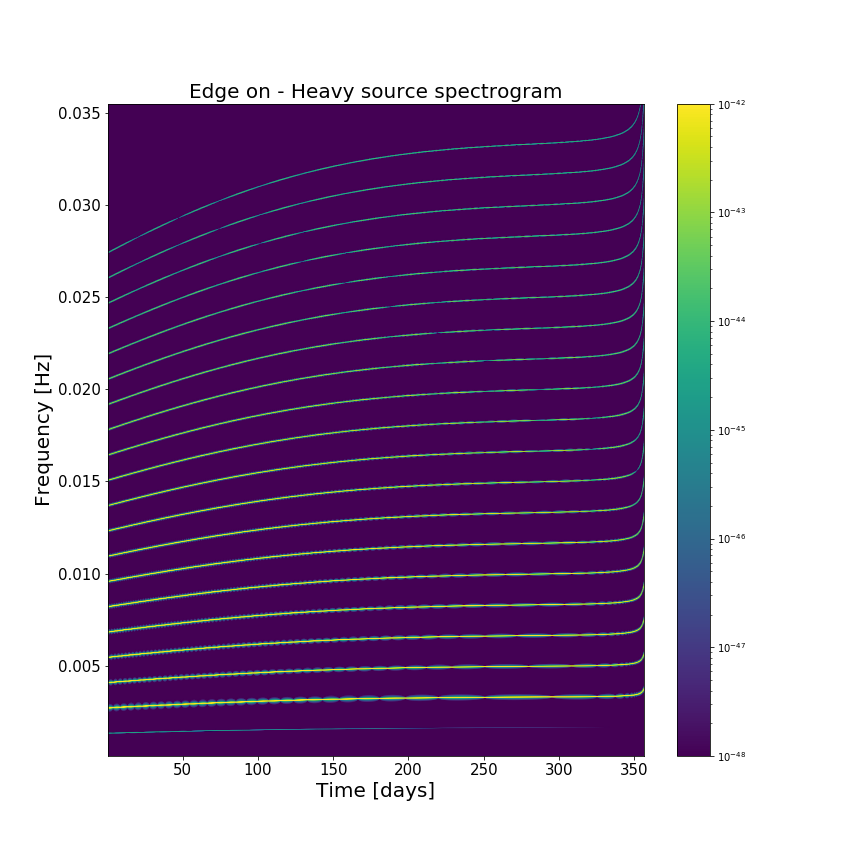}
    \caption{Here we plot the spectrogram  of $h(\boldsymbol{\theta}_{\text{heavy}};t)$ viewed edge on. We see 20 tracks in the time-frequency plane corresponding to the $m\in\{1,\ldots,20\}$ harmonics. The colorbar shows that the $m=2$ harmonic (second lowest track in frequency) is dominant, but that there are several other harmonics which contribute significantly to the radiated power}. 
    \label{fig:spectrogram}
\end{figure*}

For a non-evolving signal the SNR accumulates like $\sqrt{T_{\mt{obs}}}$, where $T_{\mt{obs}}$ is the total observation time. The pre-dampening regime lasts 308 days, and so from duration alone we would expect a fraction $\sqrt{308/365} \approx 93\%$ of SNR to be accumulated there. The difference to what we find above is explained by differences in amplitudes of the individual harmonic(s). The heavier system is within the dampening regime throughout the last year of inspiral and so all of the SNR of $\rho\sim 20$ is accumulated there. 
This may seem counter-intuitive given the exponential decay of the signal during the dampening regime. However, the exponential decay rate is relatively slow, a large number of harmonics contribute to the SNR and the emission is all within the most sensitive range of the LISA detector. This is clear from looking at the time-frequency spectrogram of the heavier signal shown in Fig.(\ref{fig:spectrogram}).
What we learn from this figure is that there are a significant number of harmonics that have comparable power to the dominant $m = 2$ harmonic. We see also that the angular velocity at each harmonic, and thus $f_{m}$, shows little rate of change for $M \sim 10^{7}$ and $\eta \sim 10^{-6}$. This is consistent with~\cite{compere2018_NHEK,2016CQGra..33o5002G}, where it was shown that a large number of $m$ harmonics is required to produce an accurate representation of the gravitational wave signal for a near-extremal EMRI, particularly for near edge-on viewing angles. For moderately spinning black holes $a \sim 0.9$ there are not as many dominant harmonics, so those waveforms are cheaper to evaluate.


We are now ready to move on to compute Fisher Matrix estimates of parameter measurement precisions. This will be the focus of the next section.

\section{Numerics: Fisher Matrix}\label{sec:Fish_Matrix_Numerics}
\begin{figure*}
    \centering
    \includegraphics[width = 0.495\textwidth]{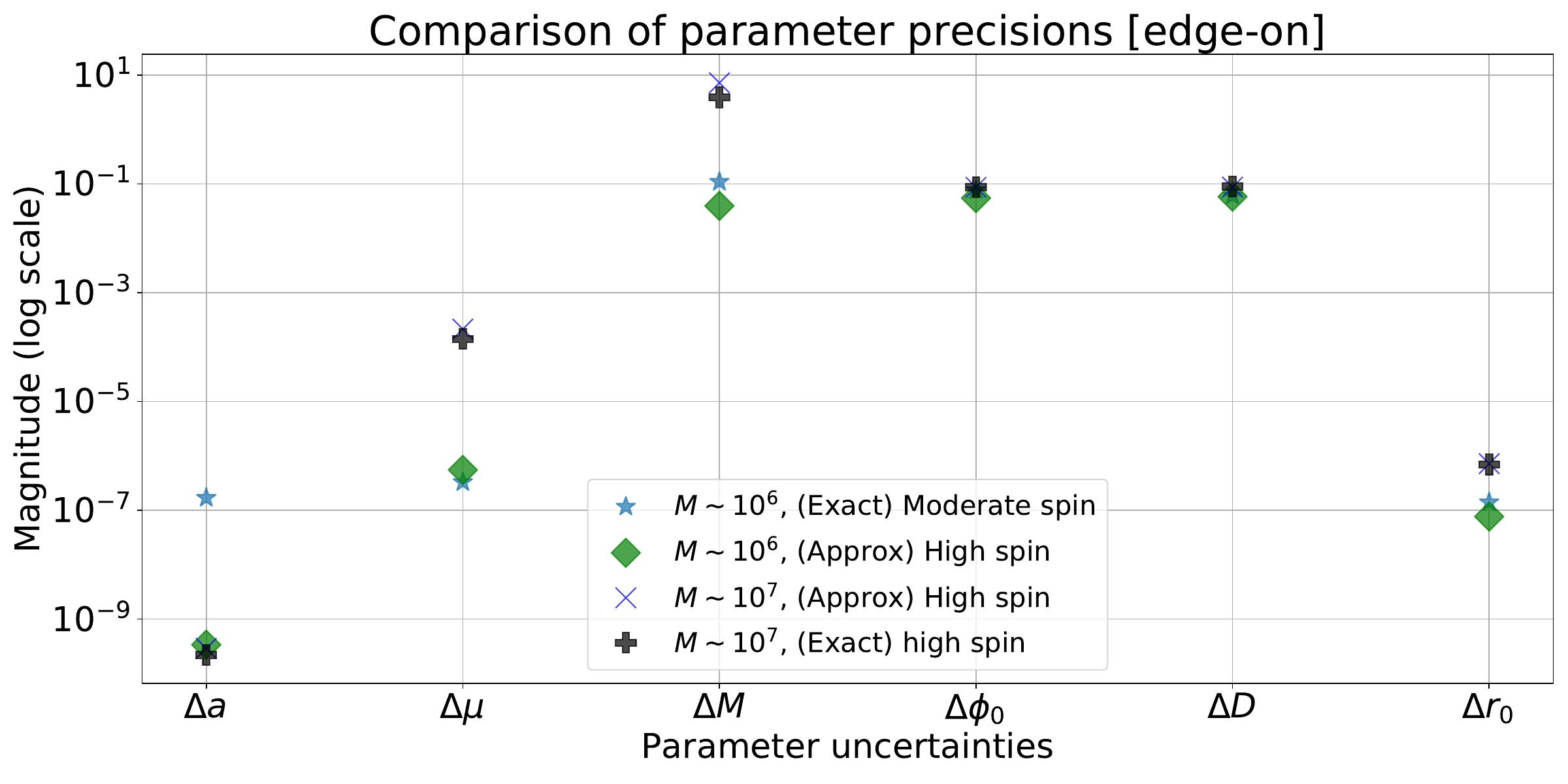}
    \includegraphics[width = 0.495\textwidth]{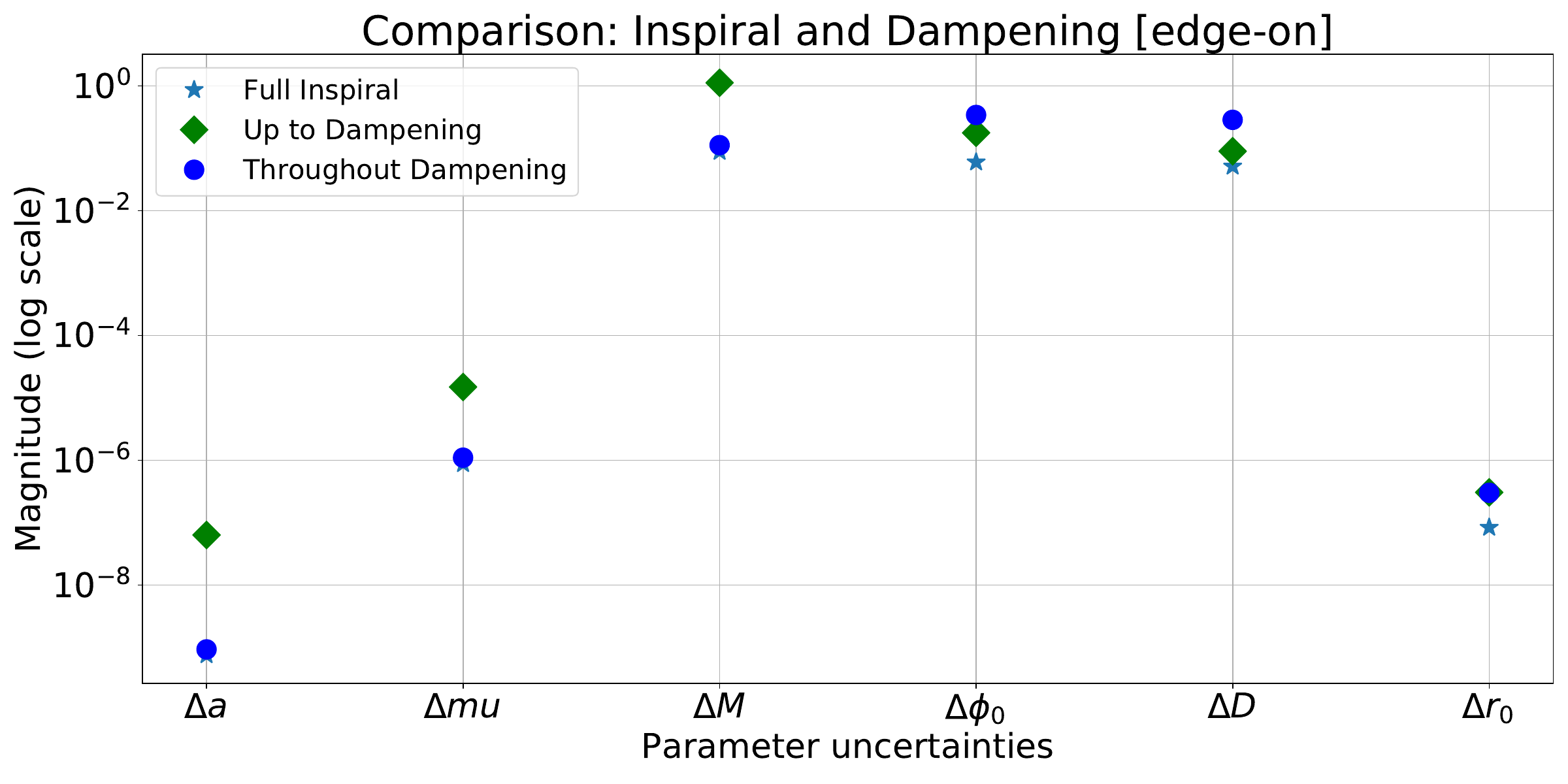}
    \caption{(Left plot) Parameter measurement precision, as estimated using the Fisher Matrix formalism, for the three reference sources, with parameters  $\boldsymbol{\theta}_{\text{light}}$ (green diamonds), $\boldsymbol{\theta}_{\text{heavy}}$ (purple crosses) and $\boldsymbol{\theta}_{\text{mod}}$ (blue asterisks). The black diamonds show the precisions obtained when including the spin-dependence of the relativistic corrections, $\Eps$ in the waveform model for the heavy source. (Right plot) Parameter measurement precisions for the source with parameters $\boldsymbol{\theta}_{\text{light}}$, computed using the full waveform (blue asterix), only the inspiral phase (blue dot) and only the dampening phase (green diamond).}   \label{fig:Fisher_Matrix_Results}
\end{figure*}

We now compute \eqref{Fisher_Matrix} numerically without making the simplifying assumptions used in Sections~\ref{subsec:analytical_traj} and~\ref{subsec:analytic_fish}. We will use one simplification, which is to ignore the spin dependence in $\Eps$, $Z^{\infty}_{lm}(\tilde{r},a)$ and $_{-2}S^{am\tilde{\Omega}}_{lm}(\theta,\phi)$  and fix these at the values computed for $a=1-10^{-9}$ using the \BHPT. We argued in Eq.~\eqref{eq:kinematics_vs_GRCs_equation} that the spin dependence of the flux correction is a sub-dominant contribution in the near-ISCO regime, and this is further justified in Appendix \ref{app:GRCs} (see Fig.~\ref{fig:spin_dependence_corrections} in particular). While $\partial_a \Eps$ does grow as the ISCO is approached, it remains sub-dominant to the spin dependence of the kinematic terms. This approximation is probably conservative in the sense that we are removing information about the spin from the waveform model and so the true measurement precision is most likely higher. Nonetheless we expect this to be a small effect, and have verified that relaxing this assumption does not significantly change the result for the heavier reference source (see Figure~\ref{fig:Fisher_Matrix_Results}). We note that we make this assumption only for computational convenience. Waveform models used for parameter estimation on actual LISA data should use the most complete results available to ensure maximum sensitivity and minimal parameter biases.

To compute the waveform derivatives required to evaluate~\eqref{Fisher_Matrix}, we use the fifth order stencil method
\begin{equation}\label{Numerical_Derivative}
\frac{\partial f}{\partial x} \approx  \frac{-f_{2} + 8f_{1}  - 8f_{-1}+ f_{-2}}{12\delta x},
\end{equation}
for $\delta x \ll 1$ and $f_{i} = f(x + i\delta x)$. To avoid numerical instability of $\partial_{a}h$ for the near-extremal spin values of $a \leq 1-10^{-9}$, we ensure that $\delta x < 1-a$ so the perturbed waveform does not have spin exceeding $a = 1$. We further assume that $\partial_{a}\tilde{t}_{\text{end}}$ is zero so there is \emph{no} spin dependence on the total observation time. 

In addition to the sources with parameters $\boldsymbol{\theta}_{\text{heavy}}$ and $\boldsymbol{\theta}_{\text{light}}$, we now consider a third source with parameters
\begin{multline}\label{params:mod}
    \boldsymbol{\theta}_{\text{mod}} = \{\tilde{r}(t_{0} = 0) = 5.01, a = 0.9,\mu = 10M_{\odot}, \\ M = 2\cdot10^{6}M_{\odot},\phi_{0} = \pi,D_{\text{edge}} = 1\text{Gpc} \},
\end{multline}
with SNR $\sim 20$.

Fisher matrix estimates of parameter measurement precisions for all three sources viewed edge-on are shown in Figure~\ref{fig:Fisher_Matrix_Results}. 
We do not present the results for a face-on observation as they are near-equivalent to the measurements presented in figure \ref{fig:Fisher_Matrix_Results} for equivalent SNR. 

We see from this figure that we should be able to constrain the spin parameter of near-extremal EMRI sources to a precision as high as $\Delta a \sim 10^{-10}$, even when accounting for correlations amongst the waveform parameters. This is true for both the lighter and the heavier sources viewed edge-on and face-on, with a constraint a factor of a few better for the heavier source. The right panel of the figure compares the contribution to the measurement precision for the lighter source from the two different phases of the signal. We see that the high spin precision comes almost entirely from the observation of the dampening regime and this phase of the signal contributes much more information than we would expect based on its contribution to the total SNR\footnote{In \eqref{eq:fish-estimation}, the growth of $\partial_{a}\tilde{r}$ exceeds the growth of $S_{n}(f) \sim \text{const}$ in the dampening regime.  This sources the high precision measurement.}. 

The spin measurement precision for the near-extremal systems is three orders of magnitude better than for the system with moderate spin, while all other parameter measurements are comparable. 

Comparing to the exact Fisher matrix result with spin dependence included in all the various terms, we see that the two precisions are almost \emph{identical}: the exact result offers precisions that are marginally better in comparison to our approximate result (removing spin dependence from the corrections). This Figure thus justifies ignoring the spin dependence of $\Eps$, since relaxing that assumption makes almost no difference to the results. This numerically confirms our belief that the spin dependence in the corrections to the fluxes are \emph{subdominant} in the analysis leading to \eqref{eq:fish-estimation}. 
In the same plot \ref{fig:Fisher_Matrix_Results}, we also compare results of near-extremal black holes to moderately spinning holes. A direct comparison shows an increase in the spin precision by $\sim 3$ orders of magnitude, which agrees with the intuition given by the earlier analytic analysis, Eq.~\eqref{ratio_integrals}. 

To our knowledge, these are the first circular and equatorial parameter precision studies for EMRIs that have employed Teukolsky-based \emph{adiabatic} waveforms, rather than approximate waveform models (or ``kludges''), which have been used for many studies~\cite{chua2017augmented,babak2007kludge,barack2004lisa}. 
Comparing our results for the moderately spinning system to these previous studies, we find that our results are very comparable, but a factor of a few \emph{tighter}. This could be because we are including only a subset of parameters and ignoring the details of the LISA response, or because we have a more complete treatment of relativistic effects. A more in depth study addressing both of these limitations would be needed to understand the origin of the differences. However, the agreement between our results and previous studies is sufficiently close, and considerably less than the difference we find between the moderate and near-extremal spin cases, to give us confidence that our results are not being unduly influenced by these simplifications.

\begin{figure*}
    \includegraphics[width = 0.495\textwidth]{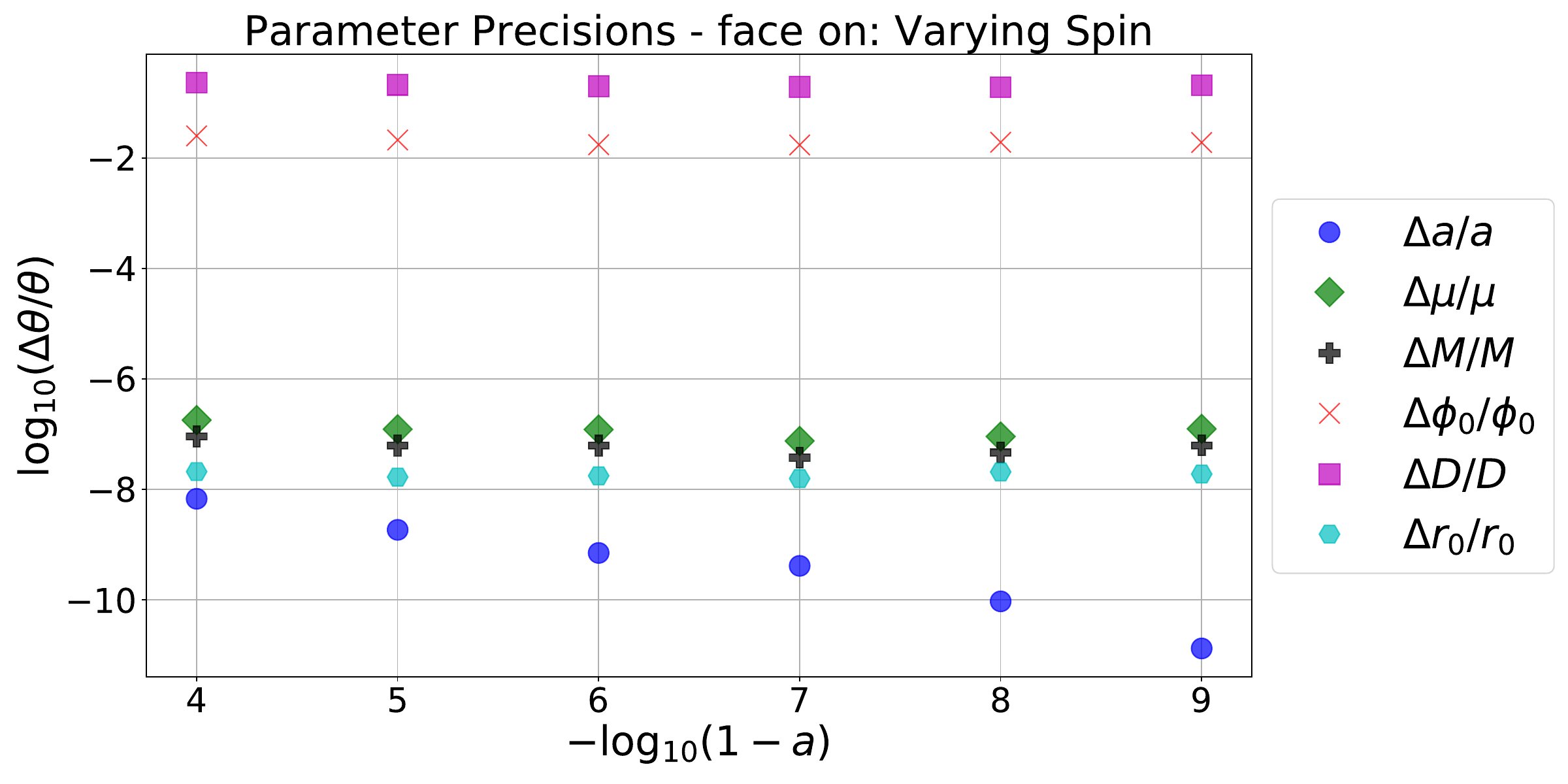}
    \includegraphics[width = 0.495\textwidth]{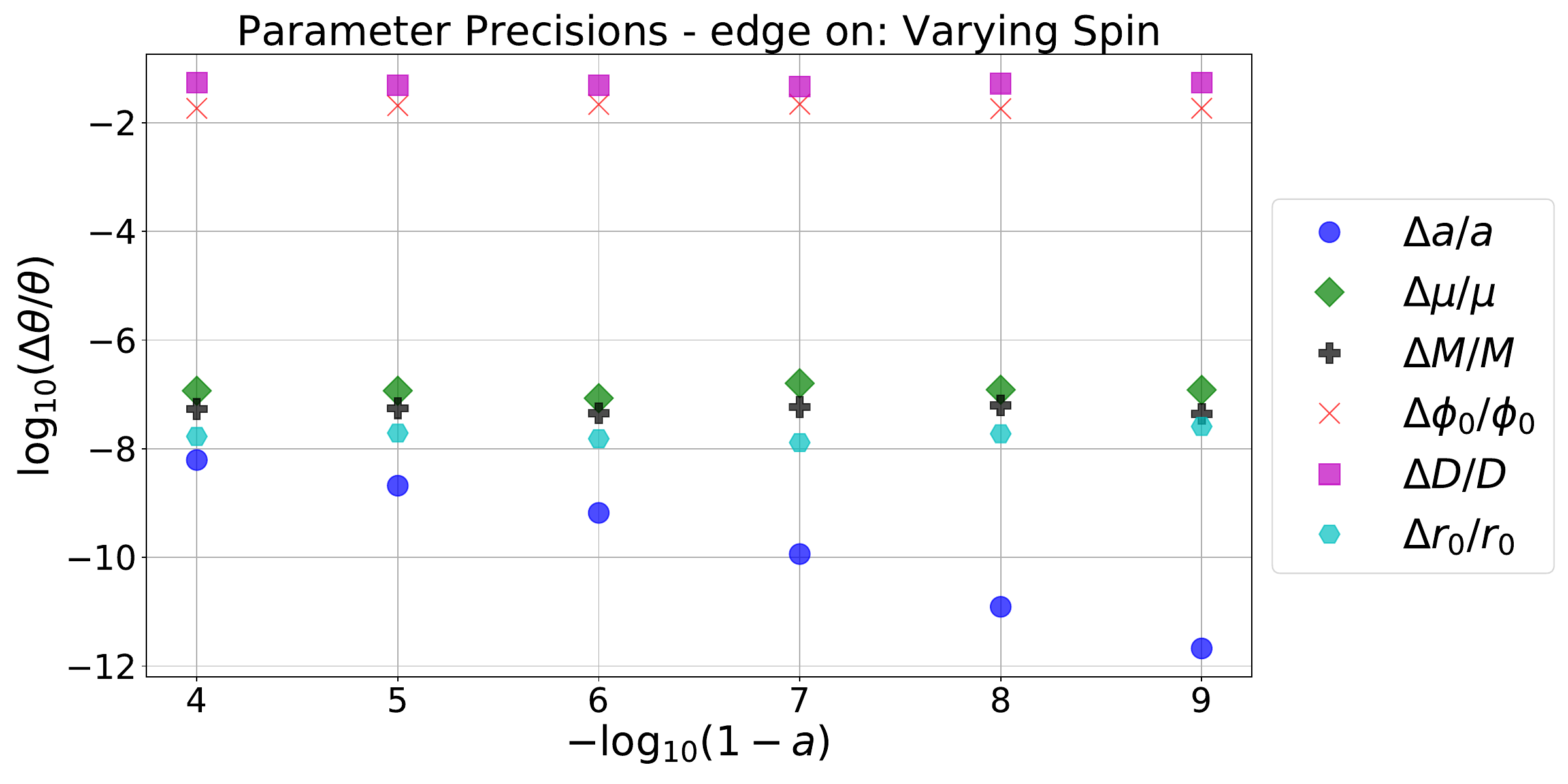}
\caption{We keep $\boldsymbol{\theta}_{\text{light}}\backslash\{a\}$  fixed and vary $a = 1-10^{-i}$ for $i \in \{4,\ldots,9\}$ while computing estimates on the precision of the measured parameters using the Fisher Matrix. Results are shown for sources viewed face-on (left) and edge-on (right).}
\label{fig:Fisher_Matrix_Whole_Param_Plots}
\end{figure*}

In Figure~(\ref{fig:Fisher_Matrix_Whole_Param_Plots}) we show how the parameter estimation precision for the source with parameters $\boldsymbol{\theta}_{\text{light}}$ changes as we vary the spin parameter, while keeping all other parameters unchanged. We present results for both face-on and edge-on viewing angles. This shows that while the measurement precision for most of the parameters is largely independent of spin in the near-extremal regime, the spin precision steadily increases as $a \to 1$. We note that even at a spin of $1-10^{-9}$, the measurement precision satisfies the constraint $\Delta a < |1-a|$ and therefore a LISA EMRI observation would be able to resolve that the system was not maximally extremal, i.e., that $a < 1$.  We stop at $1-10^{-9}$ since the derivative using Eq.\eqref{Numerical_Derivative} begins to misbehave. 

Due to large condition numbers, inverting Fisher matrices for EMRI sources is a highly non-trivial task.
In appendix \ref{app:verification}, we provide multiple diagnostic tests of our Fisher matrix algorithm and verify that, in the single parameter case, the spin parameter precision is a suitable representation of the $1\sigma$ width of the Gaussian likelihood as shown in figure \ref{fig:Likelihood_Fisher_Plot}. These single parameter tests of the Fisher matrix are useful tests to verify that a single parameter algorithm yields sensible results. However, real instabilities of the numerical procedure are prominent the moment the inverse of the Fisher matrix is performed when correlations are present. Hence, it is both necessary and sufficient to verify our Fisher matrix calculations using an \emph{independent} procedure. The next section is dedicated to performing a parameter estimation study on both near-extremal EMRIs with parameters $\boldsymbol{\theta}_{\text{light}}$ and $\boldsymbol{\theta}_{\text{heavy}}.$

\section{Numerics: Markov Chain Monte Carlo}\label{MCMC}
The Fisher Matrix is a local approximation to the likelihood, valid in the limit of sufficiently high signal-to-noise ratio. We can verify that this local approximation is correctly representing the parameter measurement uncertainties by numerically evaluating the likelihood using Markov Chain Monte Carlo. To reduce the computational cost of these simulations we use a face-on viewing profile and thus only consider the $m = 2$ harmonic. 
We have shown in figure \eqref{fig:Fisher_Matrix_Whole_Param_Plots} that parameter precision measurements are not largely dependent on the choice of viewing angle for the lighter source. We have further verified this claim for the heavier source.

\subsection{Markov Chain Monte Carlo}
Markov Chain Monte Carlo (MCMC) methods were developed for Bayesian inference to sample from the posterior probability distribution, $p(\boldsymbol{\theta}|d)$, which is given by Bayes' theorem as
\begin{equation}
    \log p(\boldsymbol{\theta}|d) \propto \log p(d|\boldsymbol{\theta}) + \log p(\boldsymbol{\theta})
\end{equation}
where $p(d|\boldsymbol{\theta})$ is the likelihood function, and $p(\boldsymbol{\theta})$ is the prior probability distribution on the parameters. In our context the likelihood is given by Eq.~\eqref{log-likelihood-function} and we will assume independent priors such that
\begin{equation}\label{log_posterior}
\log p(\boldsymbol{\theta}|d) \propto -\frac{1}{2}(d - h(t;\boldsymbol{\theta})|d - h(t;\boldsymbol{\theta})) + \\ \sum_{\theta^{i} \in \boldsymbol{\theta}} \log p(\theta^{i}).
\end{equation}
We generate a data set $d(t) = h(t;\boldsymbol{\theta}_{tr}) + n(t)$ by specifying the waveform parameters, $\boldsymbol{\theta}_{tr}$, of the injected signal and generating noise in the frequency domain
\begin{equation}\label{eq:noise_realisations}
    \tilde{n}(f_{i}) \sim N(0,\sigma^{2}(f_{i})), \ \sigma^{2}(f_{i}) \approx \frac{NS_{n}(f_{i})}{4\Delta t}.
\end{equation}
We use MCMC to sample from the posterior distribution \eqref{log_posterior}, employing a standard Metropolis algorithm with proposal distribution $q(\boldsymbol{\theta}^{\star} | \boldsymbol{\theta}^{i-1})$ equal to a multi-variate normal distribution, centred at the current point and with a fixed covariance. We take flat priors on all of the waveform parameters, since the goal is to check the validity of the Fisher matrix approximation to the likelihood. The algorithm proceeds as follows
\begin{enumerate}
    \item We start the algorithm close to the true values $\boldsymbol{\theta}_{0} = \boldsymbol{\theta}_{\text{tr}} + \boldsymbol{\delta}$ for $||\boldsymbol{\delta}|| \ll 1$. For iteration $i = 1,2,\ldots,N$
    \item Draw new candidate parameters $\boldsymbol{\theta}^{\star}  \sim  q$  and generate the corresponding signal template $h(t;\boldsymbol{\theta}^{\star})$.
    \item Using \eqref{log_posterior}, compute the log acceptance probability
    \begin{equation*}
        \log(\alpha) = \min[0,\log P(\boldsymbol{\theta}^{\star}|d,\boldsymbol{\theta}^{i-1}) - \log P(\boldsymbol{\theta}^{i-1}|d,\boldsymbol{\theta}^{\star})].
    \end{equation*}
    We note that we are using a symmetric proposal distribution and so the usual proposal ratio is not required.
    \item Draw $u \sim U[0,1]$.
    \begin{enumerate}
        \item If $\log u < \log \alpha$ we accept the proposed point and set $\boldsymbol{\theta}^{i}=\boldsymbol{\theta}^{\star}$.
        \item Else we reject the proposed point and set $\boldsymbol{\theta}^{i}=\boldsymbol{\theta}^{i-1}$. 
    \end{enumerate}
    \item Increment $i\rightarrow i + 1$ and go back to step 2 until $N$ iterations have been completed. 
\end{enumerate}
Since we know the true parameters we can start the algorithm in the vicinity of the true parameters and do not need to discard the initial samples as burn-in, allowing us to generate useful samples more quickly. 
In principle, the MCMC algorithm should converge for any choice of proposal distribution, but proposals that more closely match the shape of the posterior should lead to more rapid convergence. As we expect that the proposal should be approximated by the Fisher Matrix, we set the covariance matrix of the normal proposal distribution to be equal to the inverse Fisher matrix, evaluated at the known injection parameters.

\subsection{Results}
We compute MCMC posteriors for the two signals $h(\{\boldsymbol{\theta}_{\text{heavy}},\boldsymbol{\theta}_{\text{light}}\};t)$ for the waveform model \eqref{eq:Teuk_Waveform_gen} for the face-on case only. As before, we construct waveforms ignoring the spin dependence in $\dot{\mathcal{E}}$, the Teukolsky amplitudes $Z^{\infty}_{lm}$ and the spheroidal harmonics $_{-2}S^{a\tilde{\Omega}}_{lm}(\theta,\phi)$. We evaluate these for a fixed spin parameter of $a = 1-10^{-9}$. 

We remind the reader the main drive for the tight constraints on the spin parameter is due to the spin dependence induced through the kinematic terms present in \eqref{eq:Teuk_Waveform_gen}, as discussed in section~\ref{sec:Fish_Matrix_Numerics}, and not its dynamical terms, justifying our approximation.

The priors on $a, \phi_{0}$ and $D$ for both sources were 
\begin{align*}
    a &\sim 1 - U[10^{-4},10^{-8}] \\
    \phi_{0} &\sim U[0, 2\pi] \\
    D &\sim U[1,8]\text{Gpc}.
\end{align*}
The priors on $\mu, M$ and $\tilde{r}_{0}$ were chosen differently for the heavy and light source as
\begin{align*}
    \mu_{\text{heavy}} &\sim U[18,22] M_{\odot}\\
    \mu_{\text{light}} & \sim U[8,12]M_{\odot} \\ 
    M_{\text{heavy}} &\sim U[0.9,1.1] \times 10^{7}M_{\odot} \\
    M_{\text{light}} & \sim U[1.9,2.1] \times 10^{6}M_{\odot}. \\
    \tilde{r}_{0}^{\text{heavy}} & \sim U[1.2,1.3] \\
    \tilde{r}_{0}^{\text{light}} & \sim U[4.2,4.4]
\end{align*}
The prior on $a$ ensures that we do not move outside the range in which our approximations are valid, $a \gtrsim 0.9999.$ The tight priors on the individual component masses helped to improve the computational efficiency of our algorithm. However, there was no evidence of the MCMC chains reaching the edges of the priors in our simulations, so we are confident these restrictions are not influencing the results.

Evaluating the likelihood for EMRI waveforms is an expensive procedure. In order to obtain a sufficient number of samples from the posterior, we used high performance computing facilities and ran 20 unique chains for $N = 40,000$ iterations. 
All chains analysed the same input data set, but with different initial random seeds. This ensures that the dynamics of the chains are different but the noise realisations are the same for each MCMC procedure. 

\begin{figure*}
\centering
\includegraphics[height=14cm,width = 16cm]{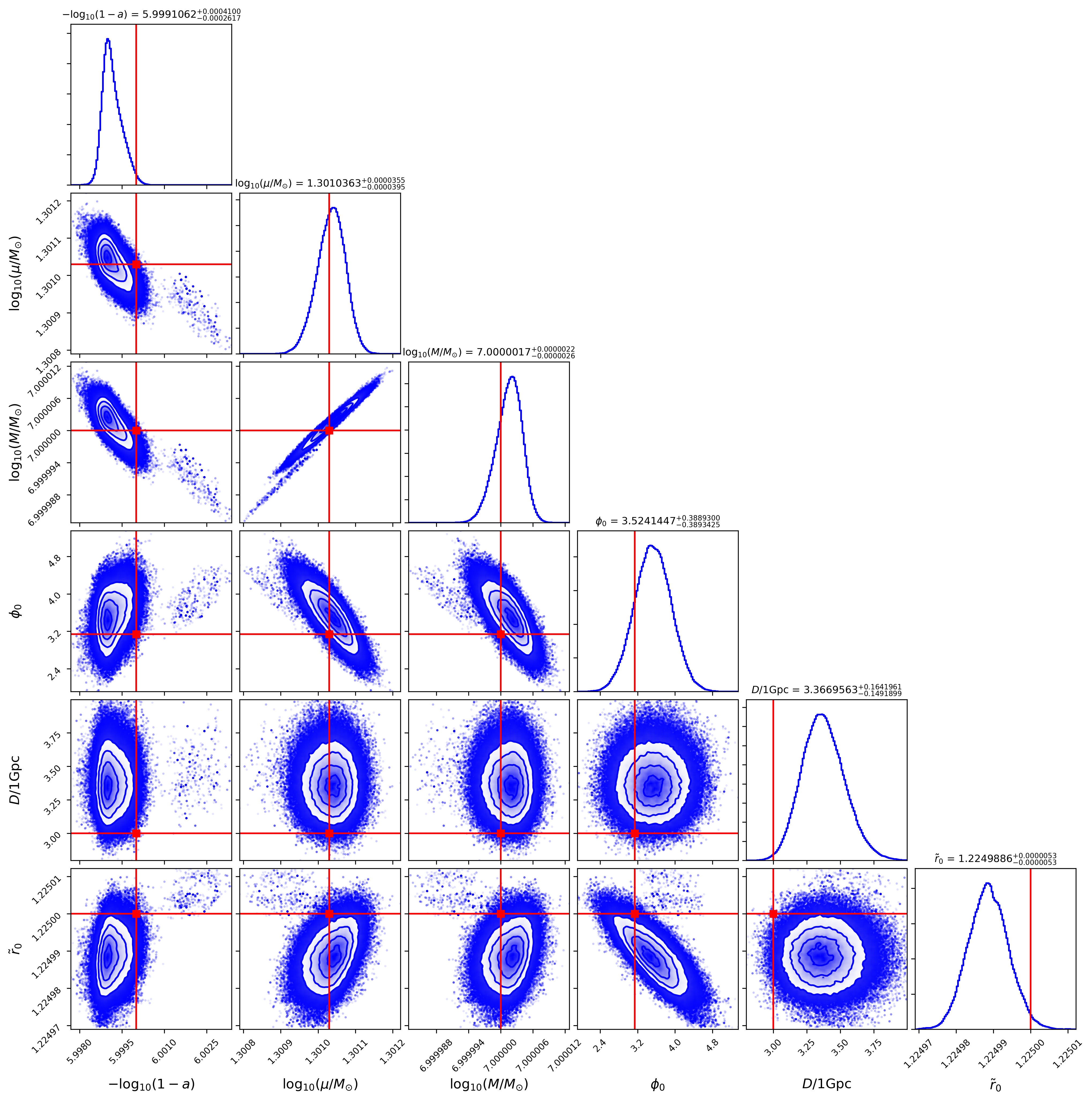}
\caption{The diagonal plots represent the marginalised posterior distributions on the parameters $\boldsymbol{\theta}_{\text{heavy}}.$ The plots below the diagonal are the joint two-dimensional posterior distributions. The red lines indicate the true values of the injected signal.}
\label{Corner_Plot_heavy}
\end{figure*}

\begin{figure*}
\centering
\includegraphics[height=14cm,width = 16cm]{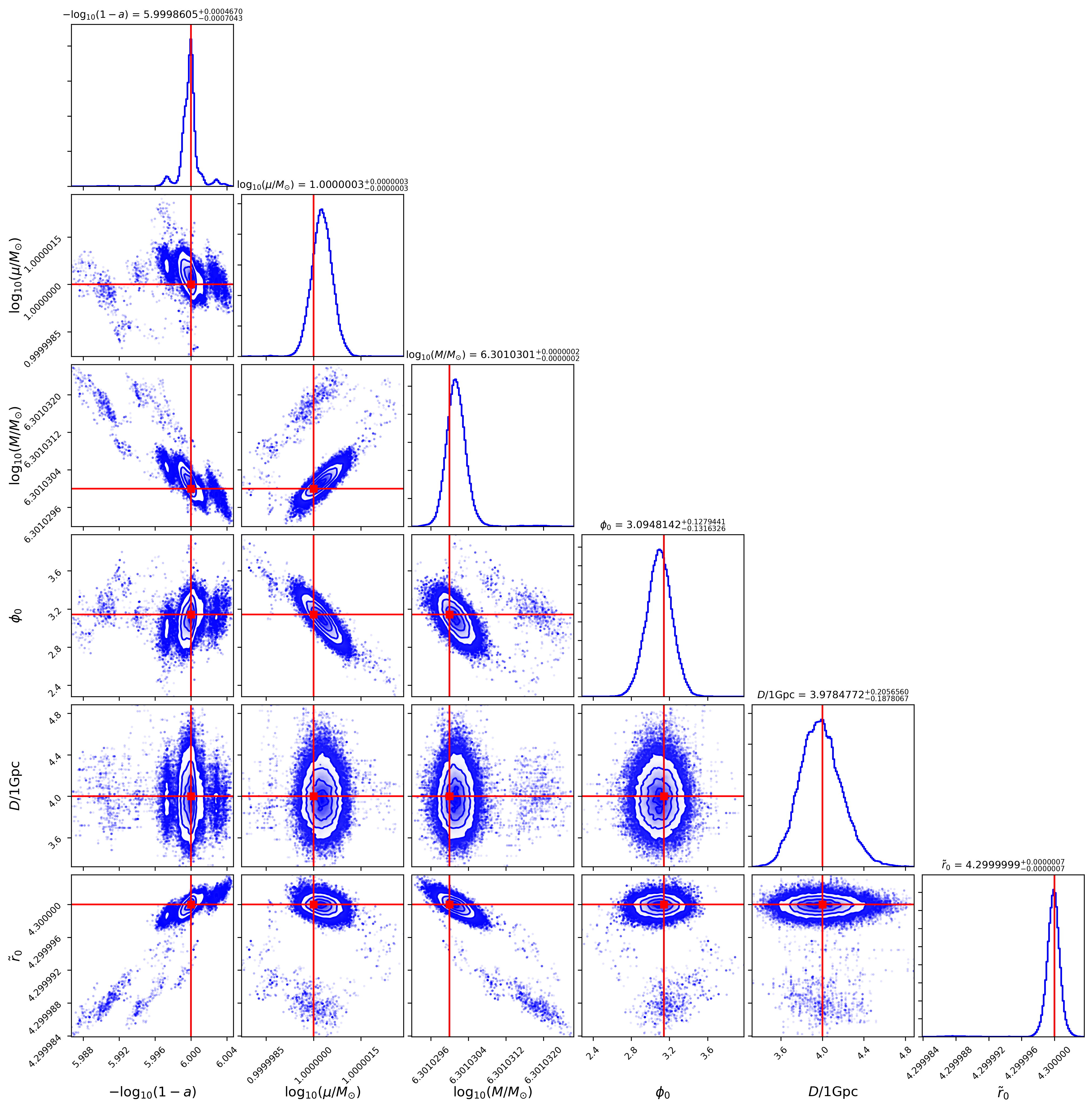}
\caption{As Figure~\ref{Corner_Plot_heavy}, but now for the source with parameters $\boldsymbol{\theta}_{\text{light}}.$}
\label{Corner_Plot_light}
\end{figure*}

The marginal posterior distributions and two-dimensional contour plots for the two sources are shown in figures (\ref{Corner_Plot_heavy}) and (\ref{Corner_Plot_light}).
These plots confirm the high precisions of parameter measurements that were seen with the Fisher Matrix. The relative uncertainties $\Delta \theta/\theta$ are similar for the two sources, although we can measure the spin parameter more precisely for the heavier source. For the most part the posteriors are unimodal, apart from the spin posterior of the lighter source. We have verified that the secondary modes are real features of the likelihood, and correspond to the waveform phase shifting by one cycle within the late dampening regime. We also note that shifts in the peak of the posterior away form the true value are larger for the heavier source than for the lighter source. This appears to be due to the particular noise realisation. For other noise realisations the noise-induced biases for the heavier source are smaller. For noise-free data sets, we find posterior distributions peaked at the true parameters, as expected.
%

\begin{figure*}
    \centering 
\begin{subfigure}{0.25\textwidth}
  \includegraphics[height =4cm, width = 5cm]{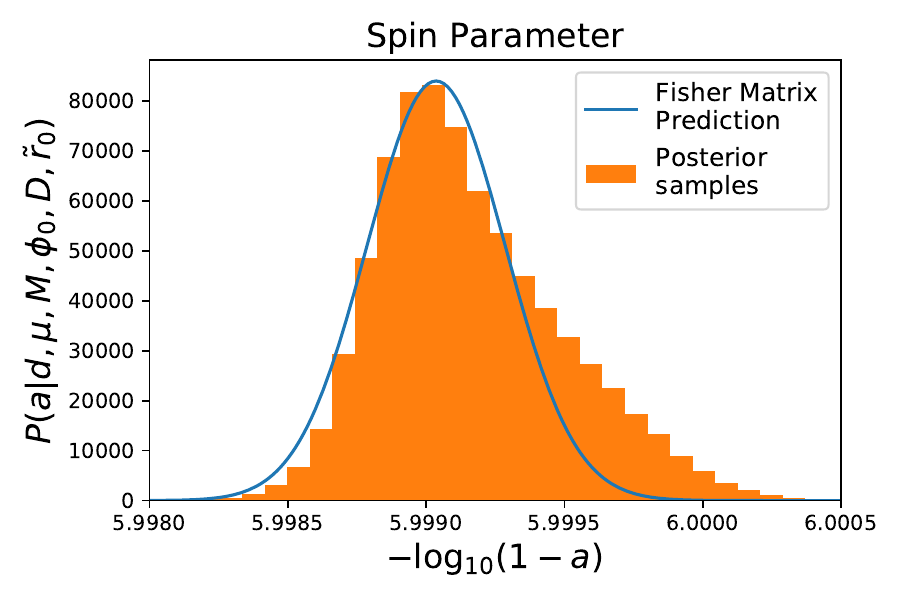}
\end{subfigure}\hfil 
\begin{subfigure}{0.25\textwidth}
  \includegraphics[height =4cm, width = 5cm]{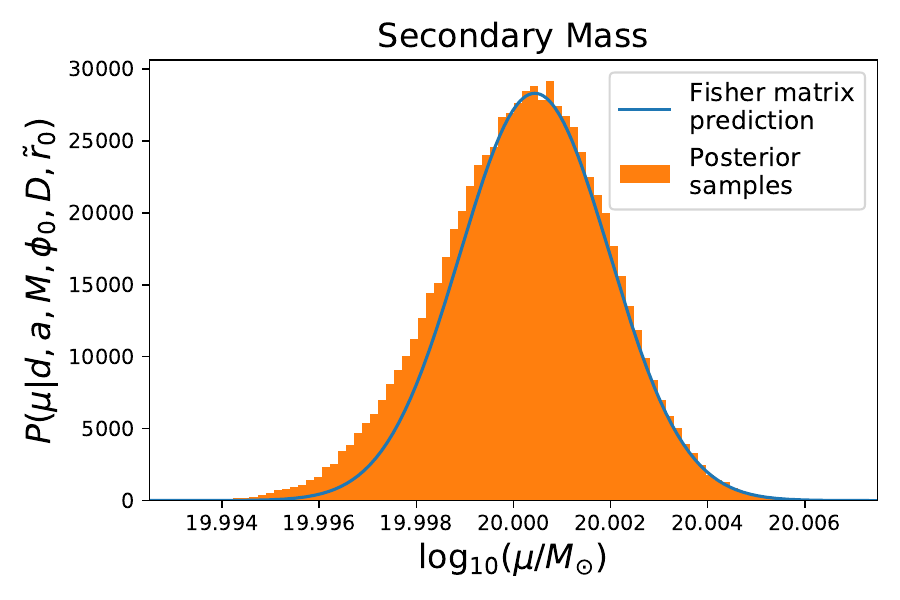}
\end{subfigure}\hfil 
\begin{subfigure}{0.25\textwidth}
  \includegraphics[height =4cm, width = 5cm]{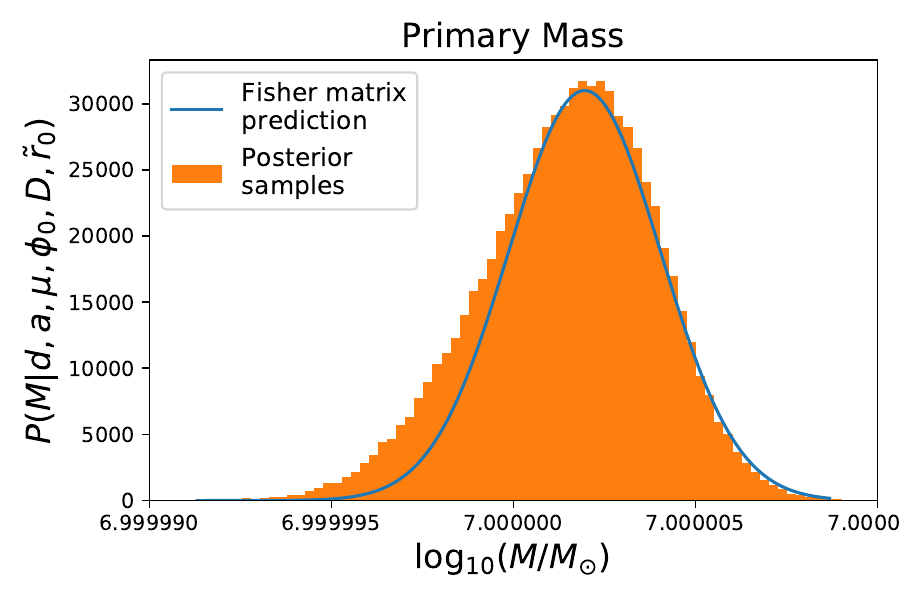}
\end{subfigure}

\medskip
\begin{subfigure}{0.25\textwidth}
  \includegraphics[height =4cm, width = 5cm]{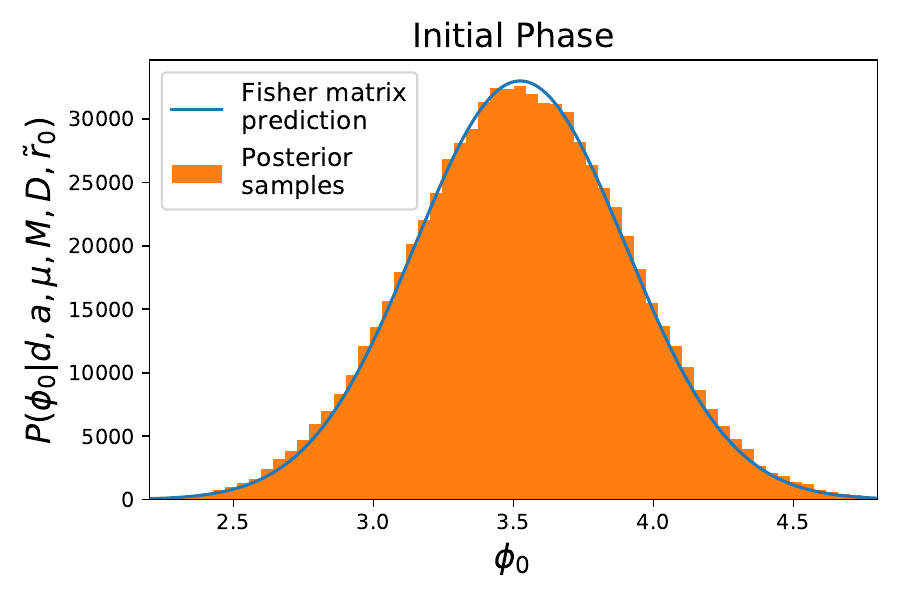}
\end{subfigure}\hfil 
\begin{subfigure}{0.25\textwidth}
  \includegraphics[height =4cm, width = 5cm]{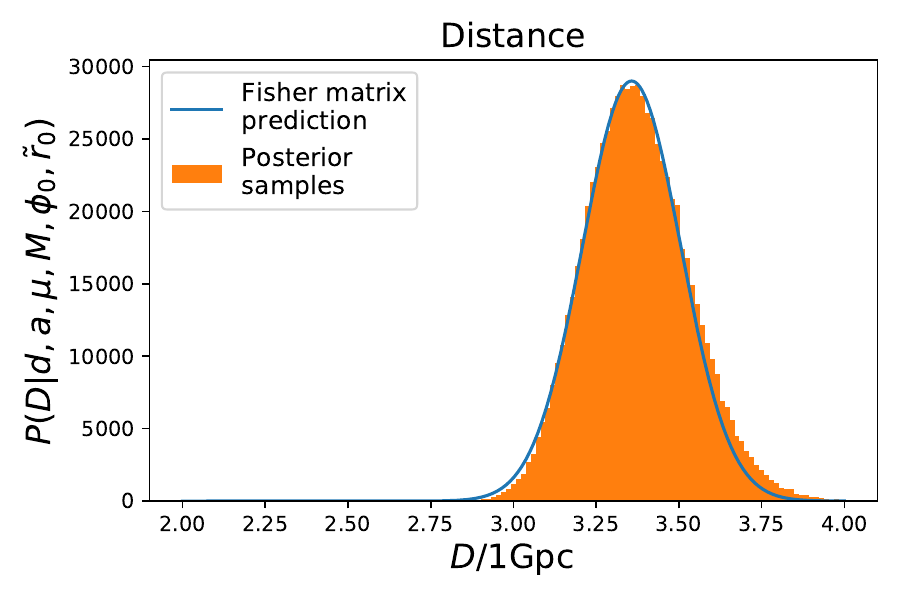}
\end{subfigure} 
\begin{subfigure}{0.25\textwidth}
  \includegraphics[height =4cm, width = 5cm]{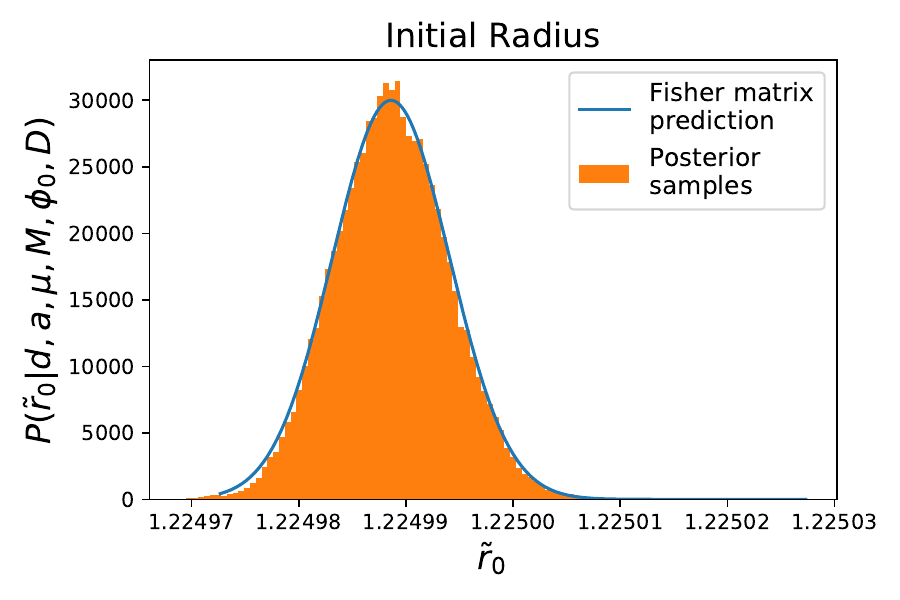}
\end{subfigure}
\caption{For the source with parameters $\boldsymbol{\theta}_{\text{heavy}}$,we compare the one-dimensional marginalised posterior distributions (orange histograms) to a Gaussian distribution (blue solid line), centred at the posterior mean, and with standard deviation set to the prediction of the Fisher matrix.}
\label{fig:Fishy_Posterior_heavy}
\end{figure*}

\begin{figure*}
    \centering 
\begin{subfigure}{0.25\textwidth}
  \includegraphics[height =4cm, width = 5cm]{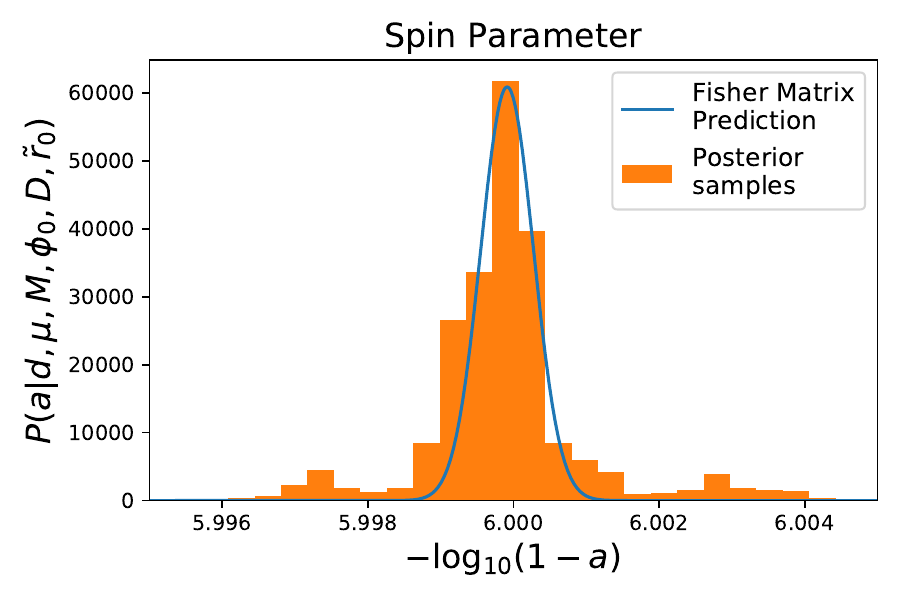}
\end{subfigure}\hfil 
\begin{subfigure}{0.25\textwidth}
  \includegraphics[height =4cm, width = 5cm]{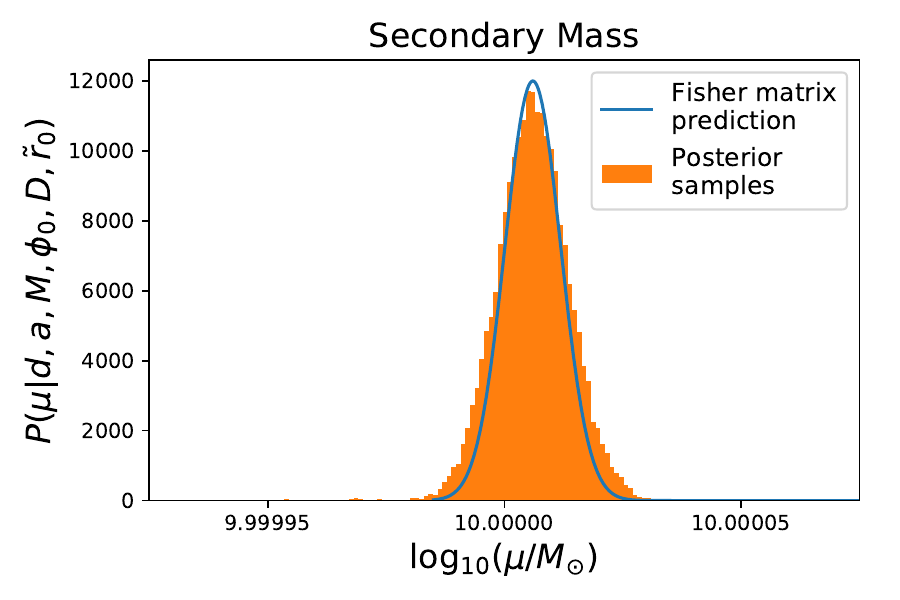}
\end{subfigure}\hfil 
\begin{subfigure}{0.25\textwidth}
  \includegraphics[height =4cm, width = 5cm]{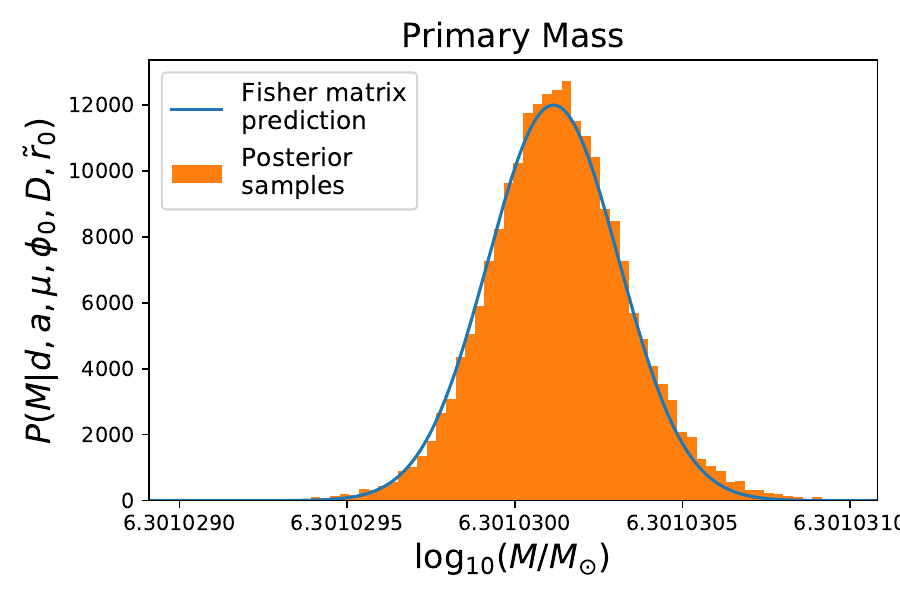}
\end{subfigure}

\medskip
\begin{subfigure}{0.25\textwidth}
  \includegraphics[height =4cm, width = 5cm]{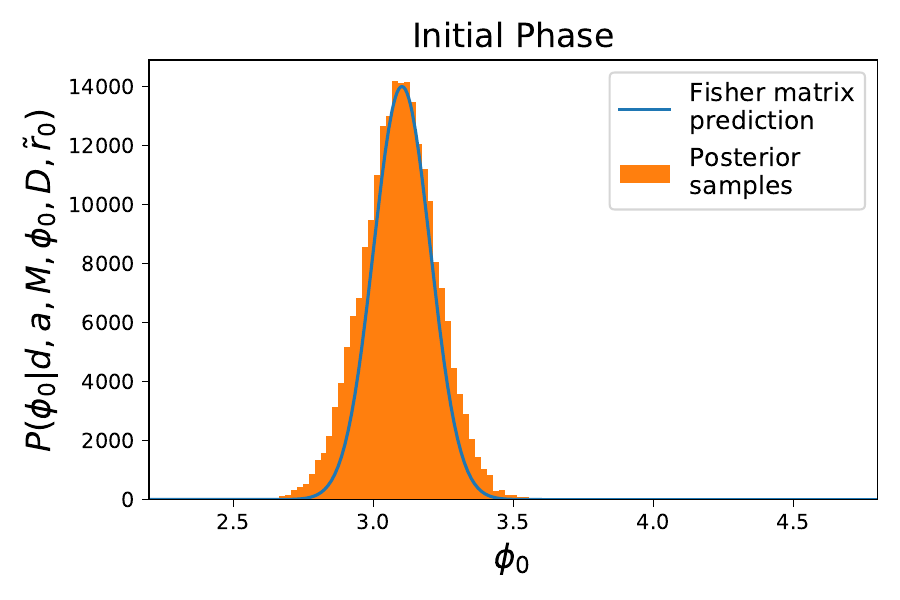}
\end{subfigure}\hfil 
\begin{subfigure}{0.25\textwidth}
  \includegraphics[height =4cm, width = 5cm]{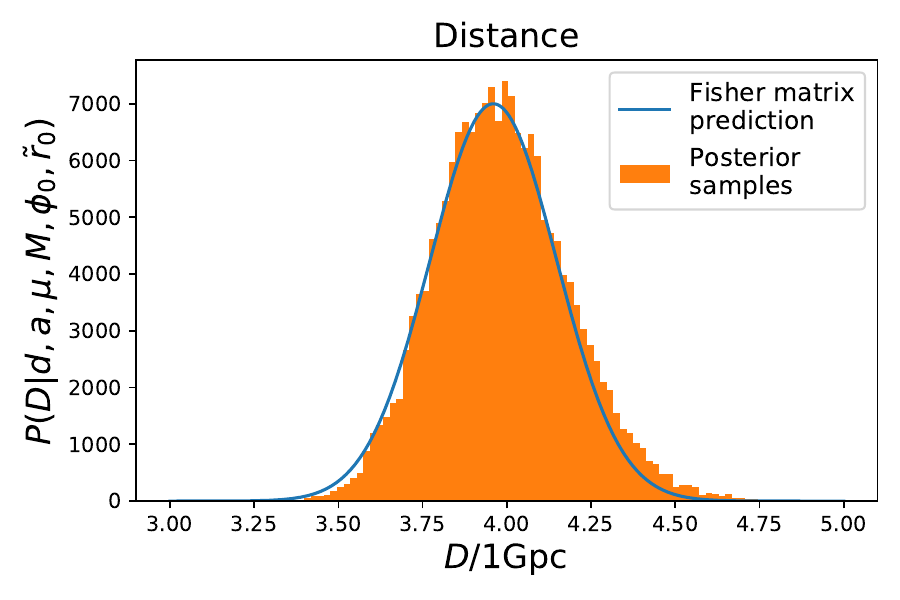}
\end{subfigure} 
\begin{subfigure}{0.25\textwidth}
  \includegraphics[height =4cm, width = 5cm]{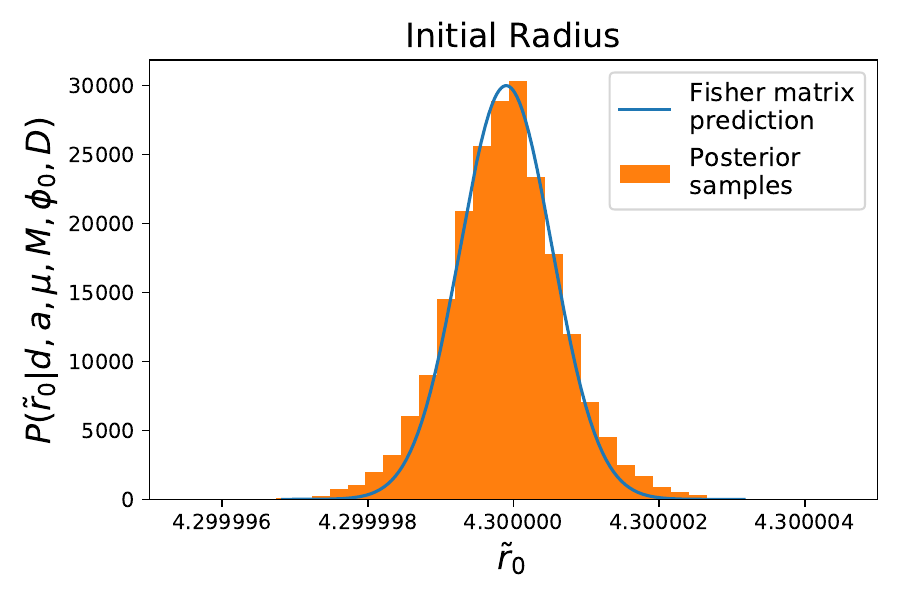}
\end{subfigure}
\caption{As Figure~\ref{fig:Fishy_Posterior_heavy}, but now for the lighter source with parameters $\boldsymbol{\theta}_{\text{light}}$.}
\label{fig:Fishy_Posterior_light}
\end{figure*}

The primary reason for doing the MCMC simulations was to verify the Fisher Matrix results found earlier. In figures \ref{fig:Fishy_Posterior_heavy} and \ref{fig:Fishy_Posterior_light}, we plot the marginalised posteriors on the parameters $\{ \boldsymbol{\theta}_{\text{light}},\boldsymbol{\theta}_{\text{heavy}}\}$ alongside a Gaussian distribution with variance given by the Fisher matrix and centred at the mean value of the posterior distributions $p(\boldsymbol{\theta}|d)$. 

These results nicely confirm the accuracy of the Fisher matrix results for these sources. In each case, the $1\sigma$ precision predicted by the Fisher matrix is slightly smaller than the width of the numerically computed posterior. This is to be expected as the Fisher matrix also provides the \emph{Cramer-Rao lower bound} on parameter uncertainties. However, the difference is very small. We are thus confident that all of our Fisher matrix predictions are accurate, including the exploration of parameter space shown in Figure~\ref{fig:Fisher_Matrix_Whole_Param_Plots}. We conclude that even at the near-threshold SNR of this source, $\rho \sim 20$, the Fisher matrix can be used to confidently estimate the precision of parameters for near-extremal sources.

\section{conclusion}\label{conclusion}

This work has shown that we expect tighter spin constraints on the Kerr spin parameter for EMRIs when the primary is rapidly rotating rather than moderately rotating. We argued that the spin precision is sourced through the rapid growth of the spin dependence on the radial trajectory throughout the near-ISCO regime. It was shown that radiation-reaction effects are subdominant to geodesic motion, which is what ultimately drives the excellent constraint on the spin parameter.

This work has tried to answer the question of how well LISA can constrain the spin parameter of an EMRI source, conditioned on the primary rapidly rotating. In the earlier sections of this paper, we were able to derive a \emph{general} linear ODE whose solution describes the spin dependence on the radial trajectory. Using this ODE, we were able to argue that this spin dependence is much more significant for near-extremal primaries rather than moderately spinning primaries $a \sim 0.9$. It was also shown that the spin dependence on the radial evolution is dominated more by geodesics, rather than the radiation reactive terms.  

Using reasonable assumptions, analytic Fisher scalars were derived for both rapidly rotating and moderately spinning black holes. It was then shown, explicitly, that the precision on spin parameters for near-extreme black holes should exceed that of moderately spinning ones by a few orders of magnitude. The exceptional precision is governed by the growth of the spin dependence of the trajectory near the ISCO. 

Given that the kinematical terms dominate the precision, we were able to build a near-extremal waveform model by removing the spin dependence in corrections to the radial fluxes. This meant we could build an interpolant, valid for quasi-circular and equatorial inspirals, into rapidly spinning black holes with $a \gtrsim 0.999$. Using this approximation, we were able to build a waveform model valid for near-extremal spins that could be used for parameter precision and estimation studies.

Our analysis showed that the LISA optimum masses for parameter estimation and precision studies are  heavier mass systems $M \approx 10^{7}M_{\odot}$. For heavier sources, the emitted frequencies will lie within the minimum of the LISA PSD during the dampening regime; when the signal provides most accurate information about the source. Analysing different parts of the signal, we were able to conclude that the end part of the signal will be ``loud" in the data stream for spins approaching near-extremality $a \rightarrow 1.$ For lighter mass sources $M\sim 10^{6}M_{\odot}$, the accumulated SNR of the dampened part of the signal is weak. Fully numerical Fisher matrix analysis revealed that we can constrain the spin parameter of a near-extremal EMRI $\sim 3$ orders of magnitude higher than moderately spinning black holes $a \sim 0.9$. For very near-extremal primaryes $a \sim 1-10^{-9}$, one is able to constrain the spin parameter with precision $\Delta a \sim 10^{-10}$. This is, as far as we know, the tightest constraint on astrophysical orbital parameter found in the literature. 

In the final section, with the view of verifying our Fisher matrix calculations, we performed a parameter estimation study on these near-extremal sources. We showed that the signal parameters are able to be extracted with precision comparable to our Fisher matrix estimates. 

To conclude, we can safely say that the spin parameter of near-extremal EMRIs can be measured with \emph{excellent} precision. We are also confident that the increase in precision is governed by the spin dependence on the radial trajectory. If the dampened part of the signal is not observed (where $\partial_{a}\tilde{r} \gg 1$), then we cannot make such precise statements on the spin of the primary black hole. 

There are a few extensions to this work. The first would be to include the results of the transition from inspiral to plunge in \cite{burke2019transition} to our analysis here. We suspect that, although the SNR accumulation would be weak, the extra contribution from $\partial_{a}\tilde{r}$ could only \emph{improve} the precision measurement. Another interesting direction to explore would be the effect of systematic errors (through innacurate waveform modelling) in our EMRI PE studies. Through the Cutler and Vallisneri formalism ~\cite{cutler2007lisa}, one could investigate whether parameter estimation/precision studies on near-extremal waveforms could be dominated by systematic errors rather than statistical ones induced purely by noise realisations from the detector. Given that we have shown precisions on the spin parameter on the order $\Delta a \sim 10^{-10}$, it would be of utmost importance to have immensely accurate waveform models in order to keep the systematic error to a minimum.  This would be to ensure that parameter estimates are not affected by (deterministic) biases due to waveform modelling errors rather than (probabilistic) biases due to noise fluctuations of the LISA detector.

Another obvious extension would be to introduce more parameters in the parameter space, such as eccentricity and inclination. In this case, we are unsure whether the kinematic terms would dominate over the flux corrections. As such, we would require a more complex and general waveform model. As seen in \cite{2016CQGra..33o5002G}, eccentric trajectories of secondaries in the vicinity of near-extremal black holes exhibit \emph{inverse} zoom-whirl behaviour. That is, the ``zoom" part of the inspiral (near aperiastron) outputs larger amplitude radiation than the ``whirl" phase. Perhaps this extra information, unique to near-extreme EMRIs, could provide even tighter constraints on the spin parameter $a$. It would be interesting to see whether the larger number of parameters will aid or hinder parameter precision/estimation studies. Finally, due to the precision on both $\Delta a$ and $\Delta M$, near-extremal EMRIs would provide stringent tests of GR theories. One of which being the no-hair theorem which states that the Kerr black hole can be uniquely parametrized in terms of two charges; mass and spin. We believe that it should be possible to measure the multipolar moments more accurately when the primary is near-extremal than if it were only moderately spinning. A paper exploring potential tests of general relativity using the near-extremal Kerr spacetime would be useful for the literature.

\begin{acknowledgments}
We give thanks to Maarten van de Meent for both providing high spin Teukolsky amplitudes and fluxes used throughout this work and for advice on the waveform generation. We also give thinks to Niels Warburton for advice on how to use the \BHPT and general comments on the results. Finally, we wish to acknowledge Lorenzo Speri and Andrea Antonelli for useful comments on the manuscript and for many fruitful discussions/support leading to the completion of this work. This work made use of data hosted as part of the Black Hole Perturbation Toolkit at \href{http://bhptoolkit.org/}{bhptoolkit.org}.  ME's and JG's work was supported by UK Space Agency Grant ST/R001901/1. We also give thanks to the AEI for use of their high performance computing facilities.
\end{acknowledgments}

\appendix

\section{Analytic estimation of the Fisher matrix}
\label{app:fish-estimation}

The derivation of the Fisher matrix estimate \eqref{eq:fish-estimation} is provided below. 

Consider the gravitational wave amplitude
\begin{equation}
  h(t) = \sum_m h_m(t) \approx \sum_m \frac{2\sqrt{\dot{E}^{\infty}_{ m}}}{m\tilde{\Omega}\tilde{D}}\,\sin (m\tilde{\Omega}\tit)\,,
\end{equation}
where we chose $\phi=0$ for simplicity. The spin dependence of each individual amplitude equals
\begin{equation}
  \partial_a h_m(t) = |h_m(t)|\left\{\sin(m\tilde{\Omega}\tit)\,{\cal B}_m
  + (m\tit\partial_a\tilde{\Omega})\,\cos(m\tilde{\Omega}\tit)\right\}
\end{equation}
where $|h_m(t)|$ stands for the amplitude without oscillatory factor and we defined
\begin{equation}
    {\cal B}_{m} \equiv \frac{\partial_a\dot{E}^{\infty}_{m}}{2\dot{E}^{\infty}_{m}} - \frac{\partial_a\tilde{\Omega}}{\tilde{\Omega}}\,.
\end{equation}
Using Parseval's identity, the integrand in the Fisher matrix can be decomposed
\begin{equation}
  |\partial_a h(t)|^2 = \sum_m (\partial_a h_m)^2 + 2\sum_{n<m} \partial_a h_n\,\partial_a h_m\,,
\end{equation}
into \emph{diagonal} and \emph{off-diagonal} contributions. Consider the diagonal ones, first. Using $\sin^2x = (1-\cos 2x)/2$ and $\cos^2 x=(1+\cos 2x)/2$, any such contribution equals
\begin{multline}
  (\partial_a h_m)^2 = \frac{|h_m|^2}{2}\left\{(m\tit\,\partial_a\tilde{\Omega})^2 + \left({\cal B}_m\right)^2 \right. \\
  \left.+ \cos (2m\tilde{\Omega}\tit) \left[(m\tit\,\partial_a\tilde{\Omega})^2 - \left({\cal B}_m\right)^2\right] \right. \\
  \left. +2\sin(2m\tilde{\Omega}\tit)\,{\cal B}_m\,(m\tit\,\partial_a\tilde{\Omega})\right\}\,.
\end{multline}
The crucial observation is $\tit\sim \mathcal{O}(\eta^{-1})$. This follows from integrating \eqref{ODE:radial_evolution} and is an inherent consequence of adiabatic inspirals evolving on the orbital timescale. Hence, we expect the dominant term to be the one of order $\mathcal{O}(\tit^2)$. Furthermore, the second and third lines above have oscillatory behaviour, which when integrated in \eqref{eq:fish-approx} will give rise to subleading contributions. Despite the robustness of these arguments, let us state the precise condition under which the
order $\mathcal{O}(\tit^2)$ term is dominant
\begin{equation}
  \left|\frac{\partial_a \dot{E}^{\infty}_{ m}}{\tit\,\dot{E}^{\infty}_{m}}\right| \ll \partial_a\tilde{\Omega}\,.
\end{equation}
Notice that both partial derivatives should be understood as in the left hand side of our equations \eqref{eq:partiala}. Since $\dot{E}^{\infty}_{m}$ typically requires numerical evaluation, we have indeed numerically checked this assumption holds for both, moderately and near-extremal primary holes. Thus, the contribution from the diagonal terms can be approximated by
\begin{equation}\label{app:eq:assumption}
  \sum_m (\partial_a h_m)^2 \approx \frac{|h_m|^2}{2}\,(m\tit\,\partial_a\tilde{\Omega})^2\,.
\end{equation}
It is worth stressing that for sources lying entirely in the close to ISCO region of a near extremal primary, this condition is under more analytical control using the NHEK flux \eqref{eq:approx:NHEK}. Whenever $\partial_a \log \tilde{C}_{\infty m} \ll \tit\,\partial_a\tr$ holds (which we checked numerically), the full condition reduces to $x\,\tit \gg 1$, which is satisfied for all secondary locations $x$ before entering the transition regime.

The discussion of the off-diagonal terms is very similar. Any such term can be written as
\begin{multline}
  \partial_a h_m\,\partial_a h_n = \frac{|h_m||h_n|}{2}\,\\
  \left\{ \cos((m-n)\tilde{\Omega}\tit)\,\left[{\cal B}_m\,{\cal B}_n + mn(\tit\,\partial_a\tilde{\Omega})^2\right] \right. \\
  \left.+ \cos((m+n)\tilde{\Omega}\tit)\,\left[-{\cal B}_m\,{\cal B}_n
   + mn(\tit\,\partial_a\tilde{\Omega})^2\right] \right. \\
  \left. + \sin((m+n)\tilde{\Omega}\tit) \left[n\tit\,\partial_a\tilde{\Omega}\,{\cal B}_m + m\tit\,\partial_a\tilde{\Omega}\,{\cal B}_n \right]\right. \\
  \left. + \sin((m-n)\tilde{\Omega}\tit) \left[n\tit\,\partial_a\tilde{\Omega}\,{\cal B}_m - m\tit\,\partial_a\tilde{\Omega}\,{\cal B}_n \right]\right\}\,.
\end{multline}
The key point is that \emph{all} such terms involve oscillatory functions, even the ones having $\mathcal{O}(\tit^2)$ dependence. Hence, when compared to the contribution from the diagonal ones, they are subleading. This justifies our claim in \eqref{eq:fish-estimation}.

\section{General Relativistic Corrections}\label{app:GRCs}

\begin{figure*}
\includegraphics[width = 0.49\textwidth]{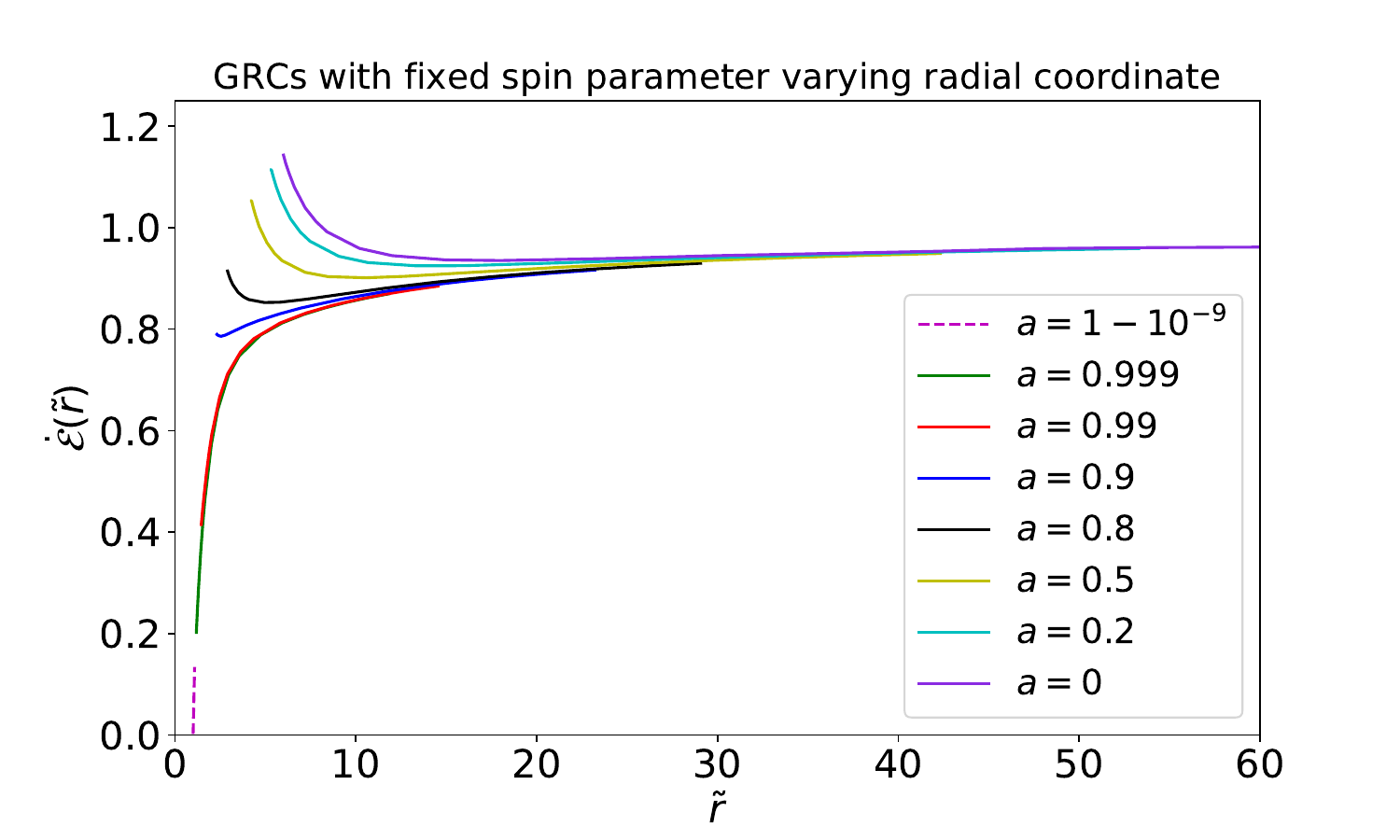}
\includegraphics[width = 0.49\textwidth]{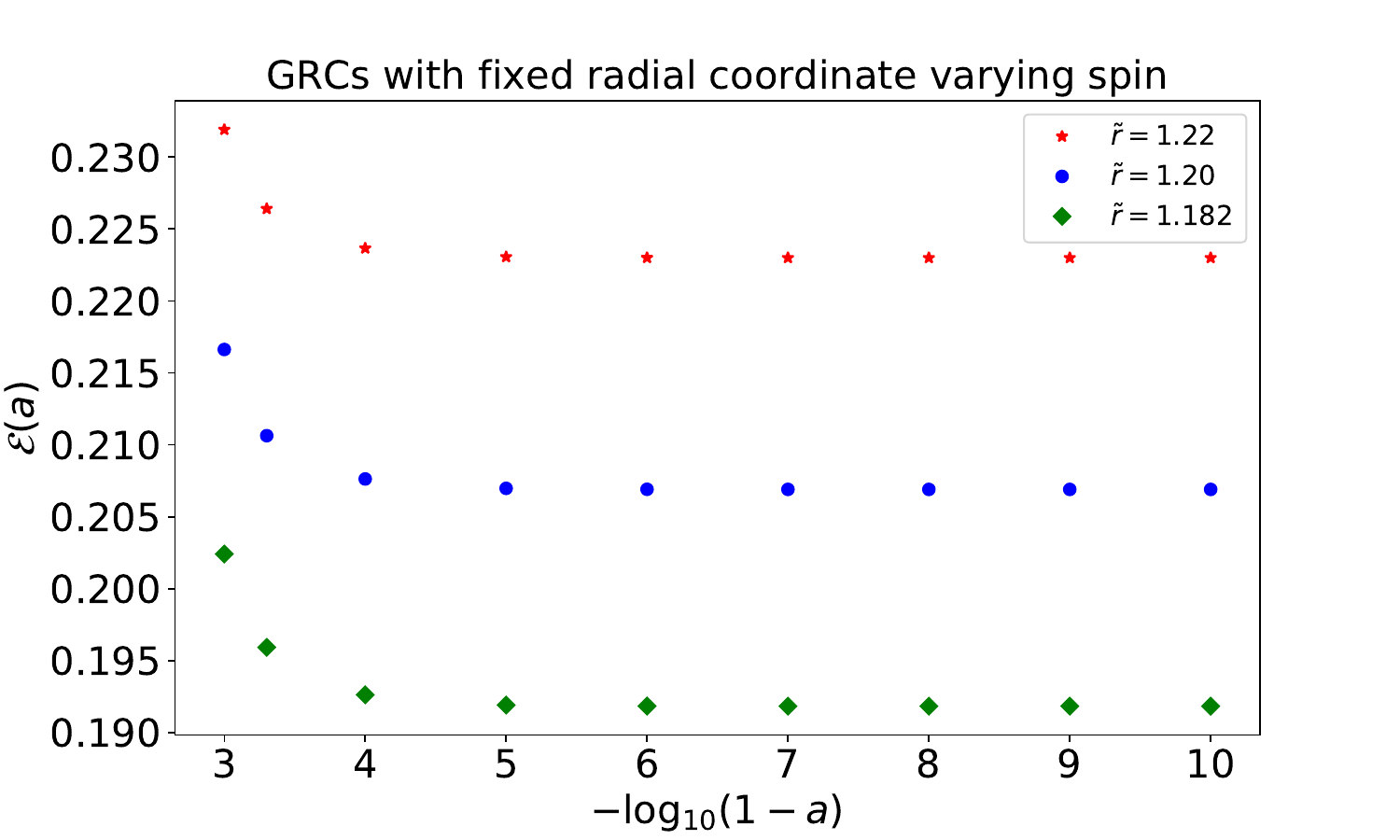}
\caption{The figure on the left shows the values of the GRCs $\dot{\mathcal{E}}(\tilde{r})$ evaluated at a fixed spin parameter. The purple dashed line (bottom left of leftmost plot) uses the near-extremal approximation to the flux \eqref{NielsEnergy}. The other colours are the tabulated values GRCs presented in Table IV in Ref.\cite{2000PhRvD..62l4021F}. The figure on the right shows the GRCs evaluated at a fixed radial coordinate $\tilde{r}$ whilst varying the spin parameter $a$. The GRCs for $a > 0.999$ were computed using the \href{http://bhptoolkit.org}{BHPT} \cite{BHPToolkit}.}
\label{fig:EpsFixedSpin}
\end{figure*}

We begin by focusing on the total energy flux $\dot{\tilde{E}}.$ The left most panel of figure \ref{fig:EpsFixedSpin} shows $\Eps(\tilde{r})$ at fixed $a$ and the right most panel $\Eps(a)$ at fixed $\tilde{r}$.
The left plot of Fig.(\ref{fig:EpsFixedSpin}), shows that for moderately rotating holes with $a \lesssim 0.99$, the corrections $\Eps$ are not strongly dependent on the radial coordinate $\tilde{r}.$ However, approaching an extremal spin parameter $a$ indicates that $\Eps$ rapidly goes to zero as the ISCO is approached. This matches the description given by both NHEK fluxes \eqref{eq:approx:NHEK} and \eqref{NielsEnergy}.

Using a Teukolsky solver from the \BHPT, we generated our own energy flux values for spin parameters $a \geq 0.999$. The corresponding corrections are shown in the right panel of figure \ref{fig:EpsFixedSpin} for fixed coordinate radii. This shows that at a fixed coordinate radius $\tilde{r}$, the corrections approach a constant as $a \rightarrow 1$. What we learn from this is that the behaviour of the relativistic corrections becomes somewhat universal as $a \to 1$. In other words, as the near extremal parameter approaches unity, the corrections approach a constant value. This means that the correction values at some $\tilde{r}$ will not differ much as the spin parameter is changed providing the spin parameter is close to one. That is, spin dependence on $\Eps$ is weak ``far" from ISCO. 

\begin{figure}[!ht]
    \centering
    \includegraphics[width = 0.49\textwidth]{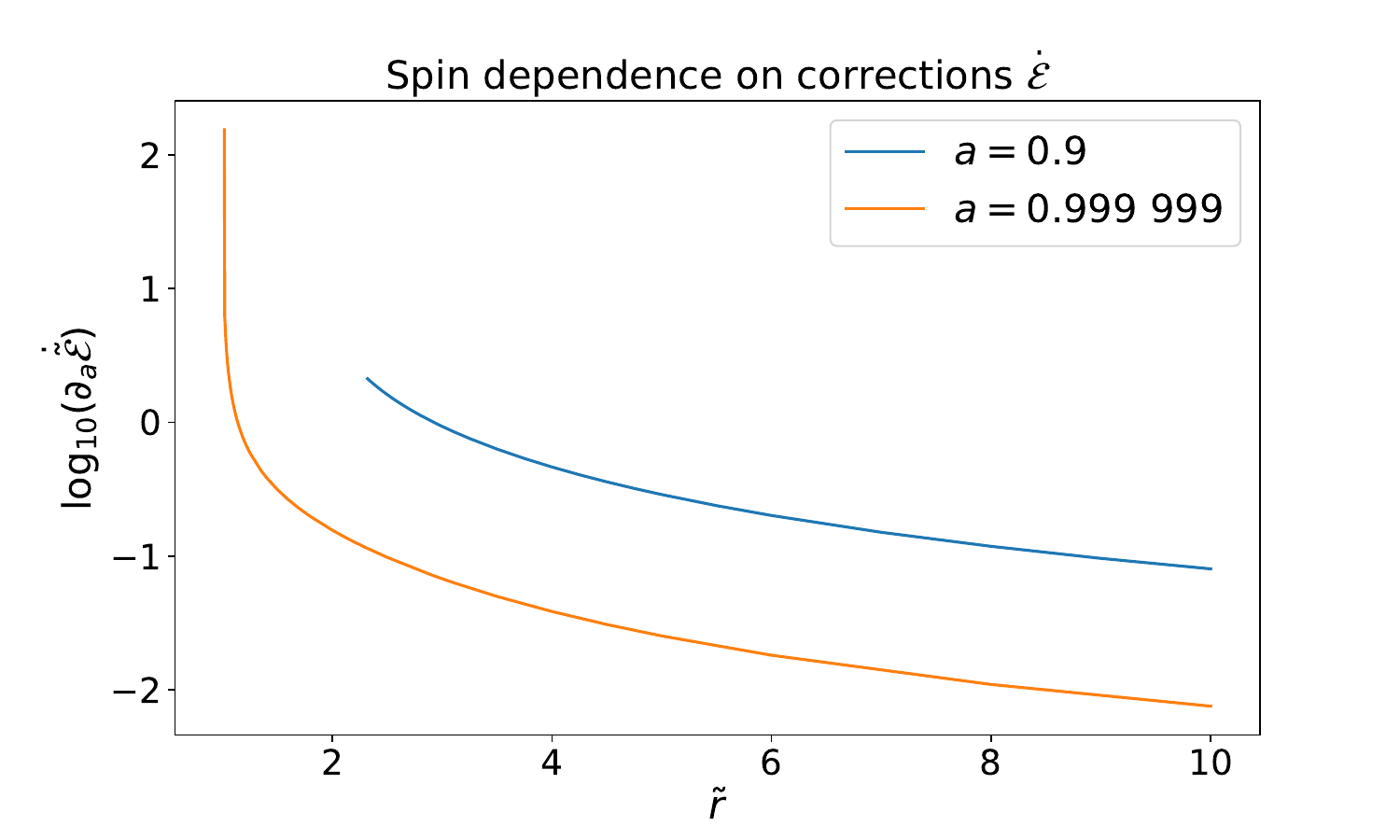}
    \caption{Here we compute $\partial_{a}\Eps$ for the two spin parameters $a = 0.999999$ (orange) and $a = 0.9$ (blue). The growth in $\partial_{a}\dot{\mathcal{E}}$ for rapidly rotating holes is larger closer to ISCO than for moderately spinning holes.}
    \label{fig:spin_dependence_corrections}
\end{figure}

We emphasise that this does \emph{not} mean that $\partial_{a}\Eps\approx 0$ throughout the entire inspiral. Here we have shown three radial coordinates which are considerably far away from the ISCO in the near-extreme spin parameter case. The closer in radial coordinate to the ISCO, the larger the spin dependence in the relativistic corrections. This can be observed in Fig.(\ref{fig:spin_dependence_corrections}).
For a weak gravitational field, the spin dependence on the flux corrections for near-extreme inspirals is small, approaching $\dot{\tilde{E}}\rightarrow 0$ for $a \rightarrow 1$ as the ISCO is approached. Much smaller than when compared to moderately spinning holes, where the total energy flux is approximately constant. Only when very close to the ISCO does the spin dependence on $\Eps$ grow significantly for near-extremal holes. 

\begin{figure}[!ht]
\includegraphics[width = 0.49\textwidth]{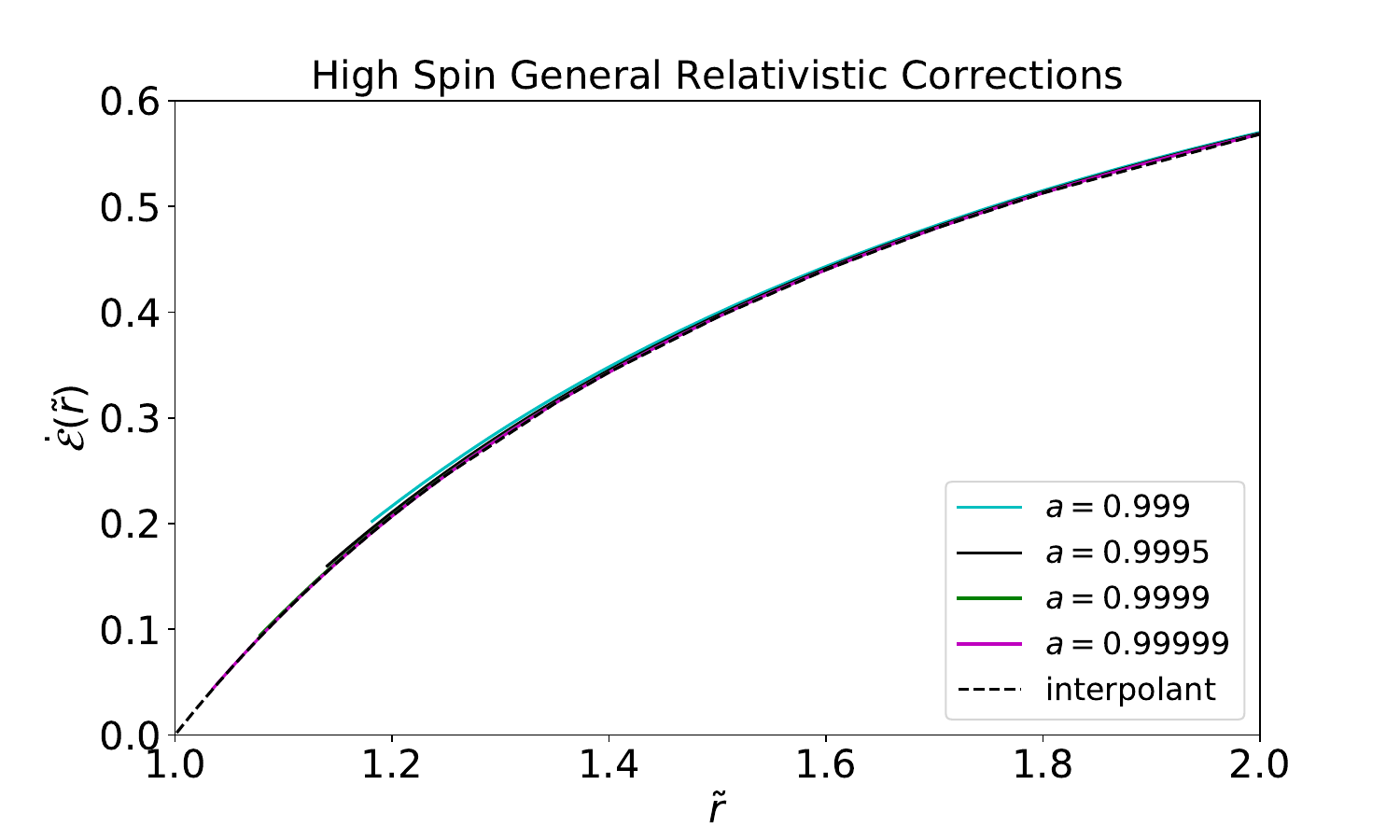}
\includegraphics[width = 0.49\textwidth]{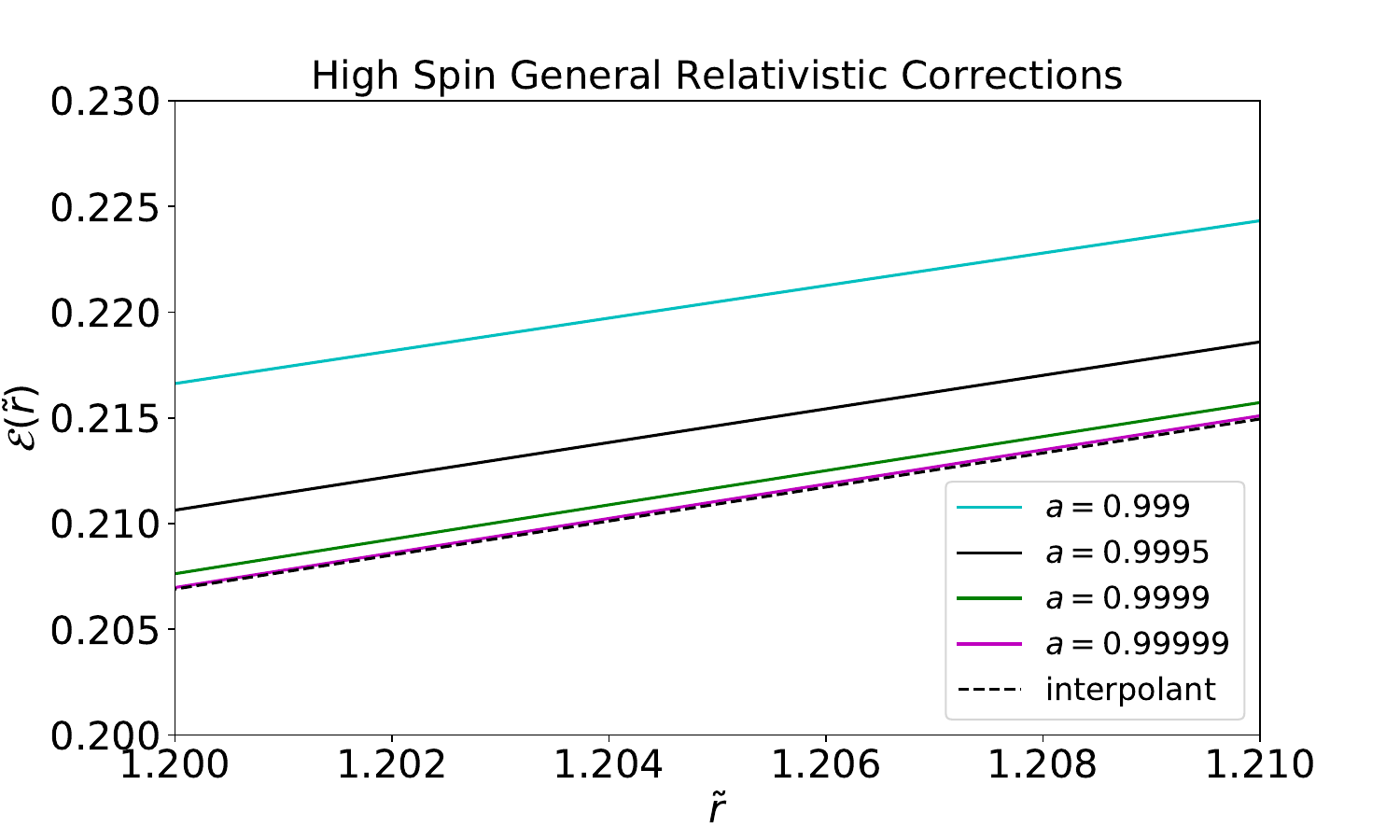}
\caption{On the left panel, we have used the \href{http://bhptoolkit.org}{BHPT} to compute the total energy flux for near-extremal spin parameters with our interpolant (black dashed line) overlaid. The right panel is a zoom in of the left plot giving visual aid as to why our interpolant can be used to approximate a larger regime of spin parameters.}
\label{fig:GRC_Far_Away}
\end{figure}

A comparison of our interpolant, $\Eps$ for $a = 1-10^{-9}$ with high spinning fluxes computed from the \BHPT \ are given in Fig.(\ref{fig:GRC_Far_Away}).
A useful approximant for the relativistic corrections $\Eps(r)$ is given by 
\begin{equation}
    \Eps(x) = ax\arctan(bx^{c}), \quad x = \tilde{r} - 1
\end{equation}
with constants $(a,b,c) = (0.6897912,1.084803,-0.936685).$ This approximates the high spin corrections $\Eps(r,1-10^{-9})$ for $\tilde{r} \lesssim 6$ with largest fractional error $\sim 1\%$ at the ISCO.

\section{Verification of Fisher Matrix Calculations}
\label{app:verification}
\subsection{Ill-Conditioned Fisher Matrix}
It is well known throughout the literature~\cite{porter2015fisher,porter2009overview,abuse_fisher,wen2005detecting,amaro2018relativistic,gair2013testing,rodriguez2012verifying} that a Fisher matrix analysis, when not handled carefully, has the potential to under-estimate (or, worse, over-estimate) precision measurements. One first has to be sure that the regime of SNR is high enough so that the linear signal approximation applies when truncating the perturbed waveform at first order. Numerical derivatives also exhibit convergence problems and are very problematic if your waveforms are not smooth. One must also make sure to taper each derivative so that  ``hard cut-offs" do not feature in the waveform~\cite{mandel2014parameter}.  Finally, one has to invert the Fisher matrix and, if the Fisher matrix is ill conditioned (as normally the case for EMRIs), it could lead to catastrophic errors in in the elements of $\Gamma^{-1}$. The condition number of a matrix is defined through
\begin{equation}\label{condition_number}
    \text{Cond}(\Gamma) =\bigg\rvert \frac{\max_{i}\{\lambda_{i}\}}{\min_{j}\{\lambda_{j}\}}\bigg\rvert,
\end{equation}
where $\lambda_{i}$ is an $i^{\text{th}}$ eigenvalue of the matrix $\Gamma$. In the case of EMRIs, the computation of the fisher matrix is affected by numerical instabilities. A small perturbation to the systems (intrinsic) parameters lead to a large overall change in the waves phase evolution. It is for this reason EMRI observations permit parameter constraints with such high precision. Since waveforms are sensitive to small perturbations of the source parameters, the numerical derivatives $\partial _{\theta} h(\boldsymbol{\theta};t)$ are large and so the elements of the Fisher matrix can be enormous in magnitude. In other words, there is a significant amount of ``information" about the parameters $\boldsymbol{\theta}$ encoded in the waveforms, which are then reflected by the large elements of the Fisher matrix. However, not all derivatives are large and the differences between the size of the elements for different parameters leads to significantly varying eigenvalues. In our case, the measurements of spin and distance are $\sim 8$ orders of magnitude apart. A consequence of this is that the condition number of the Fisher matrix is $\text{Cond}(\Gamma) \sim 10^{21}$. In light of these known instabilities, we spend the next few subsections providing suitable tests to verify our Fisher matrix in the single parameter case, assuming all other parameters are known perfectly. In order to verify Fisher matrices over multiple parameters, Bayesian techniques like MCMC are required (see sec. \ref{MCMC}).

We used three different methods to verify the Fisher matrix calculations without performing a Bayesian analysis:
\begin{enumerate}
\item Verification of the linear signal approximation 
\begin{equation*}
    h(a + \Delta a) \approx h(a) + \Delta a \frac{\partial h}{\partial a}.
\end{equation*}
\item Overlaps defined through \eqref{eq:overlap} --- perturbing the spin parameter by $\Delta a$ given  by the Fisher matrix should return overlaps close to one.
\item Likelihood --- The log-likelihood is maximised at the true parameters $\boldsymbol{\theta}_{0}.$ If a parameter is perturbed by the Fisher matrix estimate then it should be a measure of the $1\sigma$ width of this log-likelihood.
\end{enumerate}
As a proof of principle of these methods, we will compute the precision on the spin parameter for a heavy source $\boldsymbol{\theta}_{\text{heavy}}$ (see \eqref{params:heavy}), viewed face-on with $\rho \sim 20.$ For a source with this configuration of parameters, we found $\Delta a_{\text{NHEK}} \sim 2\times 10^{-10}$ numerically.

\subsubsection{Linear-Signal approximation}
In the derivation of the Fisher matrix, we used the linear-signal approximation so a first test would be to test whether it is valid in our analysis. To test whether the expansion is valid in the regime of SNR we are considering, we compute the overlap
\begin{equation}
    \mathcal{O}\left(h(a +\Delta a) | h(a) + \Delta a \frac{\partial h}{\partial a}\right)\bigg\rvert_{ a = a_{\text{true}}} \approx 1 - 10^{-5}
\end{equation}
Hence we conclude that our waveform model at $\rho \sim 20$ does not violate the linear-signal approximation. 

\subsubsection{Overlaps}

In the limit of high SNR, the inner product \eqref{eq:Inner_Prod} can be expanded in $\rho$. Observe for small $\Delta a$
\begin{equation*}
(h(a + \Delta a) | h(a)) \approx \rho^{2} + \Delta a\left(\frac{\partial h}{\partial a}\bigg\rvert h\right) + \frac{\Delta a^{2}}{2}\left( h\bigg\rvert\frac{\partial^{2}h}{\partial a^{2}}\right)
\end{equation*}
where $(h|h) = \rho^{2}$ as in \eqref{continuous_SNR}. Since the SNR is fixed, it's easy to show that $(\partial_{a} h | h) = 0$ and $\Gamma_{aa} = -(\partial_{a}^{2}h|h).$ It can then be shown for $\rho \gg 1$ 
\begin{equation}\label{Overlap_Estimate}
\mathcal{O}(h(a+\Delta a),h(a)) \approx 1 - \frac{1}{\rho^{2}} + O\left(\rho^{-4}\right).
\end{equation}
Substituting $\Delta a = \Delta a_{\text{NHEK}}$ into the left hand side of \eqref{Overlap_Estimate}, we numerically find an agreement of $\sim 0.01\%$. 


\subsubsection{Likelihood}
In the high SNR limit, the precision pridicted by the fisher matrix should approximate the $1\sigma$ width of the likelihood function. 
Using Eq.\eqref{log-likelihood-function}, we can write the log-likelihood as 
\begin{equation*}
    \log p(d|a) \propto (d|h) - \frac{1}{2}(h|h).
\end{equation*}
Since the noise realisation in the data stream $d(t) = h(t;a) + n(t)$ induces a bias to the maximum likelihood estimate, and does not affect the likelihood width, we shall ignore the noise in this case. As such, we will consider a zero noise approximation and use $d = h(a)$ with signal templates $h:=h(a + \Delta a;t)$. Substituting this into the likelihood above we find that 
\begin{equation}
    \log p(d|a) \approx \frac{1}{2}(\rho^{2} -1).
\end{equation}
Here we have \emph{assumed} that $\Gamma^{-1}_{aa} = \Delta a^{2}.$ Calculating $\log p(d|a)$ for $d = h(a + \Delta a)$ with $\Delta a$ our Fisher matrix estimate we found that $\log p(d|a) \approx 199.46$ which agrees with the above formula, to a precision of 0.05\%. 

\begin{figure}
\includegraphics[width=8cm,height = 5cm]{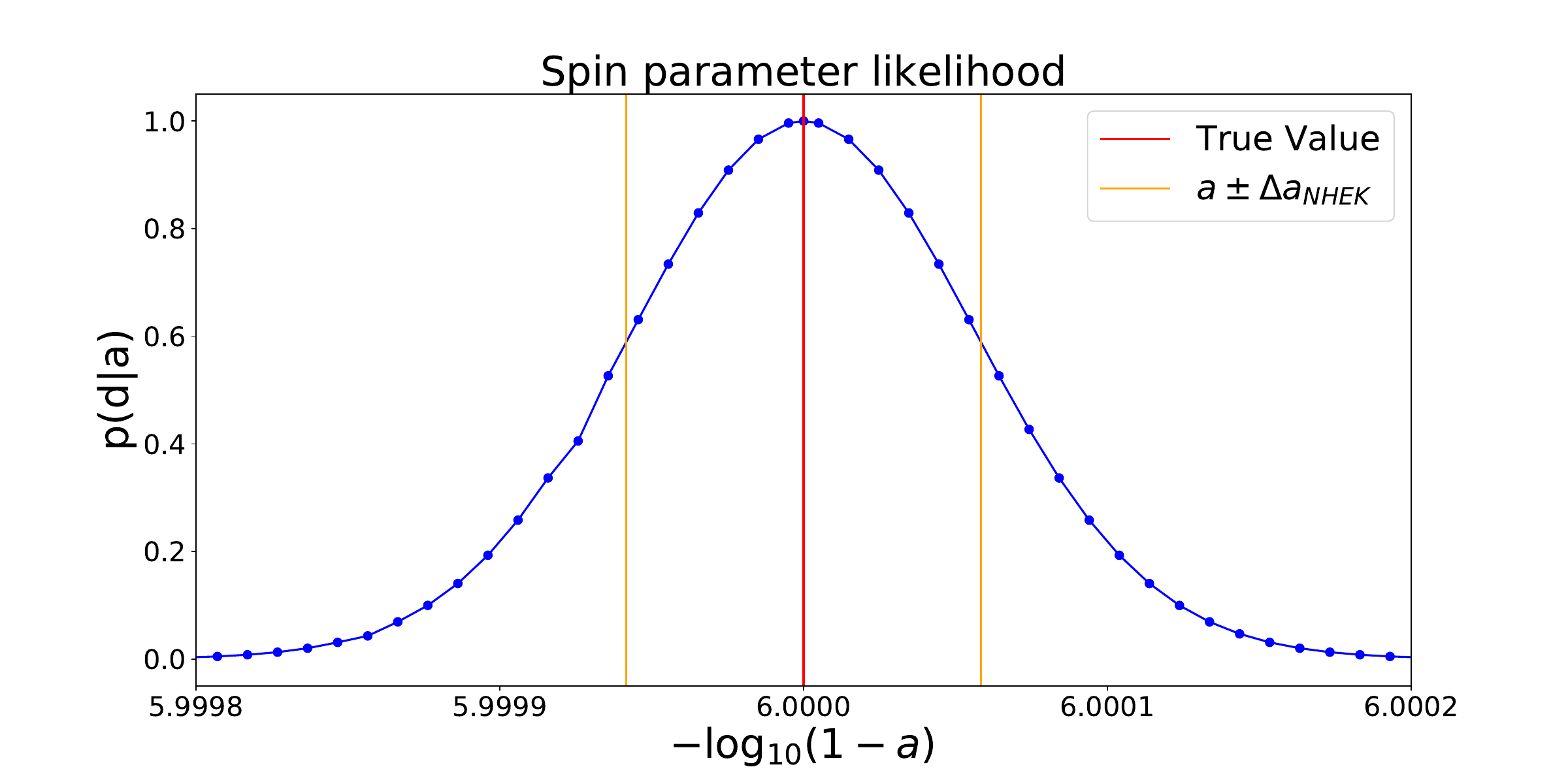}
\caption{The blue curve is the likelihood \eqref{app:likelihood} evaluated on a grid of points. The red the true value $a = 1-10^{-6}$ and orange the precision measurement predicted by the 1 parameter Fisher matrix $\Delta a_{\text{NHEK}}$.}
\label{fig:Likelihood_Fisher_Plot}
\end{figure}

Since we are interested in only one parameter, it is evaluate the likelihood function
\begin{equation}\label{app:likelihood}
p(d|a) \propto \exp[-(d-h|d-h)/2]
\end{equation}
on a grid of spin parameter values. In doing so, we find Fig.(\ref{fig:Likelihood_Fisher_Plot}). 
The area between the yellow line and red line is approximately 31.51\%, which is a reasonable approximation to the true $1-\sigma \approx 34\%$ width of likelihood. 

To conclude these subsections, we are confident that our Fisher matrix approximations in the single parameter study give a good guide to the spin parameter uncertainty.
\bibliographystyle{IEEEtran}
\bibliography{nhekpaper}

\end{document}